\documentclass[floats,aps,epsf,amsfonts,amssymb,amsmath]{revtex4}
\usepackage{multirow}
\usepackage{graphicx}
\begin{document}

\title{Gravitational self-force and the effective-one-body formalism \\ between the innermost stable circular orbit and the light ring}
\author{Sarp Akcay$^{1,2}$, Leor Barack$^1$, Thibault Damour$^2$ and Norichika Sago$^3$}
\affiliation{
$^1$School of Mathematics, University of Southampton, Southampton
SO17 1BJ, United Kingdom, \\
$^2$Institut des Hautes \'Etudes Scientifiques, 35, route de Chartres,
                91440 Bures-sur-Yvette, France \\
$^3$Faculty of Arts and Science, Kyushu University,
Fukuoka 819-0395, Japan}

\date{\today}

\begin{abstract}
We compute the conservative piece of the gravitational self-force (GSF) acting on a particle of mass $m_1$ as it moves along an (unstable) circular geodesic orbit between the innermost stable orbit and the light ring of a Schwarzschild black hole of mass $m_2\gg m_1$.  More precisely, we construct the function $h^{R, L}_{u u}(x) \equiv h^{R, L}_{\mu \nu} u^{\mu} u^{\nu}$ (related to Detweiler's gauge-invariant ``redshift'' variable), where $h^{R, L}_{\mu \nu}(\propto m_1)$ is the regularized metric perturbation in the Lorenz gauge, $u^{\mu}$ is the four-velocity of $m_1$ in the background Schwarzschild metric of $m_2$, and $x\equiv [Gc^{-3}(m_1+m_2)\Omega]^{2/3}$ is an invariant coordinate constructed from the orbital frequency $\Omega$.
In particular, we explore the behavior of $h^{R, L}_{u u}$ just outside the ``light ring'' at $x=\frac13$ (i.e., $r= 3 G m_2/c^2$), where the circular orbit becomes null. 
Using the recently discovered link between $h^{R, L}_{u u}$
and the piece $a(u)$, linear in the symmetric mass ratio $\nu \equiv m_1 m_2/(m_1+m_2)^2$, of the main radial potential  $A(u,\nu)= 1-2u + \nu \, a(u) +O(\nu^2)$
of the effective one body (EOB) formalism, we compute from our GSF data
 the EOB function $a(u)$ over the entire
domain $0<u<\frac13$ (thereby extending previous results limited to $u\leq \frac{1}{5}$).
We find that $a(u)$ {\it diverges} like $a(u)\approx 0.25 (1-3u)^{-1/2}$ at the light-ring limit, $u \to \left(\frac{1}{3}\right)^-$, explain the physical origin of this divergent behavior, and discuss its consequences for the EOB formalism.  We construct accurate global analytic fits for $a(u)$, valid on the entire domain $0<u<\frac13$ (and possibly beyond), and give accurate numerical estimates of the values of $a(u)$ and its first {\it three} derivatives at the innermost stable circular orbit, as well as the $O(\nu)$ shift in the frequency of that orbit. 
In previous work we used GSF data on slightly eccentric orbits to compute a certain linear combination 
of $a(u)$ and its first two derivatives, involving also the $O(\nu)$ piece of a second EOB radial potential ${\bar D}(u)= 1 + \nu \, {\bar d} (u) +O(\nu^2)$.  Combining these results with our present global analytic representation of $a(u)$, we numerically
 compute $\bar d (u)$ on the interval $0<u\leq \frac{1}{6}$. 
\end{abstract}


\maketitle



\section{Introduction}

For much of its long history, the two-body problem in general relativity has been studied primarily within two analytical approximation frameworks, one built around the weak-field limit and the other around the test-particle (geodesic) limit. The first analytical framework, formalized in post-Newtonian (PN) and post-Minkowskian theories, is ({\it a priori}) applicable only  when the two components of the two-body system are sufficiently far apart. The second analytical framework is ({\it a priori}) relevant only when one of the masses is much larger than the other, in which case the dynamics can be described, at first approximation, as a geodesic motion on a fixed curved background.  
Recently, rapid developments (mixing theoretical and numerical methods) in the field of {\em gravitational self-force} (GSF) calculations (see \cite{Barack:2009ux} for a review) have allowed one to go one step beyond the geodesic approximation, giving access to new information on  strong-field dynamics in the extreme-mass-ratio regime. In addition, since 2005 it has been possible
to  accurately describe the coalescence of  two black holes of comparable masses by using three-dimensional numerical simulations based on the fully nonlinear Einstein equations.
The progress in interferometric gravitational-wave 
detectors has brought with it the imminent prospect of observing gravitational radiation from inspiralling and coalescing astrophysical binaries, and with it the need to compute, in an efficient
and accurate way, the form of the many possible gravitational-wave signals emitted by
generic binary systems (having arbitrary mass ratios and spins, and moving on
generic orbits).  It has become clear over the past few years that the best way to meet the latter
theoretical challenge will be to combine knowledge from all available approximation
methods: PN theory, post-Minkowskian theory, GSF  calculations, and full
numerical simulations.

Within this program, the effective one body (EOB) formalism \cite{Buonanno:1998gg,Buonanno:2000ef,Damour:2000we,Damour:2001tu} was proposed as a flexible analytical framework for describing the motion and radiation of coalescing binaries over the entire merger process, from the early inspiral, right through the eventual plunge and final ringdown (see Ref.\ \cite{Damour:2009ic} for a review). 
The central posit of the EOB formulation is a mapping between the true dynamics and an effective description involving an {\em effective metric},  together with an extra ``mass-shell deformation''
phase-space function  $Q$ involving (effective) position and momentum variables. If the two objects are nonspinning black holes with masses $m_1$ and $m_2$, then in the extreme-mass-ratio limit [i.e., when the symmetric mass ratio $\nu \equiv m_1 m_2/(m_1+m_2)^2$ tends to zero] the effective metric is expected to reduce smoothly to the Schwarzschild metric, while $Q$ must vanish.
For a general mass ratio, i.e.~for a non-zero value of $\nu$ in the interval $0<\nu\leq\frac14$,
the effective metric involves two initially unspecified functions  of two variables (``EOB potentials''), denoted $A(u;\nu)$ and $\bar D(u;\nu)$. Here $u$ is the dimensionless ``inter-body gravitational potential" $u \equiv G M/(c^2 r_{\rm EOB})$, where $M\equiv m_1+m_2$ denotes the total mass, and $r_{\rm EOB}$ is the (EOB-defined) radial separation between the two objects. In the current, ``standard" formulation of the EOB formalism, the motion in strictly circular binaries is governed by the potential $A(u,\nu)$ alone. The (conservative) dynamics of slightly eccentric binaries involves, besides $A(u,\nu)$, the second EOB potential $\bar D(u;\nu)$. More generally, the 
conservative dynamics of arbitrary orbits (described by the full EOB Hamiltonian) involves, besides $A(u;\nu)$ and $\bar D(u;\nu)$, the third EOB function  $Q(u,p_{\varphi},p_r;\nu)$, which {\it a priori} depends on the four variables  $(u,p_{\varphi},p_r,\nu)$, where $p_{\varphi}$ is the angular momentum and $p_r$ 
is the radial momentum canonically conjugated to the radial variable $r_{\rm EOB}$. 

Post-Newtonian theory only gives access to the expansions of the EOB potentials in powers of  the inter-body gravitational potential $u$, while keeping the exact dependence upon $\nu$. For instance, PN calculations at the {\it third} PN (3PN)
approximation lead to the exact knowledge of the coefficients $A_2(\nu), A_3(\nu)$ and $A_4(\nu)$ in 
$A(u,\nu) = 1- 2 \, u +  A_2(\nu) \, u^2+ A_3(\nu) \, u^3 +  A_4(\nu) \,u^4 + O(u^5 \ln u)$ , with the remarkably simple result \cite{Damour:2000we} that the 1PN coefficient
$A_2(\nu)$ {\it vanishes}, and that  the 2PN, $A_3(\nu)$, and 3PN, $A_4(\nu)$, coefficients are both {\it linear in} $\nu$ [thanks to some remarkable cancellations; the function $A_4(\nu)$, e.g., is {\it a priori} a cubic polynomial in $\nu$].  In order to apply the EOB formalism to the description of the final stages of coalescing binaries, it is necessary to somehow  improve the behavior of these ({\it weak-field};  $u \ll 1$) PN expansions, and to extend the knowledge of the functions $A(u), \bar D(u)$ into the {\it strong-field} regime $u =O(1)$. Two different methods have been proposed to perform such a strong-field extension. Both methods exploit the flexibility of the EOB framework, which naturally allows for either the introduction of unknown parameters (parametrizing  higher-order PN terms), or for the introduction of unknown functions (linked to GSF theory).
 
The first method used for ``upgrading'' the PN expansions of $A(u;\nu)$ and $\bar D(u;\nu)$ into functions which are (tentatively) valid in the strong-field regime $u =O(1)$, was to replace them with suitably resummed expressions, namely some Pad\'e approximants of either the currently known PN expansions~\cite{Damour:2000we}, or of PN expansions
incorporating some undetermined coefficients parametrizing as-yet-unknown higher-order PN terms~\cite{Damour:2002qh,Buonanno:2007pf,Damour:2009kr,Buonanno:2009qa}. As results from strong-field numerical relativity (NR) simulations started to emerge, it became possible to ``calibrate'' some of these unknown parameters, by finding the values that ``best fit'' the NR data \cite{Damour:2002qh,Buonanno:2007pf,Damour:2009kr,Buonanno:2009qa}.
The resulting NR-fitted EOB formalisms have been found to provide a useful analytic approach to the two-body problem in both the weak- and strong-field regimes and across all mass ratios \cite{Damour:2009kr,Buonanno:2009qa,Damour:2007xr,Yunes:2009ef,Bernuzzi:2010ty,LeTiec:2011bk,Damour:2011fu}.
 
 The second method for extending the validity of the PN expansions of $A(u;\nu)$ and $\bar D(u;\nu)$ is to use information from GSF theory \cite{Damour:2009sm}. Essentially, while PN theory (in the EOB context) involves the expansion of $A(u;\nu)$, $\bar D(u;\nu)$ and  $Q(u,p_{\varphi},p_r;\nu)$ in powers of $u$ (for fixed $\nu$), GSF theory involves the expansion of  these functions in powers of $\nu$ (for fixed $u$). For instance, the GSF expansion of the $A$ potential is of the form $A(u;\nu) = 1- 2 \, u +  \nu\, a(u) +  \nu^2 \, a_2(u) +O(\nu^3 )$,
 while that of   $\bar D(u;\nu)$ starts as $\bar D(u;\nu) = 1 + \nu \, \bar d(u) +  \nu^2 \, \bar{d}_2(u) +O(\nu^3)$,
 where  we suppressed, for notational simplicity, the index 1 on the
 coefficients $a(u)$ and $\bar d(u)$ of the
 first power of $\nu$ (``first GSF level''). Note that all the GSF coefficients $a(u), a_2(u), \bar d(u), \bar{d}_2(u)$
  are functions of $u$, and are {\it a priori} defined for arbitrary
 values of $u$, including strong-field values $u =O(1)$.
Since 2008, calculations of the GSF in Schwarzschild geometry are providing valuable information on various invariant aspects of the post-geodesic dynamics in binaries of extreme mass-ratios. This offers a new opportunity for improving  the EOB formalism by acquiring knowledge on the strong-field behaviour of the
various functions   $a(u), a_2(u), \bar d(u), \bar{d}_2(u), \dots$ . The GSF data are particularly useful for this purpose since they are highly accurate (GSF calculations involve only linear differential equations), and because they give access to a portion of the parameter space inaccessible to either PN or NR: strong-field inspirals in the extreme-mass-ratio domain. Furthermore, in GSF calculations (unlike in NR) it is straightforward to extract the conservative (time-symmetric) aspects of the dynamics separately from the dissipative ones. This is an advantage because the two aspects are dealt with separately in EOB. 

The promise of such a GSF-improved EOB formalism was first highlighted  in Ref.\ \cite{Damour:2009sm}. That work suggested several concrete gauge-invariant quantities characterizing the {\em conservative} dynamics of the binary, which can be constructed (in principle) using knowledge of the GSF, and would provide accurate information
about the $\nu$-linear  EOB  functions $a(u), \bar d(u)$. As a first example, Ref.\ \cite{Damour:2009sm} used the GSF computation \cite{ISCOLetter} of the $O(\nu)$ shift in the value of the frequency of the innermost stable circular orbit (ISCO) of the Schwarzschild black hole,  to determine the value of the combination 
$\mathsf a(u) \equiv a(u) + u\, a'(u) +\frac12 u (1-2u) \,a''(u)$ (where a prime denotes $d/du$) at the ISCO potential value $u=\frac16$. Ref.\ \cite{Damour:2009sm} also proposed that a GSF computation of the frequency and angular momentum of a marginally bound zoom-whirl orbit could be used to determine the separate values of $a(u)$ and $a'(u)$ at the much stronger-field point $u=\frac14$ (the ``whirl'' radius), but such a computation is yet to be performed. 

More importantly, Ref.\ \cite{Damour:2009sm} has shown that a computation of the GSF-induced correction to the periastron advance of slightly eccentric orbits along the one parameter sequence of circular orbits, would allow one to  compute the combination \footnote{For conceptual clarity we use here the function $\bar \rho(u)$ related to the function $\rho(u)$ of Ref.\ \cite{Damour:2009sm} via $\bar \rho(u) \equiv \rho(u) - 4 u [1-(1-2u)/ \sqrt{1-3u}$].} 
$ \bar \rho(u) \equiv \mathsf a(u) + (1-6u) \, \bar d(u)$ as a {\it function of $u$} 
over the entire range where  circular orbits exist, i.e.\ $0<u\leq \frac13$. The calculation of the EOB function $ \bar \rho(u) $ was then performed in Ref.\ \cite{Barack:2010ny} along the sequence of {\it stable} circular orbits (i.e.\  $0<u\leq \frac16$), when computational tools for the GSF in eccentric binaries became available \cite{Barack:2010tm}. Ref.\ \cite{Barack:2010ny} also made the following important point. By combining PN information about the
behaviour near $u=0$ of functions such as $ a(u)$ or $ \bar \rho(u)$,
together with the GSF-computed values of these functions at a (possibly sparse) sample of strong-field points $u=u_1,u_2,\ldots$, one can construct simple (Pad\'e-like) analytic representations which can provide accurate  {\it global} fits for the corresponding EOB functions. Then, in turn, these global representations can be used to analytically represent other GSF functions of direct dynamical significance. Ref.\ \cite{Barack:2010ny} demonstrated this idea by constructing a simple, yet accurate, global analytic model for the periastron advance in slightly eccentric orbits, using only a small set of strong-field GSF data in conjunction with available weak-field PN information. In subsequent work \cite{LeTiec:2011bk} this model was successfully tested against results from fully nonlinear numerical simulations of inspiralling binaries. 

Unfortunately, knowledge of the GSF-induced periastron advance only gives access to the combination
$ \bar \rho(u)$ involving the functions $a(u), a'(u), a''(u)$ and $\bar d(u)$, and it is not sufficient for determining the individual potentials $a(u)$ and $\bar d(u)$ separately. This situation was cured in recent work by Le Tiec and collaborators. In Ref.\ \cite{Tiec:2011ab} Le Tiec {\it et al.}\  have ``derived'' (using a mixture of plausible
arguments)  a ``first law of binary black hole mechanics'', relating infinitesimal variations of the total energy ${\mathcal E}$ and angular momentum $J$ of the binary system to variations of the individual black hole ``rest masses'', and (for $m_1\ll m_2)$ involving  Detweiler's red-shift  variable $z_1$ associated with $m_1$ \cite{Detweiler:2008ft}. (The validity of this relation was established rigorously only through 3PN order.) Based on this relation, further work  \cite{Tiec:2011dp} about the functional link between ${\mathcal E}$, $J$  and the dimensionless orbital frequency parameter  $x \equiv (GM \Omega/c^3)^{2/3} $ led  Barausse, Buonanno and Le Tiec \cite{Barausse:2011dq}  to derive a simple direct relation between the $O(\nu)$ piece of the function $z_1(x)$, and the $O(\nu)$  EOB function $a(u)$ (evaluated for the argument $u=x$). This relation shows that
 GSF calculations of the $O(\nu)$  piece of the redshift function $z_1(x)$ of $m_1$, along
{\it circular orbits}, allows one to compute the function $a(u)$, {\it separately} from the second  $O(\nu)$ EOB function $\bar d(u)$. [Using the (quite simple) EOB theory of circular orbits, it is then easy to derive from $a(u)$ the functions relating ${\mathcal E}$ and $J$ to both $u$ and $x$; see Refs.~\cite{Damour:2009sm,Barausse:2011dq} and below.] By putting together the
 so-acquired knowledge of the function $a(u)$ with Ref.\ \cite{Barack:2010ny}'s GSF computation of the combination $ \bar \rho(u)$, one then has separate access to the second EOB function $\bar d(u)$, thereby completing the project initiated in Refs.~\cite{Damour:2009sm,Barack:2010ny} of using GSF data to determine the (separate) strong-field  behaviors of the two main $O(\nu)$ EOB potentials $a(u)$ and $\bar d(u)$. Note, however, that this still leaves out the third EOB function $Q(u,p_{\varphi},p_r;\nu)$. 

The analyses of Refs.\ \cite{Tiec:2011dp} and \cite{Barausse:2011dq} relied on numerical GSF data for $z_1(x)$, which have so far been available only for $x\leq 1/5$ \cite{Detweiler:2008ft,Blanchet:2010zd}. This allowed the determination of the EOB potentials (and of ${\mathcal E}$ and $J$) through $O(\nu)$ only in the restricted domain $0\leq u\leq 1/5$. The EOB potentials remained undetermined in the strong-field domain $u >\frac15 $. In the extreme-mass-ratio case, this domain corresponds to the region $r< 5 G m_2/c^2$ outside the large black hole of mass $m_2$ (where $r$ is the Schwarzschild radial coordinate associated with $m_2$, which coincides with $r_{\rm EOB}$ in the $m_1\to 0$ limit). Note that the gravitational potential varies steeply in this region, so that  the EOB functions might well vary correspondingly fast and possibly in a non-trivial way, potentially giving rise to interesting new physics. In this regard, we emphasize that (as is clear in EOB theory) it is the gravitational-potential coordinate $u$, and not $r$ itself, which best parametrizes the strength of the gravitational field. We note in this respect that the gravitational potential difference across the  seemingly ``small'' domain extending between the ISCO and the light ring (below which there exist no circular geodesic orbits), $3 Gm_2/c^2<r<6 Gm_2/c^2$
(i.e., $\frac16<u<\frac13$),  is as large as that across the entire domain $6 Gm_2/c^2<r<\infty$ (i.e., $0<u<\frac16$). This lends a strong motivation for extending the analyses of Refs.\ \cite{Tiec:2011dp} and \cite{Barausse:2011dq} to the domain $\frac15 < u < \frac13$.

In this paper we obtain numerical GSF data for $z_1(x)$ for circular geodesic orbits with radii in the range $3Gm_2/c^2<r<150Gm_2/c^2$. We use these data to compute the numerical values of the function $a(u)$ on a dense set of $u$ values, extending down to $u=\frac13$.
We then construct a global analytic fit for the function $a(u)$, valid uniformly on $0<u<\frac13$. We pay particular attention to the behavior near $u=\frac13$, which, in the limit $\nu\to 0$, represents the  light ring (LR) of the Schwarzschild black hole, where circular geodesic orbits become null. 

It should be commented immediately that the interpretation of the GSF near the LR is a subtle one: for any finite (nonzero) value of $\nu$, there are sufficiently small values of $u-\frac{1}{3}$ for which the mass-energy of the small particle becomes comparable to that of the large black hole, at which point perturbation theory breaks down and the GSF approximation ceases to be meaningful. In principle, however, it is possible to make the GSF approximation relevant arbitrarily close to the LR, simply by taking $\nu$ to be sufficiently small. This formal argument allows us to use GSF data to explore the immediate vicinity of the LR.

The structure of this paper is as follows. We start, in Sec.\ \ref{Sec:GSF}, by reviewing the formal GSF results relevant to our analysis, and then present the new sub-ISCO GSF data. The raw numerical data are given in Appendix \ref{AppA} for the benefit of colleagues interested in reproducing our analysis or studying other applications. In Sec.\ \ref{Sec:afit} we use the GSF data to construct a global analytic fit for the function $a(u)$, and in particular establish the behavior of this potential near the LR. In Sec.\ \ref{Sec:EJfit} we similarly construct global analytic models for the $O(\nu)$ pieces of ${\mathcal E}$ and $J$. In Sec.\ \ref{Sec:ISCO} we revisit the problem of determining the $O(\nu)$ shift in the ISCO frequency, and, using the method proposed in \cite{Tiec:2011dp} with our new, highly accurate $a(u)$ data, add 4 significant digits to the value obtained in previous analyses \cite{ISCOLetter,Damour:2009sm,Tiec:2011dp}. Section \ref{Sec:dfit} turns to discuss the determination of the second $O(\nu)$ EOB potential, $\bar d(u)$: by combining the new analytic $a(u)$ model with our previously obtained data for $\bar\rho(u)$, we determine $\bar d(u)$ (numerically) on the domain $0<u<\frac{1}{6}$. Section \ref{Sec:photosphere} then focuses on the LR behavior, explaining the physical origin of the observed divergent behavior of $a(x)$, and discussing its consequences for the EOB formalism. We summarize our main results in Sec.\ \ref{Sec:conclusions} and discuss future directions. 

\subsection{Setup and notation}

Henceforth, we shall use units such that $G=c=1$. We will consider a circular-orbit binary of black holes with masses $m_1\leq m_2$. Various combinations of these two masses will become relevant in different parts of our analysis: we shall use 
\begin{equation}
M\equiv m_1+m_2, \quad\quad
q\equiv \frac{m_1}{m_2} \leq 1, \quad\quad
\nu\equiv \frac{m_1 m_2}{(m_1+m_2)^2} = \frac q{(1+q)^2}
\end{equation}
to denote, respectively, the total mass, ``small'' mass ratio, and symmetric mass ratio of the system. This mass notation differs from the one used in our previous paper \cite{Barack:2010ny}, and is more in line with the notation commonly used in EOB and PN work.  It reflects the fact that in these formulations (unlike in GSF work) the two masses are treated symmetrically.

We will find it convenient, in different parts of the analysis, to use different measures of the binary separation. In the GSF-relevant limit $q\to 0$ ($\Leftrightarrow\nu  \to 0$) we will use the standard (areal) radial coordinate $r$ associated with the Schwarzschild geometry of the black hole with mass $m_2$, while in discussing EOB we will mainly use  the EOB ``gravitational potential'' (or ``inverse radius'')  $u\equiv M/r_{\rm EOB}$. A relation between the GSF and EOB descriptions can be established using the {\em invariant} frequency $\Omega$ associated with the orbit, or the dimensionless frequency parameter
\begin{equation} \label{defx}
x\equiv(M\Omega)^{2/3} =[(m_1+m_2) \Omega]^{2/3} 
\end{equation}
derived from it.  As is well-known, in the GSF limit $\nu  \to 0$, $x$ becomes equal to $u$ (``Kepler's third law''): $ x=u+O(\nu)$. When discussing the behavior near the (unperturbed) LR,  $x=\frac13=u$, it will be convenient to introduce the (invariant) coordinate 
\begin{equation}
z\equiv 1-3x.
\end{equation}
(The quantity should not be confused with $z_1$, denoting the redshift of worldline 1.) For easy reference, Table \ref{table:masses} summarizes our notation for various mass and radius quantities. 

\begin{table}[htb] 
\begin{tabular}{|c|l||c|l|}
\hline\hline
\multicolumn{2}{|c||}{Binary masses} & \multicolumn{2}{c|}{Measures of binary separation}   \\
\hline\hline
$m_1$ & particle mass &  $r$ or $r_0$ & Schwarzschild radial coordinate \\
$m_2$ & black hole mass &  $u =M/r_{\rm EOB}$ & EOB ``inverse radius'' coordinate  \\
$M=m_1+m_2$ & total mass & $\Omega$ & invariant orbital frequency \\
$q\equiv m_1/m_2$ & ``small'' mass ratio & $x=(M\Omega)^{2/3}$ &  dimensionless frequency parameter  \\
$\nu\equiv \frac{m_1 m_2}{(m_1 + m_2)^2}$ & symmetric mass ratio & $z=1-3x$  &  invariant ``distance" from light ring   \\
$\mu\equiv \frac{m_1 m_2}{m_1+m_2}$ & reduced mass  &     &   \\
\hline\hline
\end{tabular}
\caption{Various mass and separation quantities appearing in our analysis, summarized here for easy reference.}
\label{table:masses}
\end{table}

\section{Conservative GSF for (stable or unstable) circular orbits} \label{Sec:GSF}

\subsection{Redshift function and regularized self-metric perturbation}

The GSF formulation stems from a perturbative treatment of the binary dynamics. At the limit $q\to 0$ the object with mass $m_1$ becomes a ``test particle'' and its motion is described by some geodesic in a ``background" Schwarzschild geometry of  mass $m_2$. Finite-$m_1$ effects (self-force, including radiation reaction,  etc.) are incorporated, in principle, order by order in $q$, working on the fixed background of the large black hole. In this treatment, the small object experiences a GSF caused by an interaction with its own gravitational perturbation, and giving rise to an accelerated motion with respect to the Schwarzschild background. The GSF accounts for the dissipative decay of bound orbits, as well as for conservative (e.g., precessional) effects associated with the finiteness of $m_1$.
While the GSF itself is gauge-dependent, knowledge of the GSF (in a particular gauge) together with the metric perturbation due to $m_1$ (in that same gauge) gives sufficient information for quantifying the gauge-invariant aspects of the dynamics. At the foundational level the GSF is now well understood at the first order in $q$ beyond the geodesic approximation \cite{Mino:1996nk,Quinn:1996am,Gralla:2008fg,Pound:2009sm,Poisson:2011nh}, and at this order there is also a well-developed methodology and a toolkit for numerical computations, at least in the case of a Schwarzschild background \cite{BS, Barack:2010tm}. (The foundations for the second-order GSF have also been laid recently \cite{Pound:2012nt,Gralla:2012db,Pound:2012dk} but this formulation is yet to be implemented numerically.)

In the problem at hand we ignore the dissipative effect of the GSF, and the orbit is assumed to be precisely circular. We shall assume, without loss of generality, that the motion takes place in the equatorial plane $\theta=\pi/2$, where hereafter we use standard Schwarzchild coordinates $\{t,r,\theta,\varphi\}$ defined with respect to the background Schwarzschild geometry with metric $g^{0}_{\alpha\beta}(m_2)$. Detweiler and Whiting have shown \cite{Detweiler:2002mi} that the GSF-corrected worldline has the interpretation of a geodesic in a smooth perturbed spacetime with metric $g_{\alpha\beta}= g^{(0)}_{\alpha\beta}(m_2)+h_{\alpha\beta}^R$, where $h_{\alpha\beta}^R$ (the ``R field'') is a certain [$O(q)$] smooth perturbation associated with $m_1$. We let $u^{\alpha}_1=\{u^t_1,0,0,u^{\varphi}_1\}$ be the four-velocity defined with respect to proper time along this effective geodesic.  
It is straightforward to show that both $u_1^t$ and $u_1^{\varphi}$ are invariant under gauge transformations with generators $\xi^{\alpha}=O(q)$ that respect the helical symmetry of the perturbed spacetime \cite{SBD}. The azimuthal frequency (with respect to a coordinate time $t$ belonging to an ``asymptotically flat'' coordinate system),
\begin{equation}
\Omega\equiv \frac{u_1^{\varphi}}{u_1^{t}}= \frac {d \varphi}{d t},
\end{equation}
is thus also invariant under such gauge transformations. Detweiler \cite{Detweiler:2008ft,Detweiler:2009ah} proposed utilizing the functional relation $u^t_1(\Omega)$, or, equivalently the ``redshift function"
\begin{equation} \label{defz1}
z_1(\Omega) \equiv 1/u^t_1(\Omega),
\end{equation} 
as a  gauge-invariant handle on the conservative effect of the GSF in circular motion. He also discussed the physical meaning of $z_1$ as a measure of the (regularized) gravitational redshift between the worldline of $m_1$ and infinity.


The expressions derived by Detweiler for $u^t_1(\Omega)$ [or $z_1(\Omega)$] involve the double contraction of $h_{\alpha\beta}^R$ with the four-velocity $u_1^{\alpha}$, namely 
\begin{equation}\label{huu}
h_{uu}^{R, G}\equiv h^{R, G}_{\alpha\beta}u_1^{\alpha}u_1^{\beta}.
\end{equation}
Here we have introduced the extra label $G$, for ``gauge''(besides the first label $R$ referring to ``regularized''), to keep track of the coordinate gauge in which one evaluates the metric perturbation. This is important for the following reason. The prescription in Ref.\ \cite{Detweiler:2008ft} assumes that the metric perturbation is given in a gauge which is manifestly asymptotically flat (i.e., one in which the unregularized metric perturbation vanishes at infinity). This, however, happens not to be the case for the  {\em Lorenz gauge} that we shall use in our actual GSF calculations. As a consequence, a certain gauge correction term will enter our expressions for $u^t_1(\Omega)$, as we discuss below.  We shall use the label $G=F$ to refer to a manifestly asymptotically-flat gauge, and the label $G=L$ for the Lorenz gauge.

Using an {\it asymptotically flat} gauge, and the dimensionless frequency parameter 
\begin{equation}
\label{F1}
y \equiv (m_2 \, \Omega)^{2/3},
\end{equation}
Ref.\ \cite{Detweiler:2008ft} obtained the simple relation
\begin{equation}
\label{F2}
u_1^t(\Omega) = \frac{1}{\sqrt{1-3y}} \, \left[ 1+\frac{1}{2} \, h_{uu}^{R,F} + O(q^2) \right] \, .
\end{equation}
In terms of the redshift variable (\ref{defz1}) this reads
\begin{equation}
\label{F3}
z_1(\Omega) = \sqrt{1-3y}  \left[ 1 - \frac{1}{2} \, h_{uu}^{R,F} + O (q^2) \right] \, .
\end{equation}
The GSF-adapted frequency parameter $y$ [Eq.\ (\ref{F1})] is related to the more symmetric (EOB-adapted) frequency parameter $x$ [Eq.\ (\ref{defx})] through
\begin{equation}
\label{F4}
\frac{y}{x} = \left( \frac{m_2}{m_1 + m_2} \right)^{2/3} = \frac{1}{(1+q)^{2/3}} = 1 - \frac{2}{3} \, q + O (q^2),
\end{equation}
so that
\begin{equation}
\label{F5}
\sqrt{1-3y} = \sqrt{1-3x} + q \, \frac{x}{\sqrt{1-3x}} + O(q^2) \, .
\end{equation}
Substituting in Eq.~(\ref{F3}) then yields, through $O(q)$,
\begin{equation}
\label{z1F}
z_1 (x) = \sqrt{1-3x} \left[ 1-\frac{1}{2} \, h_{uu}^{R,F} + q \, \frac{x}{1-3x} \right] \, .
\end{equation}

The form of the last relation is invariant under gauge transformations within the class of asymptotically flat (and helically symmetric) gauges. However, our GSF calculations will be carried out in the Lorenz gauge, in which the metric perturbation $h_{\mu\nu}^L$ turns out {\it not} to decay at infinity  (its monopolar piece tends to a constant value there \cite{Barack:2005nr}). We need to have at hand the link between the normal ``asymptotically flat'' $h_{uu}^{R,F}$ and its Lorenz-gauge counterpart $h_{uu}^{R,L}$. The issue was discussed in Refs.\ \cite{SBD,Damour:2009sm}, and we recall here the end result. 

A simple gauge transformation away from Lorenz into a corresponding asymptotically flat gauge is obtained by rescaling 
the Lorenz-gauge time coordinate $t_L$ using 
\begin{equation}
\label{F6}
t_F = (1+\alpha) \, t_L ,
\end{equation}
with
\begin{equation}
\label{F7}
\alpha = q \, \frac{x}{\sqrt{1-3x}} \, .
\end{equation}
This defines an F-gauge with metric perturbation given [through $O(q)$] by
\begin{equation}\label{FtoL}
h_{00}^F (r) = h_{00}^L (r) + 2 \alpha \left( 1 - \frac{2m_2}{r} \right) ,
\end{equation}
with $h^F_{\alpha\beta}=h^L_{\alpha\beta}$ for all other components. Since the gauge transformation relating $h^F_{\alpha\beta}$ to $h^L_{\alpha\beta}$ is regular, the corresponding regularized fields $h^{R,F}_{\alpha\beta}$ and $h^{R,L}_{\alpha\beta}$ are related to one another in just the same way (this comes from a general result derived in \cite{Barack:2001ph}).
Evaluating on the $m_1$ worldline and contracting twice with the four-velocity, one then finds
\begin{equation}
\label{F10}
h_{uu}^{R,F} = h_{uu}^{R,L} + 2 \alpha \left( 1 - \frac{2m_2}{r} \right) (u_1^t)^2 \, ,
\end{equation}
which reads explicitly [using $(u_1^t)^2 = (1-3x)^{-1} + O(q)$]
\begin{equation}
\label{F11}
h_{uu}^{R,F} = h_{uu}^{R,L} + 2q \, \frac{x \, (1-2x)}{(1-3x)^{3/2}} \, .
\end{equation}
Inserting Eq.~(\ref{F11}) into Eq.~(\ref{z1F}) finally leads to an expression for $z_1 (x)$ in terms of $h_{uu}^{R,L}$:
\begin{equation}
\label{z1L}
z_1 (x) = \sqrt{1-3x} \, \left[ 1 - \frac{1}{2} \, h_{uu}^{R,L} - q \, \frac{x(1-2x)}{(1-3x)^{3/2}} + q \, \frac{x}{1-3x} \right] \, .
\end{equation}

In the above expressions we have not specified the argument in terms of which $h_{uu}^{R,G}$ should be expressed, or---more precisely---the specific orbit along which $h_{uu}^{R,G}$ should be evaluated. The explicit GSF computations presented below actually give $h_{uu}^{R,L}$ along an {\it unperturbed}, geodesic orbit, parametrized by the {\it unperturbed} Schwarzschild-radius variable $m_2 / r$. However, to leading order in $q$ we have $m_2 / r = y + O(q) = x + O(q)$, so that in Eqs.~(\ref{z1F}) and (\ref{z1L}) we can simply replace $h_{uu}^{R,G}(m_2/r) \to h_{uu}^{R,G}(x)$ [as, of course, $h_{uu}^{R,G}$ itself is already $O(q)$ and in our analysis we ignore terms of $O(q^2)$ or higher].

Finally, we note that $h_{uu}^{R,G}$ describes a purely {\em conservative} effect of the GSF (even though in practice we shall extract $h_{uu}^{R,G}$ from the {\em retarded} metric perturbation). To see this, it is enough to recall Eq.\ (\ref{F2}), which relates $h_{uu}^{R,G}$ to the {\em time-symmetric} function $u_1^t(\Omega)$. The property that $h_{uu}^{R,G}$ encodes a purely conservative piece of the GSF is special to circular orbits, and it does not carry over to (e.g.) eccentric orbits; cf.\ \cite{Barack:2011ed}.

\subsection{Mode-sum computation of $h^{R,L}_{uu}$}\label{Subsec:modesum}

Our method follows closely the standard strategy of mode-sum regularization \cite{Barack:1999wf,Barack:2001gx,Barack:2009ux}. 
As, in this section, we work only with the Lorenz-gauge perturbation we shall drop, for concision, the extra label $L$ on $h_{\alpha\beta}$. We begin by writing $h^{R}_{\alpha\beta}=h_{\alpha\beta}-h_{\alpha\beta}^{S}$, where $h_{\alpha\beta}$ is the full (retarded) Lorenz-gauge metric perturbation associated with the mass $m_1$, and $h_{\alpha\beta}^{S}$ is the locally defined Detweiler--Whiting Singular field (``S field'') \cite{Detweiler:2002mi}. Both $h_{\alpha\beta}$ and $h_{\alpha\beta}^{S}$ diverge at the particle, but their difference $h^{R}_{\alpha\beta}$ is perfectly smooth. We formally construct the fields $h_{uu} \equiv h_{\alpha\beta}\hat u_1^{\alpha}\hat u_1^{\beta}$ and $h_{uu}^S \equiv h^S_{\alpha\beta}\hat u_1^{\alpha}\hat u_1^{\beta}$, where $\hat u_1^{\alpha}$ is any smooth extension of the four-velocity $u_1^{\alpha}$ off the particle's worldline (so that $\hat u_1^{\alpha}= u_1^{\alpha}$ on the worldline itself).  We then consider the formal decomposition of the fields $h_{uu}(t,r,\theta,\varphi)$ and $h_{uu}^S(t,r,\theta,\varphi)$ in scalar spherical harmonics $Y^{lm}(\theta,\varphi)$, defined as usual on the spherically symmetric Schwarzschild background, and we let $h^{l}_{uu}(r)$ and $h^{S,l}_{uu}(r)$ denote the individual $l$-mode contributions to the respective fields, summed over $m$ for fixed $l$, and evaluated at the particle [i.e., in the limit $r \to r_{\rm particle}(t)$]. As shown in Appendix D of Ref.\ \cite{Barack:2011ed}, the particle limit in the above procedure is well defined, and the resulting values $h^{l}_{uu}(r)$ and $h^{S,l}_{uu}(r)$ are finite and do not depend on the direction (upwards or downwards) from which the limit $r \to r_{\rm particle}(t)$ is taken.  We thus have 
\begin{equation}\label{modesum1}
h^{R}_{uu}(r)=\sum_{l=0}^{\infty} \left(h_{uu}^{l}(r)-h_{uu}^{S,l}(r)\right),
\end{equation}
where it should be noted that while each of the individual $l$-mode sums $\sum_l h_{uu}^{l}$ and $\sum_l h_{uu}^{S,l}$ would be divergent,  the mode sum of the difference $\sum_l (h_{uu}^{l}-h_{uu}^{S,l}) $ converges exponentially fast (because the difference $h_{\alpha\beta}-h_{\alpha\beta}^{S}$ is a smooth function).  We also note that the individual $l$-mode contributions $h^{l}_{uu}$ and $h^{S,l}_{uu}$ depend on the off-worldline extension chosen for $u_1^{\alpha}$, while the sum over modes in Eq.\ (\ref{modesum1}) is, of course, extension-independent.


The formulation of the $l$-mode method proceeds by obtaining an analytic description of the large-$l$ behavior of $h_{uu}^{S,l}$. Ref.\ \cite{Detweiler:2008ft}  (see also Ref.\ \cite{Barack:2011ed}) obtained the asymptotic form (as $l\gg1$)
\begin{equation}\label{huus_asy0}
h^{S,l}_{uu}(r)= D_0(r) + O(l^{-2}),
\end{equation}
where $D_0(r)$ is an $l$-independent parameter depending only of the orbital radius $r$:
\begin{equation}
D_0(r)=\frac{4m_1 Z(r)}{\pi r}\, {\rm EllipK}(w(r)),
\end{equation}
with
\begin{equation}
Z(r)\equiv \sqrt{\frac{r-3m_2}{r-2m_2}}, \quad\quad
w(r)= \frac{m_2}{r-2m_2},
\end{equation}
and with ${\rm EllipK}(w)=\int_0^{\pi/2}(1-w\sin^2x)^{-1/2}dx$ denoting the complete elliptic integral of the first kind. [The large-$l$ behavior of $h^{S,l}_{uu}$ was analyzed in \cite{Barack:2011ed} for generic (stable) eccentric orbits, and we specialize the expressions obtained there to circular orbits; the validity of these analytic results for $r<6m_2$ will be discussed below.] It was found \cite{Barack:2011ed} that the asymptotic value $D_0$ does {\it not} depend on the $u^{\alpha}$-extension involved in the definition of the modes $h^{S,l}_{uu}$.  Furthermore it was found that (for any such extension)
\begin{equation}
\sum_{l=0}^{\infty} \left(h^{S,l}_{uu}(r)-D_0(r)\right)=0.
\end{equation}
This allows us to write Eq.\ (\ref{modesum1}) in the form 
\begin{equation}\label{modesum2}
h^{R}_{uu}(r)=\sum_{l=0}^{\infty} \left(h_{uu}^{l}(r)-D_0(r)\right),
\end{equation}
which is an operational mode-sum formula for $h^{R}_{uu}$, describing the correct mode-by-mode regularization of the fields $h_{uu}^{l}$. The latter are to be provided as input, typically in the form of numerical solutions to the mode-decomposed Lorenz-gauge metric perturbation equations with suitable ``retarded'' boundary conditions (details of our particular numerical implementation are provided below).  

Since the mode sum in Eq.\ (\ref{modesum1}) converges faster than any power of $1/l$, it follows that the retarded modes too must have the asymptotic form $h^{l}_{uu}(l\gg 1)= D_0 + O(l^{-2})$. Thus, in general, we expect the partial mode sum in Eq.\ (\ref{modesum2}) to converge with a slow power law $\sim l^{-1}$ (and this was indeed confirmed numerically in \cite{Barack:2011ed}). This is problematic from the practical point of view, and restricts the accuracy within which $h^{R}_{uu}$ can be computed. As emphasized notably in Ref.\ \cite{Detweiler:2008ft} the problem can be mitigated by including  higher-order terms in the large-$l$ expansion of $h^{S,l}_{uu}$. Recently, Heffernan {\it et al.}~\cite{Heffernan:2012su} were able to obtain analytic expressions for a couple of these:
\begin{equation}\label{huus_asy2}
h^{S,l}_{uu}(r)=D_0(r)+\frac{D_2(r)}{L_2}+\frac{D_4(r)}{L_4}
+O(l^{-6}),
\end{equation}
where \cite{,Detweiler:2002gi,Detweiler:2008ft} $L_2\equiv (l-\frac{1}{2})(l+\frac{3}{2})$, $L_4\equiv (l-\frac{3}{2})(l-\frac{1}{2})(l+\frac{3}{2})(l+\frac{5}{2})$, and where the $l$-independent (but $r$-dependent) coefficients $D_{2,4}$ are given by 
\begin{equation}\label{eq:D2}
D_2(r)= \frac{m_1}{2\pi r^2 Z(r)}\left[\left(\frac{7r^2-61 m_2 r+96 m_2^2}{r-2m_2}\right){\rm EllipK}(w(r)) -(7r-33m_2){\rm EllipE}(w(r))\right],
\end{equation}
\begin{eqnarray}\label{eq:D4}
D_4(r) &=& \frac{3m_1}{160\pi r^3 Z(r)(r-3m_2)^2}\times
\nonumber\\
&&\left[
\left(\frac{30r^5-2683 m_2 r^4+30741 m_2^2 r^3-131855m_2^3 r^2+241905 m_2^4 r-160530 m_2^5}{r-2m_2}\right){\rm EllipK}(w(r)) 
\right. \nonumber\\
&&
-\left. 2\left(\frac{15r^5-1469 m_2 r^4+13990 m_2^2 r^3-56858 m_2^3 r^2+106395m_2^4 r-71385m_2^5}{r-3m_2}\right){\rm EllipE}(w(r))\right],
\end{eqnarray}
with ${\rm EllipE}(w)=\int_0^{\pi/2}(1-w\sin^2x)^{1/2}dx$ denoting the complete elliptic integral of the second kind.
[The expressions in \cite{Heffernan:2012su} were derived for generic (stable) bound geodesics, and we specialize them here to circular orbits.] It is important to note that the values of the subleading parameters $D_2$ and $D_4$, unlike that of $D_0$, {\em do} depend on the off-worldline extension of $u_1^{\alpha}$. The above values correspond to the particular extension $\hat u_1^{\alpha}\equiv u_1^{\alpha}$ (in Schwarzschild coordinates), i.e., an extension in which the contravariant Schwarzschild components of the field $\hat u_1^{\alpha}$ are taken to have the constant values $u_1^{\alpha}$ everywhere. This is a practically useful extension and we shall refer to it as the ``constant'' extension.

The $l$-dependent factors in Eq.\ (\ref{huus_asy2}) have the important property (first exploited in Ref.\ \cite{Detweiler:2002gi} in the context of the scalar-field self-force)
\begin{equation}
\sum_{l=0}^{\infty}\frac{1}{L_2}=0, \quad\quad \sum_{l=0}^{\infty}\frac{1}{L_4}=0,
\end{equation}
which allows us to recast the mode-sum formula (\ref{modesum2}) in the more useful form 
\begin{equation}\label{modesum3}
h^{R}_{uu}(r)=\sum_{l=0}^{\infty} \left(h_{uu}^{l}(r)-D_0(r) -\frac{D_2(r)}{L_2}-\frac{D_4(r)}{L_4}\right).
\end{equation}
Once again, since the sum in Eq.\ (\ref{modesum1}) converges faster than any power of $1/l$, we have that $h^{l}_{uu}$ and $h^{S,l}_{uu}$ must share the same asymptotic power-law expansion (\ref{huus_asy2}), with the {\em same} coefficients $D_n$ (as long as $h^{l}_{uu}$ is defined and computed using the above ``constant'' $u_1^{\alpha}$-extension). Therefore, we expect the revised mode-sum formula (\ref{modesum3}) to converge like $\sim l^{-5}$---significantly faster than the original mode sum (\ref{modesum2}). This will be confirmed numerically below. The fast-converging mode-sum formula (\ref{modesum3}) forms the basis for our numerical implementation in this work.

\subsection{Behavior of the mode sum near the light ring}\label{Subsec:modesumLR}

The results presented in the previous subsection were derived in \cite{Barack:2011ed,Heffernan:2012su} for {\it stable} geodesic orbits. However, all of these results, and in particular the form of the mode-sum formula (\ref{modesum3}) and the values of the parameters $D_n$, are equally applicable for circular (timelike) geodesics below the ISCO. Subtleties begin to manifest themselves only when the orbit is sufficiently close to the LR at $r=3m_2$. There, the orbit becomes asymptotically null and beaming-type effects distort the usual $l$-mode distribution, potentially enhancing the relative contribution of higher multipoles [see, e.g., Davis {\it et al.}~\cite{Ruffini1972}, but note that their analysis concerns the distribution {\it at infinity} of {\em tensorial}-harmonic modes, while ours involves {\em scalar}-harmonic modes {\it near the $m_1$ worldline} of the particular (extension-dependent) contraction $h_{uu}$]. 

\begin{figure}[Htb]
\includegraphics[width=12cm]{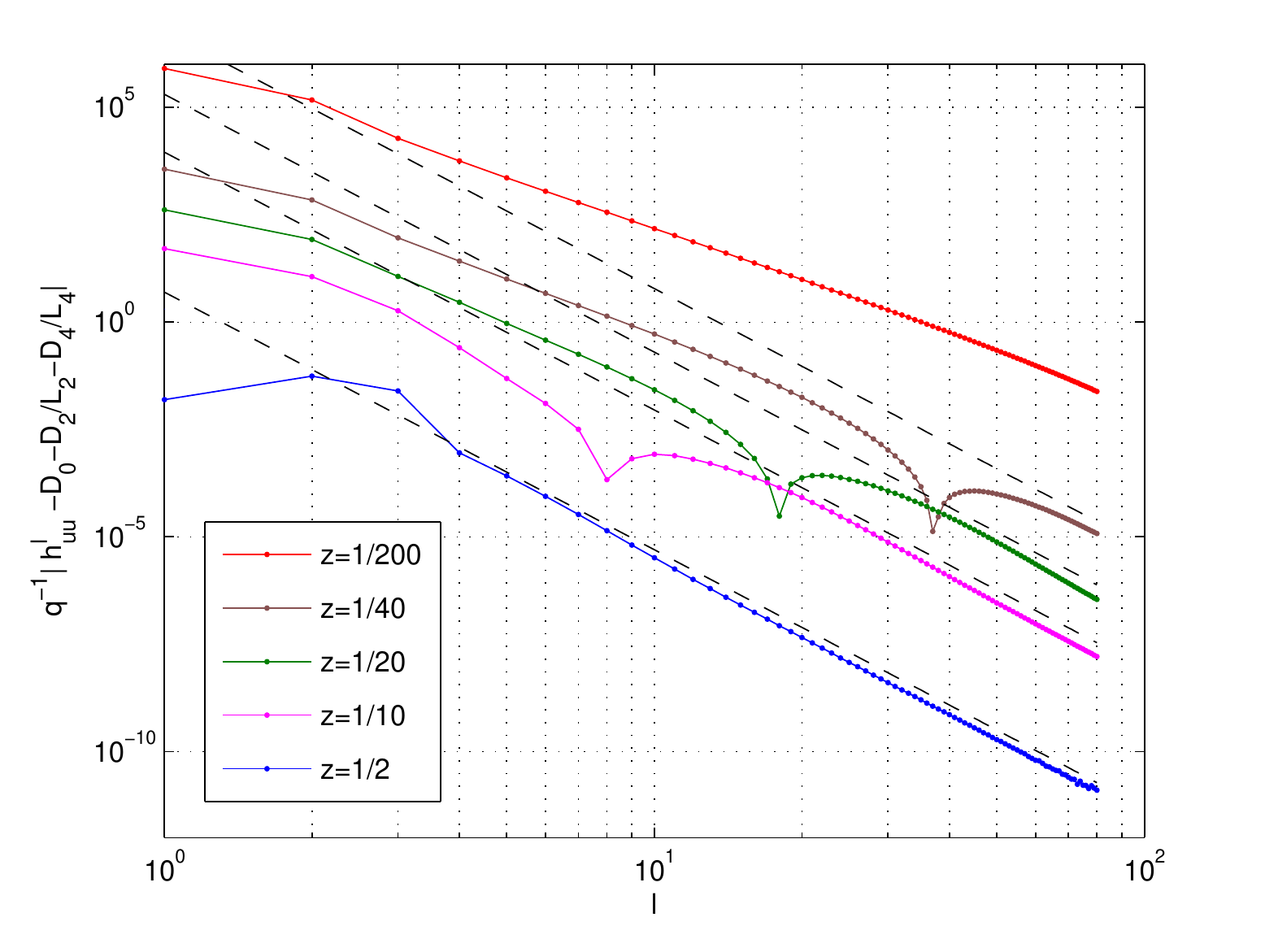}
\caption{Broadening of the $l$-mode spectrum near the light ring (LR), $z=1-3m_2/r=0$. We plot here 
the (absolute values of the) regularized modes $h_{uu}^{l}-D_0 -\frac{D_2}{L_2}-\frac{D_4}{L_4}$ [see Eq.\ (\ref{modesum3})] for $0\leq l\leq 80$, for a range of radii on and below the ISCO. 
The dashed lines are arbitrary $\propto l^{-6}$ references. Away from the LR, the regularized modes are expected to fall off at large $l$ with an $\sim l^{-6}$ tail, as is clearly manifest in the case $z=1/2$ (the ISCO, lower curve). As the radius gets closer to the LR, the onset of the $l^{-6}$ tail shifts to larger $l$-values, with the standard tail not developing until around $l\sim 1/z$ (the regularized mode contributions turn from negative to positive around that value of $l$). In the near-LR case $z=1/200$ (upper curve) no transition to power law is evident below $l=80$. How these data are obtained is described in Sec.\ \ref{subsec:numerics}. 
}
\label{fig:lz}
\end{figure}

That the $l$-mode behavior becomes subtle near the LR is evident from the asymptotic form of the parameters $D_n$. Defining $z\equiv 1-3m_2/r$ we find
\begin{equation}
D_0(z\ll 1)= -\frac{2qz^{1/2}\ln(3z/16)}{\sqrt{3} \, \pi } +O(z^{3/2}\ln z, z^{3/2}),
\end{equation}
\begin{equation}
D_2(z\ll 1)= \frac{2qz^{-1/2}[1+\ln(3z/16)]}{3\sqrt{3} \, \pi } +O(z^{1/2}\ln z, z^{1/2}),
\end{equation}
\begin{equation}
D_4(z\ll 1)= \frac{32q}{405\sqrt{3}\, \pi }\left(z^{-7/2} + 13 z^{-5/2}\right) +O(z^{-3/2}\ln z, z^{-3/2}).
\end{equation}
This suggests that successive terms in the $l$-mode series become increasingly more singular in $1/z$. Even though it is not possible to predict the leading-order singular behavior of an arbitrary term $D_{2n}$ based only on the known terms $D_{0,2,4}$ (and this behavior may anyway depend on the extension), it is clear that the limits $l\to \infty$ and $z\to 0$ are {\em not} interchangeable, and that the mode-sum series (\ref{modesum3}) becomes ill-convergent near the LR. For any given $0<z\ll 1$, we expect the series to start showing the standard power-law convergence only for $l\gtrsim \tilde l(z)$, where $\tilde l(z)$ is some monotonically increasing function of $1/z$, with ${\tilde l} (z)\to + \infty$ for $z\to 0^+$. Our numerical experiments confirm these expectations and suggest ${\tilde l}(z)\propto 1/z$---see Fig.\ \ref{fig:lz}.

The evident broadening of the $l$-mode spectrum near the LR is problematic from the practical point of view: at a given $z$, one must compute at least ${\tilde l}(z)$ modes in order to reach the power-law ``tail'' regime where the series begins to converge, and this quickly becomes computationally prohibitive as $z$ gets smaller. Assuming the empirical scaling $\tilde l \propto 1/z$ holds, we find that at least $\sim 1/z$ modes must be calculated. Current codes cannot in practice compute more than a hundred or so modes, which, {\it a priori}, restricts the reach of our analysis to $z\gtrsim 0.01$.

We should comment, in passing, about a more fundamental issue. Strictly speaking, for any (small) nonzero value of the mass ratio $q$, the GSF approximation itself ceases to be meaningful  
sufficiently close to the LR at $z\to 0$. This is because, for a given $q$, there are sufficiently small values of $z$ for which the mass-energy of the small particle becomes comparable to that of the large black hole, at which point perturbation theory clearly breaks down and the notion of GSF is no longer useful.  However, reversing the argument, it is also true that we can make the GSF approximation valid arbitrarily close to the LR simply by taking $q$ to be sufficiently small. Thus, GSF calculations (and ours in particular) can be used to explore the geometry arbitrarily close to the LR. 

\subsection{Raw numerical data for $h^{R,L}_{uu}$}\label{subsec:numerics}

We computed $h^{R,L}_{uu}$ for a dense sample of orbital radii in the range $3m_2<r\leq 150m_2$ using two independent numerical codes. The first code, presented in Ref.\ \cite{Barack:2010tm}, is based on a direct time-domain integration (in 1+1 dimensions) of the metric perturbation equations in the Lorenz gauge. The second code employs a newer algorithm based on a frequency-domain treatment of the Lorenz-gauge perturbation equations \cite{Akcay:2010dx}. Each code takes as input the orbital radius $r$, and returns the value $h^{R,L}_{uu}(r)$ computed via the mode-sum formula (\ref{modesum3}). We typically compute numerically the contributions from the modes $0\leq l\leq 80$, confirm the expected $~l^{-6}$ falloff of the regularized modes (see Fig.\ \ref{fig:lz}), and analytically fit a power-law tail to account for the remaining modes $81\leq l< \infty$. Note that the observed $~l^{-6}$ behavior comes as a result of a delicate cancellation of as many as {\it six} terms in the $1/l$ expansion of the unregularized modes $h_{uu}(r)$ [i.e., the terms of $O(l^0)$ through $O(l^{-5})$]. It thus provides an excellent cross-validation test for both our numerical computation and the analytical parameter values derived in Ref.\ \cite{Heffernan:2012su}. 

The new, frequency-domain, Lorenz-gauge algorithm offers significant computational savings as it only involves solution of {\rm ordinary} differential equations, and since, in our circular-orbit case, the spectrum of the perturbation fields is trivial (it contains only one frequency for each azimuthal $m$-mode).  This is a crucial improvement, because self-force calculations in the time-domain are extremely computationally intensive. The new, frequency-domain code allows us to obtain very accurate results at relatively small computational cost. Nonetheless, we have also used our time-domain code to check (with lower accuracy) many of our data points. 

Our raw numerical data for $h^{R,L}_{uu}(x)$, which form the basis for our analysis, are presented in Appendix \ref{AppA}. The data for $x>1/5$ are new, while our data for $x<1/5$ are much improved in accuracy, and more finely sampled, compared to previous results \cite{Detweiler:2008ft,SBD}.  For most data points the {\it fractional} accuracy of our data is around $\sim 10^{-10}$, decreasing to $\sim 10^{-9}$ at large $r$ and to $\sim 10^{-3}$ very near the LR. (The results of Ref.\ \cite {Blanchet:2010zd}, obtained by a frequency-domain, Regge-Wheeler-gauge method, are  more accurate than ours, but the data shown in that paper are restricted to the weak-field domain $1/500\leq x\leq 1/200$.)

\begin{figure}[Htb]
\includegraphics[width=12cm]{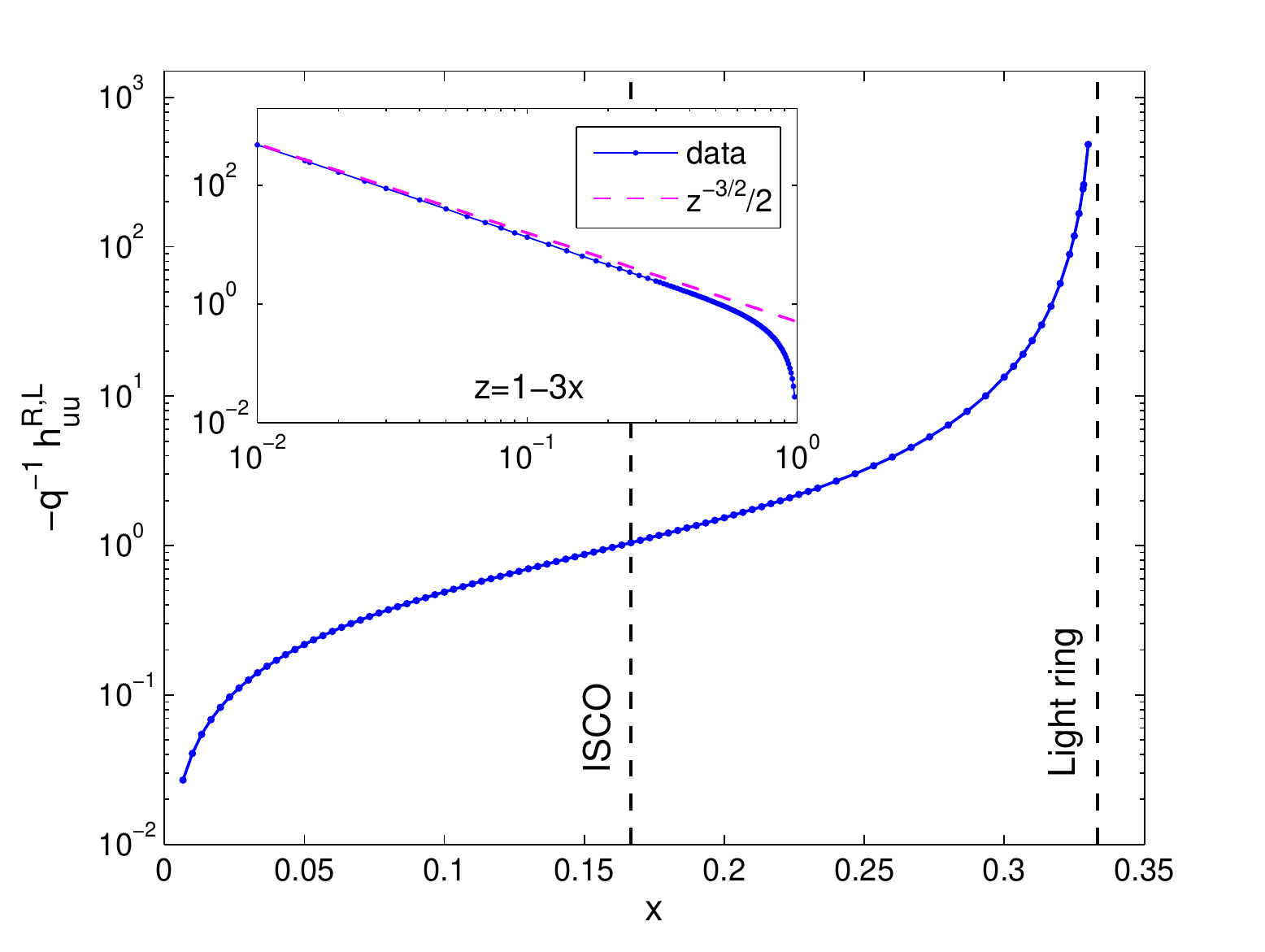}
\caption{Our raw numerical data for the Lorenz-gauge quantity $h^{R,L}_{uu}(x)$ (blue data points; the solid line is an interpolation). The numerical values are tabulated, with error bars, in Appendix \ref{AppA}. Note in the main plot the orbital radius increases to the {\it left}; the locations of the geodesic ISCO ($x=1/6$) and LR ($x=1/3$) are marked with vertical dashed lines. The inset shows the same data (in absolute value) plotted against $z=1-3x$ on a double logarithmic scale (note here the orbital radius increases to the {\it right}, and the LR limit is $z\to 0$ asymptotically far to the left). The dashed (magenta) line is a simple power-law model $h^{R,L}_{uu}\sim -\frac{q}{2}z^{-3/2}$. 
}
\label{fig:huu}
\end{figure}

Our $h^{R,L}_{uu}$ data are plotted in Fig.\ \ref{fig:huu} as a function of $x$. The inset, showing $h^{R,L}_{uu}$ as a function of $z=1-3x$ on a log-log scale, suggests the near-LR power-law behavior 
\begin{equation} \label{LRscaling}
h^{R,L}_{uu}\sim -\frac{q}{2}\, \zeta\, z^{-3/2}\quad \text{as $z\to 0$}, \quad \text{with $\zeta\approx 1$}.
\end{equation}
We will return to discuss the LR behavior in detail in Sec.\ \ref{Sec:photosphere}. 

\section{Determining the EOB potential $a(x)$ below the ISCO}\label{Sec:afit}

\subsection{$a(x)$ from $h^{R,F}_{uu}$  or  $h^{R,L}_{uu}$}

Barausse, Buonanno and Le~Tiec~\cite{Barausse:2011dq} (using the previous results of Le Tiec {\it et al.}~\cite{Tiec:2011ab,Tiec:2011dp}) have derived a simple link between the $O(\nu)$ piece $z_{SF} (x)$ of the function $z_1 (x)$, defined through
\begin{equation}
\label{F12}
z_1 (x) \equiv \sqrt{1-3x} + \nu \, z_{SF} (x) + O(\nu^2) \, ,
\end{equation}
and the $O(\nu)$ piece $a(u)$ of the EOB function $A(u;\nu)$, defined through
\begin{equation}
\label{defa}
A(u;\nu) = 1-2u + \nu \, a (u) + O(\nu^2) \, ,
\end{equation}
namely
\begin{equation}
\label{F13}
a(x) = \sqrt{1-3x} \, z_{SF} (x) - x \left( 1+\frac{1-4x}{\sqrt{1-3x}} \right) \, .
\end{equation}

Let us clarify that in Eq.~(\ref{F13}) we regard $a(\cdot)$ and $z_{SF}(\cdot)$  as {\it functions }, with $x$ merely denoting a ``dummy'' argument.  We could as well have written (\ref{F13}) using the natural notation for the EOB argument of the function $a(\cdot)$, namely
\begin{equation}
\label{F13u}
a(u) = \sqrt{1-3u} \, z_{SF} (u) - u \left( 1+\frac{1-4u}{\sqrt{1-3u}} \right) \, .
\end{equation}
In the following, we shall freely alternate between using $x$ or $u$ as independent variables for the function $a(\cdot)$. Note that the two physical variables $x$ and $u$ satisfy the functional relation $x(u) = u + O(\nu)$ (see below), so that Eqs.~(\ref{F13}) and (\ref{F13u}) would anyway have the same content (at leading order in $\nu$) even if we interpret the arguments $x$ and $u$ as physical variables. 



Putting together the definition (\ref{F12}) of $z_{SF} (x)$ and the previous links (\ref{z1F}) and (\ref{z1L}) between the redshift function $z_1 (x)$ and the metric perturbations, we have the following relations between $z_{SF} (x)$ and the two types of metric perturbations (Flat or Lorenz):
\begin{equation}
\label{F14}
z_{SF} (x) = \sqrt{1-3x} \left[ - \frac{1}{2} \, \tilde h_{uu}^{R,F} + \frac{x}{1-3x} \right],
\end{equation}
\begin{equation}
\label{F15}
z_{SF} (x) = \sqrt{1-3x} \, \left[ - \frac{1}{2} \, \tilde h_{uu}^{R,L} - \frac{x \, (1-2x)}{(1-3x)^{3/2}} + \frac{x}{1-3x} \right] \, .
\end{equation}
Here, the tilde over $h_{uu}^{R,G}$ indicates that one has factored out the mass ratio $q = \nu + O(\nu^2)$:
\begin{equation}
\label{F16}
\tilde h_{uu}^{R,G} \equiv q^{-1} \, h_{uu}^{R,G} \, .
\end{equation}
Finally, inserting Eq.~(\ref{F14}) or Eq.~(\ref{F15}) into Eq.~(\ref{F13}) yields, respectively,
\begin{equation}
\label{F17}
a(x) = -\frac{1}{2} \, (1-3x) \, \tilde h_{uu}^{R,F} - x \, \frac{1-4x}{\sqrt{1-3x}},
\end{equation}
\begin{equation}
\label{F18}
a(x) = - \frac{1}{2} \, (1-3x) \, \tilde h_{uu}^{R,L} - 2x \, \sqrt{1-3x} \, .
\end{equation}

Note in passing that some cancellations took place when replacing $z_{SF} (x)$ in terms of either $\tilde h_{uu}^{R,F} (x)$ or $\tilde h_{uu}^{R,L} (x)$. In particular, when relating $a(x)$ to the Lorenz-gauge perturbation $\tilde h_{uu}^{R,L}$, the single ``extra'' term that remains (the one not involving $\tilde h_{uu}^{R,L}$) has the property of tending towards zero at the LR limit $\left( x \to \frac{1}{3} \right)$, while the corresponding extra term in Eq.~(\ref{F17}) tends to infinity in that limit. We will come back later to a deeper discussion of the physics near the LR.

\subsection{The ``doubly rescaled'' function $\hat a_E(x)$}

Using Eq.\ (\ref{F18}) and our numerical results for $\tilde h_{uu}^{R,L}(x)$, one obtains a dense sample of numerical values for $a(x)$ over the entire range $0<x<\frac{1}{3}$. Our goal now is to obtain a global analytic fit formula that  faithfully represents this relation. 

The function $a(x)$ itself varies rapidly at both ends of the above domain (just like $\tilde h_{uu}^{R,L}$ in Fig.\ \ref{fig:huu}): as we discuss below, $a(x)$ vanishes fast at $x=0$ and blows up at $x=1/3$. Rather than fitting directly for $a(x)$, it is more convenient to fit for a new function, constructed from $a(x)$ by ``factoring out'' suitable terms representing the leading-order behavior at both ends of the domain, so that the resulting ``rescaled'' function is relatively slowly varying over the entire domain. 

Let us first consider the behavior near $x=0$. Information from PN theory determines the form of $a(x)$ in this weak-field regime. Refs.\ \cite{Damour:2000we,Damour:2009sm,Blanchet:2009sd,Blanchet:2010zd,DamourLogs,Barausse:2011dq} obtained the expansion
\begin{equation}\label{aPN}
a_{\rm PN}(x)= \sum_{n=3}^{\infty} (a_n + a_n^{\rm ln}\ln x)x^n,
\end{equation}
where $a_3^{\rm ln}=a_4^{\rm ln}=0$, and the first few nonzero coefficients that can be determined {\it analytically} from available PN expressions are
\begin{eqnarray}\label{PNanalytic}
a_3          &=& 2 , \nonumber \\
a_4          &=& \frac{94}{3}-\frac{41\pi^2}{32} ,  \nonumber \\
a_5^{\rm ln} &=& \frac{64}{5} ,  \nonumber \\
a_6^{\rm ln} &=& -\frac{7004}{105}.
\end{eqnarray}
Note that the leading-order (Newtonian) behavior is $a(x)\sim 2x^3$, and that logarithmic running  first appears at $O(x^5)$. A few more, higher-order coefficients in the expansion (\ref{aPN}) were obtained {\it numerically} in Ref.\ \cite{Barausse:2011dq} by fitting to the accurate large-radius GSF data of Ref.\ \cite{Blanchet:2009sd,Blanchet:2010zd}:  
\begin{eqnarray} \label{a57}
a_5          &=& +23.50190(5),  \nonumber \\ 
a_6          &=& -131.72(1) ,  \nonumber \\
a_7          &=& +118(2) , \nonumber \\ 
a_7^{\rm ln} &=& -255.0(5), 
\end{eqnarray}
where in each case a parenthetical figure indicates the estimated uncertainty in the last decimal place. Let us introduce the notation $\hat a(x)$ to denote the normalization of the function $a(x)$ using the leading-order PN term, i.e.,
\begin{equation}
\hat a(x) \equiv \frac{a(x)}{2x^3},
\end{equation}
so that $\hat a(0)=1$.

Consider next the behavior near the LR, $x_{LR}=\frac13$. This behavior
has not been studied so far, neither numerically, nor analytically. Note,
however, that Ref.\ \cite{Tiec:2011dp} has remarked that the extrapolation
beyond $x=\frac15$ of a 5-parameter fit to 55 data points (ranging between
$x=0$ and $x= \frac15$)  for
$z_{SF}(x)$ indicated the possible presence of a simple pole $z_{SF}(x)
\sim (x-x_{\rm pole})^{-1}$ located near the LR. (Their fit yields $x_{\rm pole}
\approx 0.335967$, which is slightly beyond the LR.)  In this work, we
study the behavior near $x=\frac13$ both numerically and analytically.
We have already mentioned that our data suggest the  scaling relation
({\ref{LRscaling}) (which corresponds to a simple pole in $z_{SF}(x)$
located exactly at the LR) .  Combined with Eq.\ (\ref{F18}), this
suggests the leading order divergent behavior
\begin{equation}\label{aLR}
a(x\to \frac{1}{3})\sim \frac{\zeta}{4}(1-3x)^{-1/2},
\end{equation}
where, recall, the ``fudge'' factor $\zeta$ is $\approx 1$.
We will discuss the analytical origin of this asymptotic behavior in Sec.\
\ref{Sec:photosphere} below.

Equation (\ref{aLR}) suggests that it would be convenient to further normalize the function $\hat a(x)$ by a factor $(1-3x)^{-1/2}$. However, as will be further discussed below, there is a more physically motivated normalization: we recall that the conserved {\it specific energy} associated with $m_1$ as it moves along a circular geodesic orbit in the Schwarzschild background of $m_2$ is given by
\begin{equation}\label{E}
E(x)=\frac{1-2x}{\sqrt{1-3x}},
\end{equation}
which has the same type of divergent behavior as $a(x)$ for $x\to \frac{1}{3}$ (but is regular elsewhere). We hence choose to use $E$ for our second normalization. Let us introduce a notation whereby a sub-index $E$ denotes normalization with respect to $E(x)$; in particular,
\begin{equation}\label{aE}
a_E(x)\equiv \frac{a(x)}{E(x)}.
\end{equation}
Note $a_E(1/3)\sim \frac{3}{4}\zeta$.

Let us finally introduce the ``doubly-rescaled'' function 
\begin{equation}\label{hataE}
\hat a_E(x)\equiv \frac{a(x)}{2x^3 E(x)},
\end{equation}
which attains finite, nonzero values at both ends of the domain $0\leq x\leq \frac{1}{3}$, namely
$\hat a_E(0)=1$ and $\hat a_E(1/3)\sim \frac{81}{8}\zeta$. Our numerical dataset for $\hat a_E(x)$ is plotted in Fig.\ \ref{fig:aE}. Evidently, the function $\hat a_E(x)$ is monotonically increasing and convex over $0\leq x\leq \frac{1}{3}$.
\begin{figure}[Htb]
\includegraphics[width=12cm]{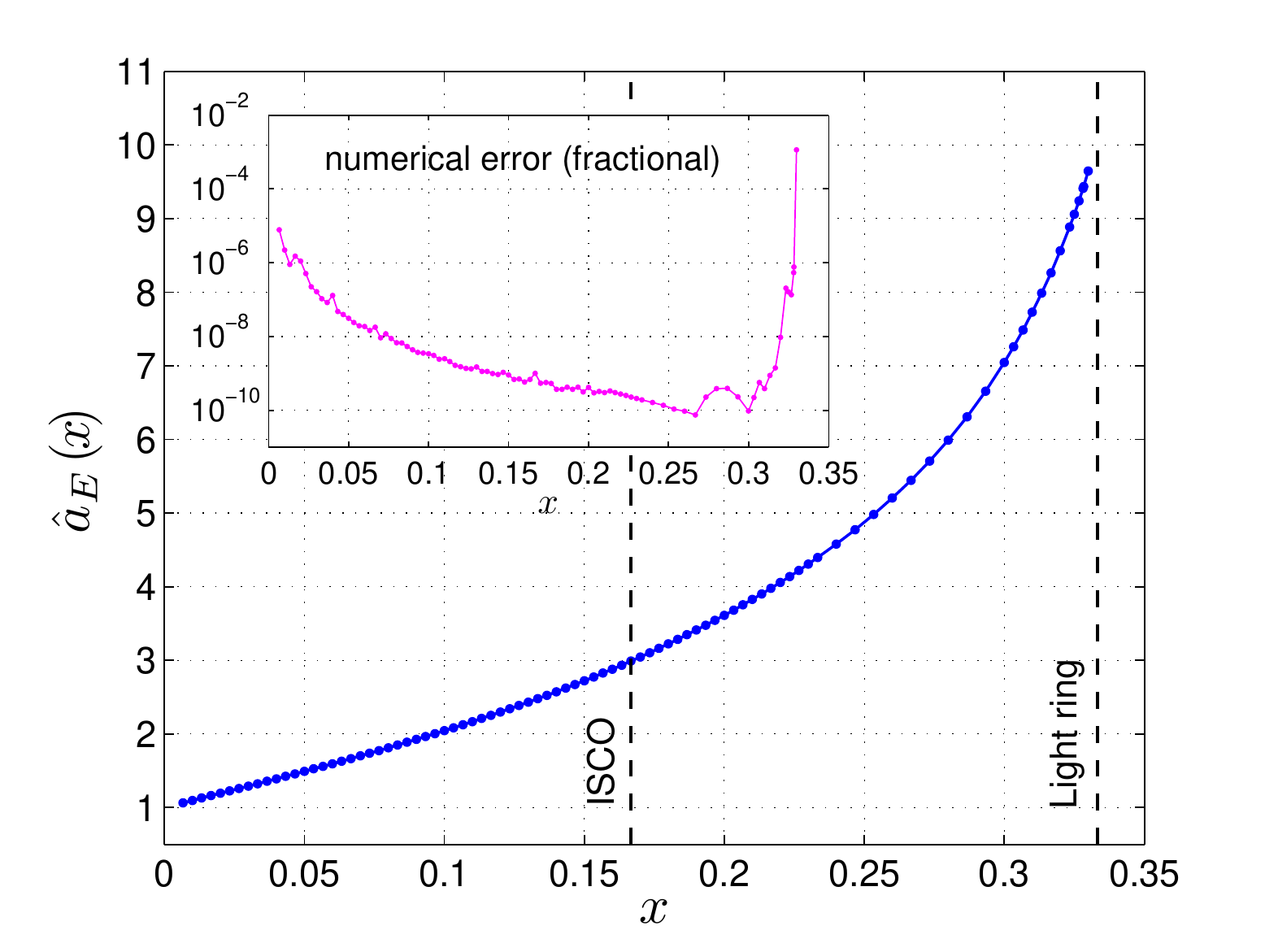}
\caption{Numerical data for the doubly-rescaled function $\hat a_E(x)$ [see Eq.\ (\ref{hataE})]. The solid line is a cubic interpolation of the numerical data points (beads). The inset shows, on a semi-logarithmic scale, the relative numerical error in the $\hat a_E$ data, computed based on the estimated errors tabulated in Appendix \ref{AppA}. Note that the relative error is between $10^{-8}$ and $10^{-10}$ over most of the domain, and it never exceeds $10^{-5}$ (except at a single point, closest to the LR, where it is $\sim 0.1\%$). 
}
\label{fig:aE}
\end{figure}

\subsection{Accurate global analytic model}

We have explored a large set of global analytic models for the function $\hat a_E(x)$. In each case we used a least-squares approach, i.e.~given the data (sampled at a discrete set ${x_1, x_2, \cdots}$ of $x$ values), together with an estimate of the corresponding standard data errors $\sigma_{\rm data}(x_i)$, 
we minimized (using 
\textsc{Mathematica}'s function \texttt{NonlinearModelFit[\,]}) the standard  $\chi^2$ statistic 
\begin{equation}\label{chi2}
\chi^2({\rm parameters}) = \sum_i \left( \frac{{\rm data}(x_i)-{\rm model}(x_i, {\rm parameters})}{\sigma_{\rm data}(x_i)}  \right)^2 \, 
\end{equation}
over the model parameters to determine the best-fit parameters. In addition, we 
evaluated the faithfulness of the fit by recording the minimum value of  $\chi^2$: $\chi^2_{\rm min}=\chi^2(\rm{best\, fit \, parameters})$.
 If numerical errors were normally distributed without a systematic bias (which we assume here), and if the model were ``true'', then we would expect $\chi^2_{\rm min}$ to equal approximately 
the number of degrees of freedom (DoF) in our model. As a second measure of the fit quality we also considered the norm $\left|\left|a_E^{\rm fit}-a_E^{\rm data}\right|\right|_{\infty}$, i.e., the maximal absolute difference between the best-fit model and the data over all data points. We are using here $a_E(x)$ [rather than $a(x)$ or $\hat a_E(x)$] because this is the relevant quantity entering the EOB expressions, as we discuss in Sec.\ \ref{Sec:photosphere} below.

We report here some of our results, and present two selected models: a 16-parameter high-accuracy model with $\chi^2$/DoF of order unity; and (in the following subsection) a simpler, 8-parameter model, which is less accurate (has a very large $\chi^2$ value) but has a sufficiently small norm $\left|\left|a_E^{\rm fit}-a_E^{\rm data}\right|\right|_{\infty}$ to be useful in some foreseeable applications. 

Let us focus, for ease of presentation, on the following restricted class of analytic models, which employ a Pad\'e-like approximant for $\hat a_E$:
\begin{equation} \label{fit}
\hat a_E^{\rm fit}=
\frac{1 + \sum_{i=1}^{p}\left(c_i+c_i^{\rm log}\ln x\right)x^i
+x^3 z\ln|z| \left(c_{z0} + c^{\rm log}_{z0}\ln|z| + c_{z1}z \right)  }
     {1 + \sum_{j=1}^{q} d_j x^j}.
\end{equation}
Here $p\geq 3$ and $q\geq 1$ are constant integers (see below), and $\{c_{i\geq 2},c_{i\geq 4}^{\rm log},d_i ,c_{z0},c_{z0}^{\rm log},c_{z1}\}$ are the model parameters to be fitted. The first few $c$ parameters are constrained so as to reproduce all analytically available PN information:
\begin{eqnarray}\label{PNconstraints}
c_1&=&\frac{97}{6}-\frac{41\pi^2}{64}+d_1 ,  \nonumber \\
c_1^{\rm log} &=& 0 ,  \nonumber \\
c_2^{\rm log} &=& \frac{32}{5} , \nonumber \\
c_3^{\rm log} &=& -\frac{3166}{105}+\frac{32}{5}d_1.
\end{eqnarray}
We do not constrain the remaining parameters to agree with the additional PN information available through numerical fit [Eq.\ (\ref{a57})], but rather allow our model to ``re-fit'' some of these high-order PN terms. We find, in general, that this leads to improved global fits.

Our model family $\hat a_E^{\rm fit}(x)$ is designed (heuristically) to capture all global features of $\hat a_E(x)$ from $x=0$ down through the LR and (potentially) beyond. We use a Pad\'e-type expansion in $x$ (with logarithmic running terms), augmented with $z$-dependent terms which are aimed at capturing the behavior near the LR. The latter terms are multiplied by $x^3$ to suppress their support in the weak field, where the known PN behavior should apply (we have tried various powers of $x$ and found that $x^3$ generally works best). 

The form of the $z$-dependent terms in (\ref{fit}) is motivated as follows. We have initially experimented with simple polynomials in $z$ (without logarithmic terms), but found that these always yielded best-fit models that possessed poles (singularities) immediately behind the LR (i.e., just above $x=\frac{1}{3}$). This suggested to us that the true function $\hat a_{E}(x)$ has a remaining non-smoothness at $x=\frac{1}{3}$, and the form of the function suggested a weakly divergent derivative. In our model family (\ref{fit}) we have attempted to represent this type of non-smoothness with a term of the form $\sim z\ln|z|$, which indeed seemed to have the effect of removing the undesired pole. To allow more freedom in fitting the correct LR behavior we have added a few higher-order terms in $z$ and $\ln|z|$. We experimented with a large variety of such higher-order term combinations, and found that the form shown in (\ref{fit}) worked well (while minimizing the number of extra model parameters).

Each member of our model family $\hat a_E^{\rm fit}(x;p,q)$ has $2p+q-1$ fitting parameters. In Table \ref{Table:afit} we show fitting results for a variety of $p,q$ values (and also for models in which we remove some of the $\ln x$ terms). For each fitting model we compute the $\chi^2$ statistic using as weights the estimated numerical errors from Tables \ref{table:data1} and \ref{table:data2}. For each best-fit model we also display in Table \ref{Table:afit} the value of the norm 
$\left|\left|a_E^{\rm fit}-a_E^{\rm data}\right|\right|_{\infty}$. Some of the models presented in Table \ref{Table:afit} have remaining poles on $x>1/3$, and we indicate in the table the location of the first pole below the LR if any occurs. Finally, the table shows the predicted value of the fudge factor $\zeta$ for each of the models.

\begin{table}[Htb]
\begin{tabular}{||c||c|c|c||c|c|c|c||c||}
\hline\hline
model  &
\multicolumn{3}{|c||}{Fit model [Eq.\ (\ref{fit})]} & 
$\#$ model  &
\multirow{2}{*}{$\chi^2$/DoF}  & 
\multirow{2}{*}{$\left|\left|a_E^{\rm fit}-a_E^{\rm data}\right|\right|_{\infty}$} & 
\multirow{2}{*}{pole?} &
\multirow{2}{*}{$\zeta$} \\
\cline{2-4}
$\#$ & $p$ \quad &   $q$ \quad & params.\ set to zero & parameters  & & & &
\\ \hline\hline
1 & \multirow{2}{*}{4} & \multirow{2}{*}{4} 
	&	$c_4^{\rm log}$	&	10	&	$4.45\times 10^6$	&	$1.62\times 10^{-2}$	& --- &	0.991785 \\
\cline{4-9}
2 & &	&	---	&	11	&	$6.71\times 10^4$	&	$4.01\times 10^{-3}$	& --- &	1.00984 \\
\hline
3 & \multirow{2}{*}{4} & \multirow{2}{*}{5}
	&	$c_4^{\rm log}$	&	11	&	$3.81\times 10^4$	&	$2.65\times 10^{-4}$	& --- &	1.00791 \\
\cline{4-9}
4 & &	&	---	&	12	&	$2.81\times 10^{3}$	&	$1.46\times 10^{-4}$	& --- &	1.00192 \\
\hline
5 & \multirow{3}{*}{5} & \multirow{3}{*}{5}
 &	$c_4^{\rm log}$, $c_5^{\rm log}$	&	12	& $5.55\times 10^3$	&	$4.53\times 10^{-5}$	& --- &	1.00408 \\
\cline{4-9}
6 & & &	$c_5^{\rm log}$	&	13	& $2.79\times 10^{3}$	&	$1.37\times 10^{-4}$	& --- &	1.00214 \\
\cline{4-9}
7 & & &	---	&	14	& $1.52\times 10^3$	&	$1.48\times 10^{-4}$	& $x\approx 0.45$ &	0.999680 \\
\hline
8 & \multirow{3}{*}{5} & \multirow{3}{*}{6}
 &	$c_5^{\rm log}$ &	14	& $1.83\times 10^3$	&	$1.41\times 10^{-4}$	& --- &	1.00137 \\
\cline{4-9}
9 & & &	---	&	15	& $1.03\times 10^3$	&	$1.98\times 10^{-4}$	& $x\sim 0.35$ &	0.988620 \\
\hline
10& \multirow{4}{*}{6} & \multirow{4}{*}{7}
& 	$c_4^{\rm log}$, $c_5^{\rm log}$, $c_6^{\rm log}$	&	15	& $19.2$	&	$1.12\times 10^{-6}$	&  --- &	1.00750 \\
\cline{4-9}
11 & & &	$c_5^{\rm log}$, $c_6^{\rm log}$	&	16	& $9.97$	&	$1.08\times 10^{-5}$	& $x\approx 0.42$ &	1.00536 \\
\cline{4-9}
12 & & &	$c_6^{\rm log}$	&	17	& $3.37$	&	$2.52\times 10^{-6}$	& $x\approx 0.375$ &	1.00525 \\
\cline{4-9}
13 & & &	---	&	18	& $4.94$	&	$2.01\times 10^{-5}$	& --- &	1.00907 \\
\hline
\bf 14 & \multirow{5}{*}{7} & \multirow{5}{*}{7}
& 	$c_4^{\rm log}$, $c_5^{\rm log}$, $c_6^{\rm log}$, $c_7^{\rm log}$	&	{\bf 16}	& {\bf 4.77}	&	$\mathbf{1.97\times 10^{-5}}$	&  --- &	\bf 1.00899 \\
\cline{4-9}
15 & & &	$c_5^{\rm log}$, $c_6^{\rm log}$, $c_7^{\rm log}$	&	17	& 3.08	&	$3.81\times 10^{-6}$	& $x\approx 0.36$ &	1.00453 \\
\cline{4-9}
16 & & &	$c_6^{\rm log}$, $c_7^{\rm log}$	&	18	& 3.03	&	$5.81\times 10^{-6}$	& $x\approx 0.35$ &	1.00345 \\
\cline{4-9}
17 & & &	$c_7^{\rm log}$	&	19	& 2.87	&	$4.28\times 10^{-6}$	& $x\approx 0.575$ &	1.00968 \\
\cline{4-9}
18 & & &	---	&	20	& 2.93	&	$2.62\times 10^{-6}$	& $x\approx 0.58$ &	1.00918 \\
\hline
19 & \multirow{5}{*}{7} & \multirow{5}{*}{8}
& 	$c_4^{\rm log}$, $c_5^{\rm log}$, $c_6^{\rm log}$, $c_7^{\rm log}$	&	17	& 4.79	&	$1.79\times 10^{-5}$	&  --- &	1.00882 \\
\cline{4-9}
20 & & &	$c_5^{\rm log}$, $c_6^{\rm log}$, $c_7^{\rm log}$	&	18	& 3.04	&	$5.71\times 10^{-6}$	& $x\approx 0.35$ &	1.00350 \\
\cline{4-9}
21 & & &	$c_6^{\rm log}$, $c_7^{\rm log}$	&	19	& 3.01	&	$5.46\times 10^{-6}$	& $x\approx 0.35$ &	1.00373 \\
\cline{4-9}
22 & & &	$c_7^{\rm log}$	&	20	& 2.87	&	$2.51\times 10^{-6}$	& $x\approx 0.53$ &	1.00775 \\
\cline{4-9}
23 & & &	---	&	21	& 2.87	&	$2.16\times 10^{-6}$	& $x\approx 0.5$ &	1.00732 \\
\hline\hline
\bf 24 & 4 & 4 & $c_4^{\rm log}$, $c_{z0}^{\rm log}$, $c_{z1}$ & \bf 8 & $\mathbf{1.08\times 10^7}$ & $\mathbf{1.21 \times 10^{-5}}$ & $x\sim 0.7$ & $\mathbf{1.00554}$ \\
\hline\hline
\end{tabular}
\caption{Model fitting for the doubly-rescaled function $\hat a_E(x)$.  Each row describes best-fit results for the model $\hat a_E^{\rm fit}(x)$ given in Eq.\ (\ref{fit}), with particular values of $p$ and $q$; in some of the models we have eliminated some of the fitting parameters, as indicated in the fourth column. [In the last row we show best-fit results for the model $\hat a_E^{\rm fit,simp}(x)$ given in Eq.\ (\ref{fitsimple}), to be discussed in Sub.\ \ref{subsec:simp} below].
The fifth column shows the total number of fitting parameters for each model, and the sixth columns displays the value of $\chi^2$/DoF for the best fit parameters. In the seventh column we show the maximal difference between the model (with the best fit parameters) and the data, for the physically relevant quantity $a_E(x)\equiv a(x)/E(x)$. In the penultimate column we indicate the location of the first pole of $\hat a^{\rm fit}_E(x)$ [which is the same as for $a^{\rm fit}_E(x)$ or $a^{\rm fit}(x)$].
The last column presents the value of the fudge factor $\zeta=\frac{4}{3}a_{E}(1/3)$ [see Eq.\ (\ref{aLR})], as predicted by each of the best-fit models.
Highlighted in boldface are values for the two selected models ($\#14$ and $\#24$) whose parameters are given, respectively, in Tables \ref{Table:fit model} and \ref{Table:simple_model} below.
}
\label{Table:afit}
\end{table}

The data in Table \ref{Table:afit} suggest that, at least within the model family (\ref{fit}), one cannot obtain a good fit for $\hat a_E(x)$ with just a handful of model parameters. At least $14$ parameters are needed to achieve $\chi^2/{\rm DoF}\sim 1000$ and at least 16 for $\chi^2/{\rm DoF}<10$. However, the value of $\chi^2$/DoF becomes saturated at around $3$ or $4$ for $\gtrsim 16$ parameters, and does not decrease much further upon adding extra parameters (this may indicate that our quoted numerical errors $\sigma_i$ are slightly too optimistic, consistent with our estimated factor $\sim 2$ uncertainty in the values of the quoted $\sigma_i$; see Appendix \ref{AppA}). We have experimented with several other model families but were not able to achieve $\chi^2/{\rm DoF}$ values of order unity with less than 16 parameters. 

We choose to present here the accurate 16-parameter model highlighted in Table \ref{Table:afit} (model $\#14$), which has $\chi^2$/DoF=4.77 and $\left|\left|a_E^{\rm fit}-a_E^{\rm data}\right|\right|_{\infty}= 1.97\times 10^{-5}$. Some of the other models in the table have slightly lower values of $\chi^2$/DoF but in all such cases the models are more complicated (have more parameters) and, more importantly, they present undesired poles below the LR. The best-fit parameters for our selected model are shown in Table \ref{Table:fit model}. Note that we are giving the parameter values accurate to many decimal places: this accuracy is necessary for our model to reproduce the data down to the level of the numerical noise, which is generally as small as $10^{-9}$--$10^{-10}$ (fractionally). We have checked that all decimal figures shown are significant, in the sense that omitting any of the figures would result in $\chi^2$/DoF values larger than 4.77.

\begin{table}[Htb]
\begin{tabular}{c|ccc}
\hline\hline
$i$\quad		&	$c_i$	& $c_i^{\rm log}$	&	$d_i$	 \\
\hline\hline
$1$	\quad 	&	$+7.48610059021$	&	0							&	$-2.357850757006$\\
$2$	\quad	& 	$-8.81722069138$	&	$+\frac{32}{5}$				&	$-1.889967139293$\\
$3$	\quad 	& 	$-227.6806641934$	&	$-45.2426257972$			&	$-109.86788081837$\\	
$4$	\quad 	&	$-1336.5402672986$	&	0							&	$+535.45853874191$\\
$5$	\quad 	&	$+8044.588011262$	&	0							&	$-53.572041734$\\
$6$	\quad 	&	$-5643.745303388$	&	0							&	$-3030.7781195456$\\
$7$	\quad 	&	$-7744.83943928$	&	0							&	$+4106.962599268$\\
\hline
\multicolumn{4}{c}{$c_{z0}=-32.8395937428$}\\
\multicolumn{4}{c}{$c_{z0}^{\rm log}=-4.34430971904$}\\
\multicolumn{4}{c}{$c_{z1}=-365.569972774$}\\
\hline\hline
\end{tabular}
\caption{Parameter values for the 16-parameter model highlighted in Table \ref{Table:afit}. The model belongs to the family (\ref{fit}), with $c_1$ and $c_{1,2,3}^{\rm log}$ constrained in accordance with (\ref{PNconstraints}) so as to impose all PN information available analytically. This 16-parameter model has $\chi^2{\rm /DoF}\sim 4.77$ and it reproduces all of our numerical data points for $a_E(x)$ to within an absolute difference of $1.97\times 10^{-5}$. (The difference for most data points is actually much smaller---see Fig.\ \ref{fig:diff}.) All decimal figures are significant, in the sense that removal of any figure would lead to $\chi^2{\rm /DoF}>4.77$.
}
\label{Table:fit model}
\end{table}

Figure \ref{fig:diff} shows a performance diagnostic for our selected model. From the left panel it is evident that our model for $a_E(x)$ reproduces most data points to within mere differences of $10^{-10}$--$10^{-12}$. Larger differences appear only at $x\gtrsim 0.3$, in which domain our data is less accurate.  The right panel compares the difference between the model and the data with the estimated numerical error in the data points. We observe that most data points are reproduced by the model at the level of the numerical noise, as desired. 
\begin{figure}[Htb]
\includegraphics[width=8.5cm]{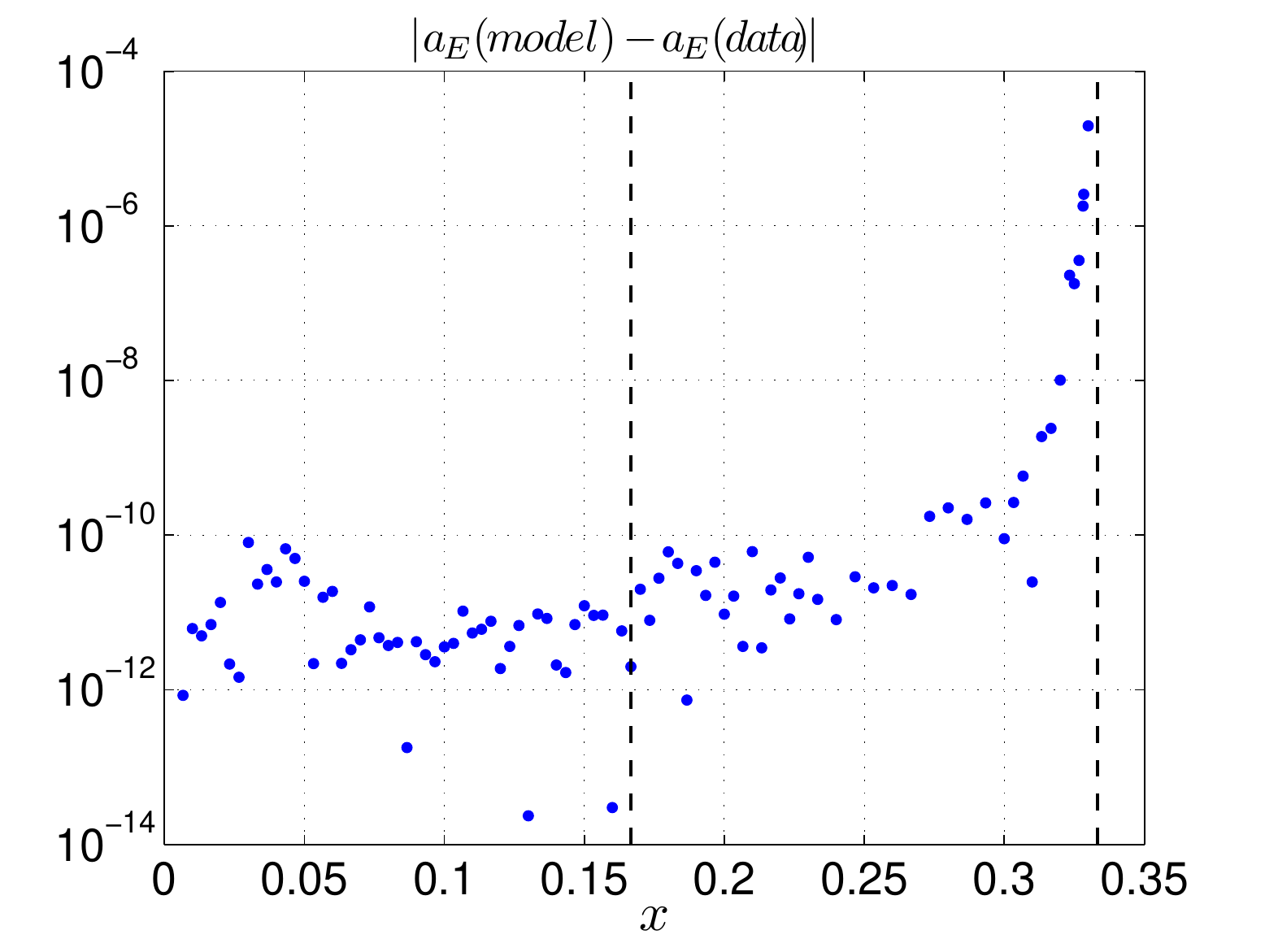}
\includegraphics[width=8.5cm]{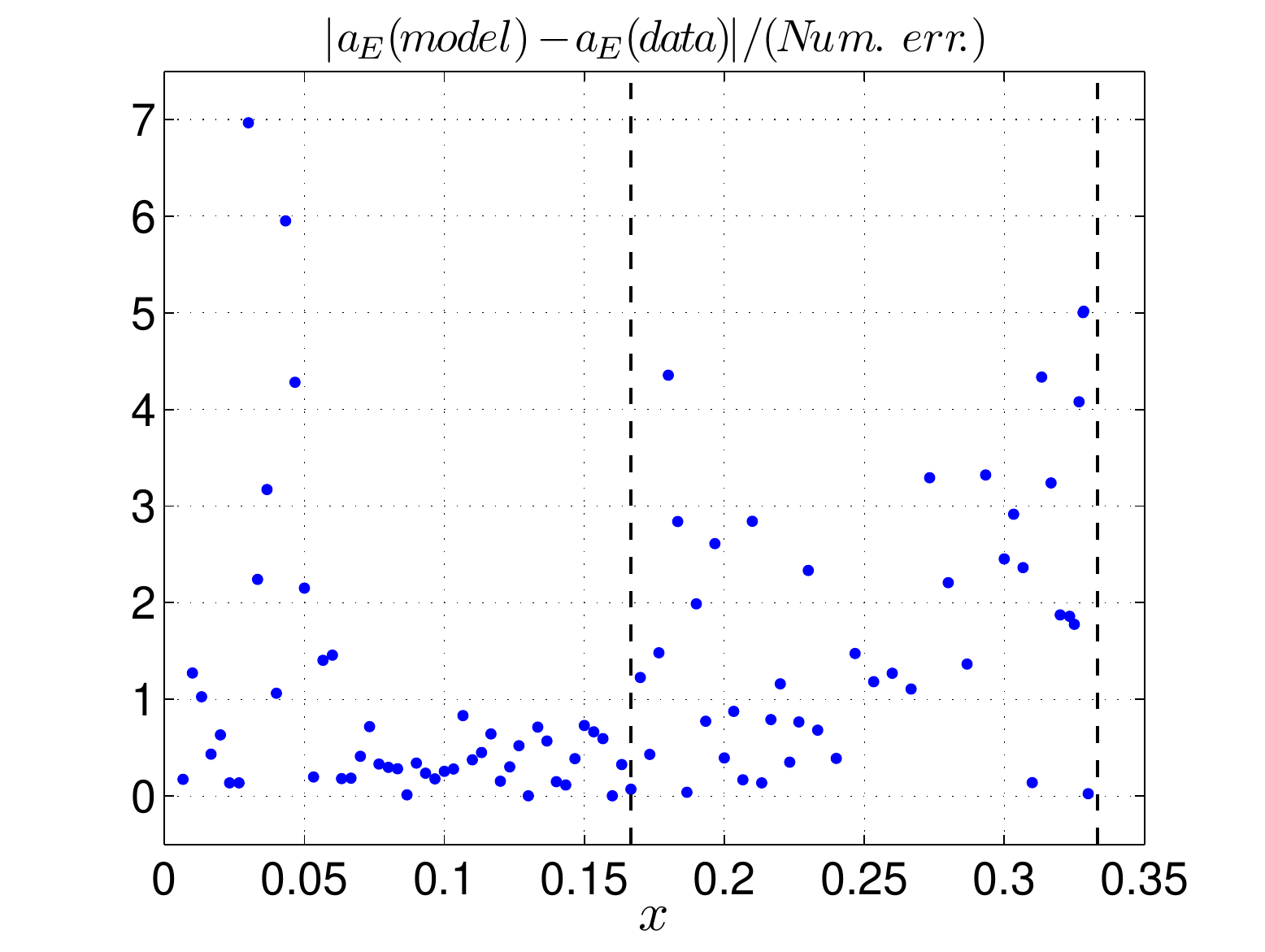}
\caption{Faithfulness of the analytic best-fit model (\ref{fit}), with parameters as given in Table \ref{Table:fit model}. The {\it left panel} shows, on a semi-logarithmic scale, the magnitude of the absolute difference between the model and the data; we use here the variable $a_E(x)$ [rather than $\hat a_E(x)$], which is the relevant one entering the EOB potential. The {\it right panel} shows (now on a linear scale) that same difference divided by the estimated numerical error for each data point. For most data points the model reproduces the data down to the level of our numerical noise.
}
\label{fig:diff}
\end{figure}

In Fig.\ \ref{fig:global} we plot our selected model for $\hat a_E(x)$ over the {\it entire domain} $0<x<1$, showing how it extends beyond the LR. We observe that the function $\hat a_E(x)$ peaks closely below the LR, then drops and changes its sign around $x\sim \frac{1}{2}$ (the location of the event horizon in the background geometry). To assess the robustness of these features we have also plotted in Fig.\ \ref{fig:global} the global extensions of a handful of other models---all models from Table \ref{Table:fit model} with $\left|\left|a_E^{\rm fit}-a_E^{\rm data}\right|\right|_{\infty}<10^{-4}$ admitting a smooth behavior with no poles below the LR. Remarkably, the above basic features seem to be preserved: a peak right below the LR, followed by a change of sign. The function $\hat a_E(x)$ generally turns negative at $x$ values in the range $\sim 0.5$--$0.6$ (more towards $0.5$ for the more accurate models). Note that the function $a_E(x)$ [as well as $a(x)$ itself] turns negative at precisely the same location as $\hat a_E(x)$. Note also that $a_E(x)$ vanishes at the proximity of $x=\frac{1}{2}$ {\it despite} having $E(x)$ (which vanishes at $x=\frac{1}{2}$) being factored out in its definition [recall $a_E(x)\equiv a(x)/E(x)$]. This may (heuristically) point to a rather rapid vanishing of $a(x)$ at the horizon. Whether or not the above features are indeed robust remains to be verified. 

\begin{figure}[Htb]
\includegraphics[width=12cm]{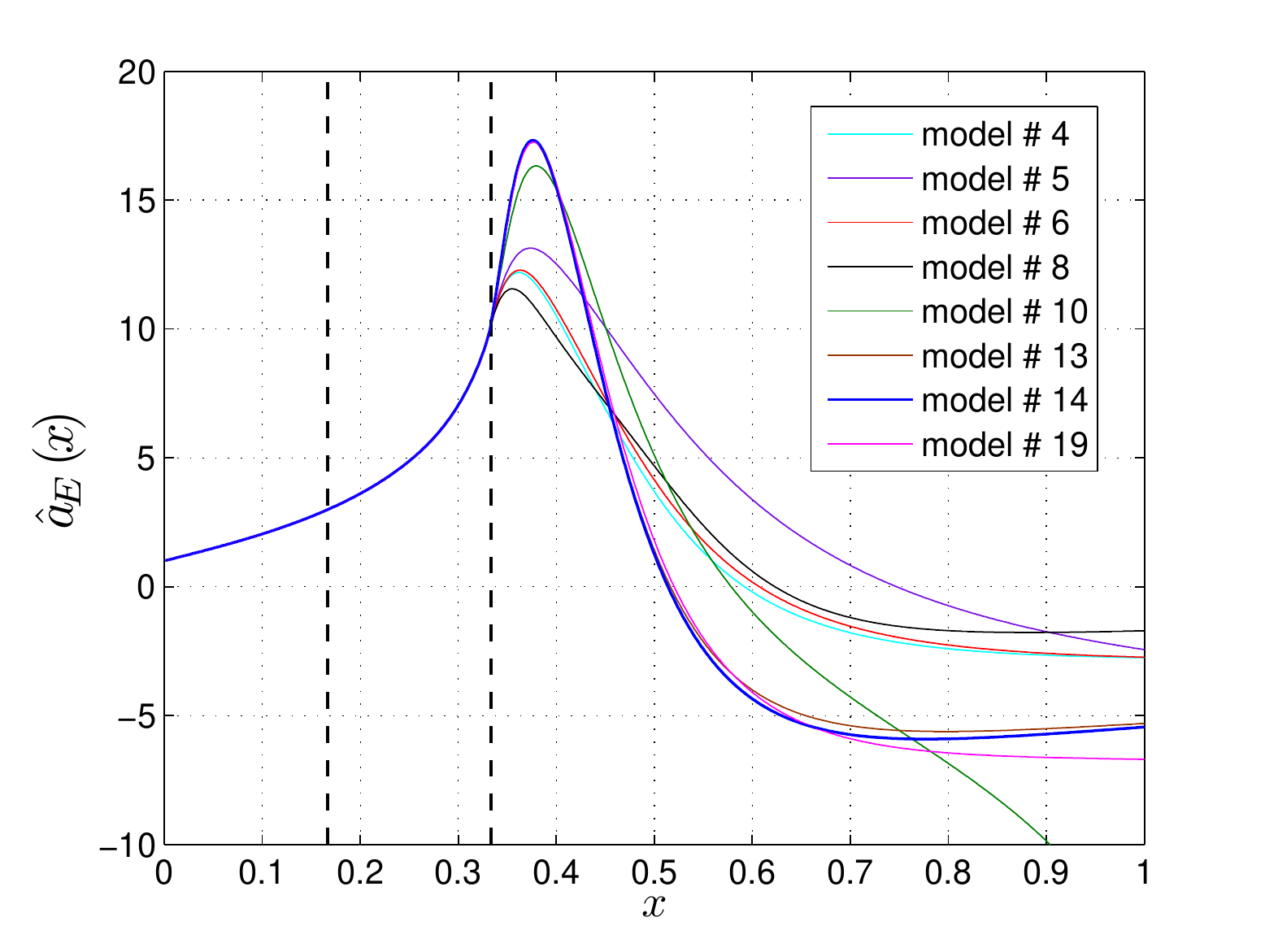}
\caption{Extension of our analytic $\hat a_E(x)$ models below the LR. The thick (blue) curve shows the behavior of our selected model ($\#14$ in Table \ref{Table:afit}, and Table \ref{Table:fit model}) over the entire domain $0<x<1$. Other curves, labelled by model numbers from Table \ref{Table:afit}, show the behavior of other models for comparison. Shown, from top to bottom at $x=0.8$, are models number 5, 8, 6, 4, 13, 14, 19 and 10. 
}
\label{fig:global}
\end{figure}


\subsection{Simpler global analytic model}\label{subsec:simp}

In the above ``high fidelity'' 16-parameter model, the maximum norm of differences $\left|\left|a_E^{\rm fit}-a_E^{\rm data}\right|\right|_{\infty}$ is determined by the data point nearest the LR, which is our least accurate point. As can be seen from Fig.\ \ref{fig:diff} (left panel), removing just a few near-LR points from our sample would result in a norm of a mere $\sim 10^{-10}$. This high standard of accuracy may not be necessary for some applications. Indeed, in designing analytical waveform templates for comparable-mass binaries one usually has no reason to require the model to be locally accurate at that level. It is therefore both convenient and useful to have at hand a simpler, less unwieldy model, which reproduces all numerical data points to within a prescribed accuracy (say, a fiducial $\lesssim 10^{-5}$), but not necessarily to within the (very high) accuracy of our numerical computation; in other words, a model in which we relax the requirement $\chi^2/{\rm DoF}\sim 1$ and replace it with an upper bound on the norm $\left|\left|a_E^{\rm fit}-a_E^{\rm data}\right|\right|_{\infty}$.

Through experimentation, we were able to devise an 8-parameter model with a norm as small as $\left|\left|a_E^{\rm fit}-a_E^{\rm data}\right|\right|_{\infty}= 1.2\times 10^{-5}$ (that is, even slightly smaller 
than the norm for our 16-parameter model). The model is given by 
\begin{equation} \label{fitsimple}
\hat a_E^{\rm fit,simp}=
\frac{1 + c_1 x+x^2(c_2+c_2^{\rm log}\ln x)+x^3(c_3+c_3^{\rm log}\ln x) + c_4 x^4 +c_{z0}x^4 z\ln|z|}
     {1 + d_1 x+d_2 x^2+d_3 x^3+d_4 x^4},
\end{equation}
where $c_1$ and $c_{2,3}^{\rm log}$ are again PN-constrained as in Eq.\ (\ref{PNconstraints}), and $\{c_2,c_3,c_4,c_{z0},d_1,d_2,d_3,d_4\}$ are 8 independent model parameters. The best-fit values of these parameters are given in Table \ref{Table:simple_model}, and the residual differences between the model and the data are plotted in Fig.\ \ref{fig:diff_simp}. We see that the model reproduces the numerical $a_E(x)$ data to within $10^{-7}$ for $x\lesssim 0.28$ and to within $10^{-6}$ for $x\lesssim 0.32$. 

We note that the above simple model has a very large value of $\chi^2$/DoF ($\sim 1.08\times 10^7$). Also, its behavior beyond the LR is problematic and rather different from that shown in Fig.\ \ref{fig:global}: the function $\hat a_E(x)$ does not reach a maximum but instead it grows monotonically with $x$ and eventually blows up at a pole located at $x=0.695694\ldots$ (i.e., below the background horizon).  However, we emphasize, the model reproduces all of the numerical data points for the function $a_E(x)$ to within a maximal absolute difference of a mere $1.2\times 10^{-5}$ over the entire domain $0<x<\frac{1}{3}$.
\begin{table}[Htb]
\begin{tabular}{c|ccc}
\hline\hline
$i$\quad		&	$c_i$	& $c_i^{\rm log}$	&	$d_i$	 \\
\hline\hline
$1$	\quad 	&	$+2.0154525$	&	0						&	$-7.82849889$\\
$2$	\quad	& 	$-39.57186$	    &	$+\frac{32}{5}$			&	$+20.84938506$\\
$3$	\quad 	& 	$-24.30744$		&	$-80.254774$			&	$-20.5092515$\\	
$4$	\quad 	&	$+103.93432$		&	0					&	$+5.383192$\\
\hline
\multicolumn{4}{c}{$c_{z0}=+10.22474$}\\
\hline\hline
\end{tabular}
\caption{Parameter values for the ``simple'' model (\ref{fitsimple}). The parameters $c_1$ and $c_{1,2,3}^{\rm log}$ are constrained in accordance with Eq.\ (\ref{PNconstraints}), and the other 8 parameters are found by model fitting. This 8-parameter model reproduces all of our numerical data points for $a_E(x)$ to within an absolute difference of $1.2\times 10^{-5}$. All decimal figures are significant, in the sense that removal of any figure would lead to a larger difference.
}
\label{Table:simple_model}
\end{table}

\begin{figure}[Htb]
\includegraphics[width=9cm]{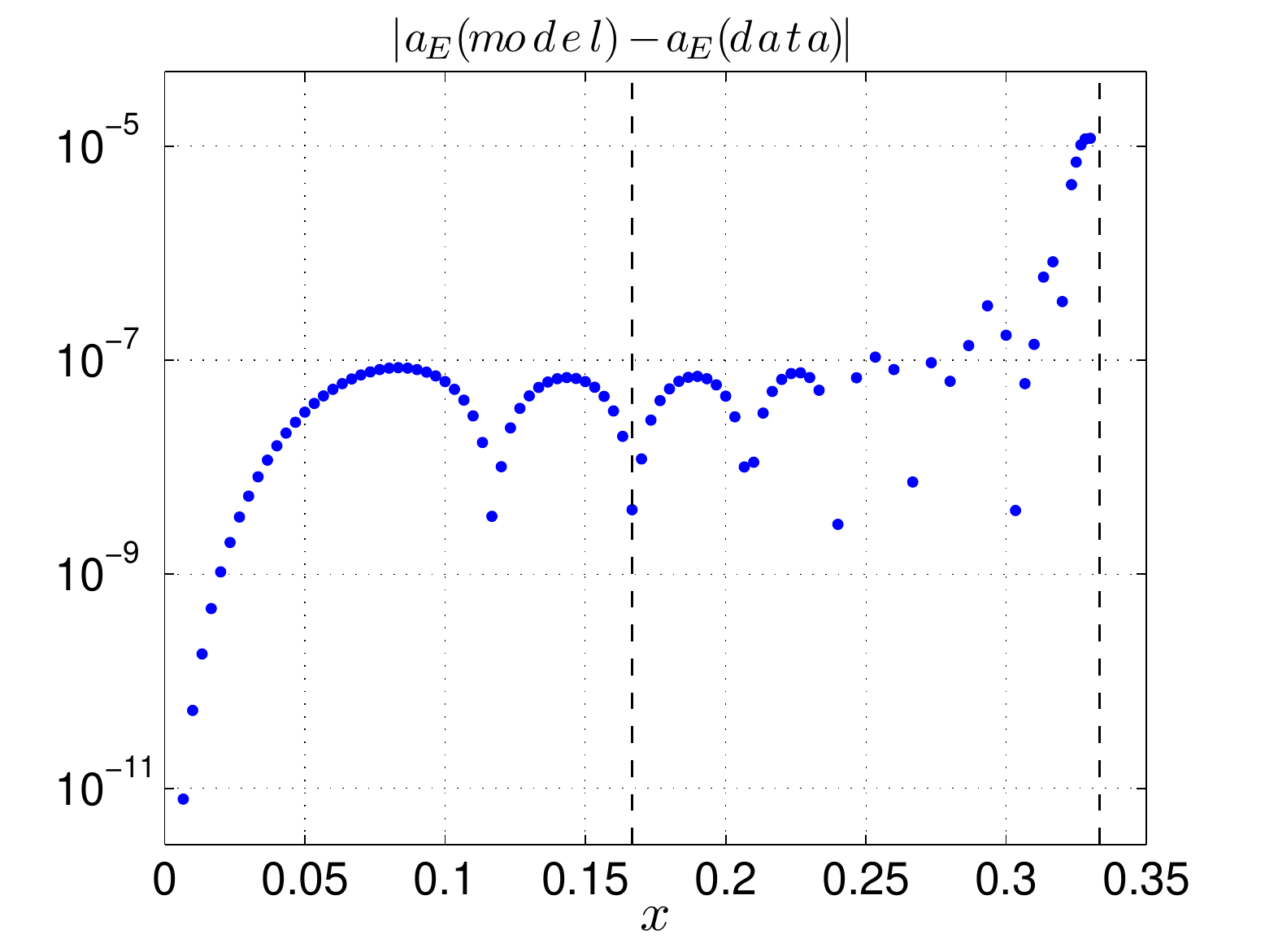}
\caption{Performance of the simpler analytic model (\ref{fitsimple}), with the parameters given in Table \ref{Table:simple_model}. The plot shows, on a semi-logarithmic scale, the magnitude of the absolute difference between the model and the numerical data for the function $a_E(x)$.
}
\label{fig:diff_simp}
\end{figure}

\subsection{PN limit of the global analytic models}

It may be useful to study here the PN expansion of our global models. Recall that in our treatment we have imposed all PN information known {\it analytically} [Eq.\ (\ref{PNanalytic})] but refrained from imposing the numerical values of the higher-order PN coefficients $a_5,a_6,\ldots$ [Eq.\ (\ref{a57})] obtained in \cite{Barausse:2011dq} by fitting to large-radius numerical data from Refs.\ \cite{Blanchet:2009sd,Blanchet:2010zd}. We may now check how these numerically specified high-order PN coefficients compare with the ones entailed by our global analytic models. By considering the PN expansion of our 16-parameter analytic model ($\#14$), we obtain, for the two leading coefficients,
\begin{eqnarray}\label{PN14}
a_5^{\#14} &=&  23.47267\ldots  [23.50190] , \nonumber\\
a_6^{\#14} &=& -127.154\ldots  [-131.72] ,
\end{eqnarray}
where in square brackets we recall the values from \cite{Barausse:2011dq} [Eq.\ (\ref{a57}) above]. Higher-order coefficients agree only in sign:  $a_7^{\#14}=701.092\ldots  [118]$ and $a_7^{{\rm ln},\#14}-83.0457\ldots  [-255.0]$. Similarly, our simple, 8-parameter model ($\#24$) yields 
\begin{eqnarray}\label{PN24}
a_5^{\#24} &=&  24.19028\ldots  [23.50190] , \nonumber\\
a_6^{\#24} &=& -163.396\ldots  [-131.72] ,
\end{eqnarray}
with $a_7$ and $a_7^{\rm ln}$ agreeing much less well (and $a_7$ not even agreeing in sign).

It should come as no surprise that the values extracted from our global fits differ from those obtained using large-$r$ data only, and that the discrepancy increases rapidly with PN order. Our goal here was not to obtain accurate values for the coefficients of the asymptotic PN series (as in Refs.\ \cite{Blanchet:2009sd,Blanchet:2010zd}) but rather to devise a globally accurate model for the function $a(x)$. The latter goal could be achieved more ``economically''  (i.e., with a simpler analytic model) by relaxing (and thus effectively re-fitting) the values of some of the high-order PN coefficients. 

Finally, comparing between the results of Eqs.\ (\ref{PN14}) and (\ref{PN24}) provides a rough idea of the  uncertainty within which  the values of PN parameters can be extracted from any global fit. We see that, while different global models roughly agree on the value of $a_5$, they predict rather different values for $a_6$ (and for $a_7,a_7^{\rm ln},\ldots$). One should keep this uncertainty in mind when comparing the ``effective values'' of the PN coefficients extracted from any single global fit.

\section{GSF correction to the energy and angular momentum of circular orbits}\label{Sec:EJfit}

\subsection{Energy and angular momentum in terms of $a(x)$ and $a'(x)$}

Reference \cite{Tiec:2011dp} has used the results of \cite{Tiec:2011ab} to derive links between the function $z_{SF} (x)$, Eq.~(\ref{F12}), and the GSF corrections to the {\it functions} $e (x)$ and $j(x)$, where $e \equiv ({\mathcal E} - M) / \mu$ is the binding energy per unit reduced mass, and $j \equiv J/(M\mu) = J/(m_1 m_2)$ is the rescaled total angular momentum. Here ${\mathcal E}$ represents the (invariant) total gravitational energy of the binary, as it is defined in PN or EOB theories \footnote{One should carefully distinguish the various energy measures we are using: 
${\mathcal E}$, $e$,  $E_{\rm eff}=H_{\rm eff}$ and $E$. The first three are gauge-independent, while the last is the perturbatively defined, gauge-dependent energy $E$ of Eq.\ (\ref{E}); the gauge-invariant EOB-defined effective energy $E_{\rm eff}/\mu=H_{\rm eff}/\mu$  coincides with the GSF-defined $E$ in the test-particle limit $\nu\to 0$.}. By inserting the link (\ref{F13}) between $z_{SF} (x)$ and $a(x)$ into their results one can derive the corresponding links between $(e (x) , j(x))$ and the function  $a(x)$. However, we find it simpler, and conceptually more transparent, to use (more general) known results from EOB theory to directly derive the latter links. Let us start by recalling some basic results from EOB theory of circular orbits and its GSF expansion (see \cite{Damour:2009sm} for details).


The total energy ${\mathcal E}$ (including the rest mass contribution) of a (circular) binary system is simply given by the value of the EOB Hamiltonian $H$. Given the EOB main radial potential $A(u;\nu)$, it is easy to derive the exact link between $H$ and the EOB radial variable $u = M/r_{\rm EOB}$. (This is done, exactly as in the textbook treatments of circular orbits around a Schwarzschild black hole, by extremizing an effective potential; see below.) Before describing the result of this extremization let us recall the explicit structure of the EOB Hamiltonian: it is given by 
\begin{equation}
\label{H}
H (u,j,p_r) = M \, h (u,j,p_r) \, ,
\end{equation}
with
\begin{equation}
\label{h}
h = \sqrt{1+2\nu (\widehat H_{\rm eff} - 1)} 
\end{equation}
and
\begin{equation}
\label{Heff}
\widehat H_{\rm eff} (u,j,p_r) = \sqrt{A(u;\nu) \left( 1 + j^2 \, u^2 + \frac{p_r^2}{B(u;\nu)} + Q(u,j,p_r ; \nu) \right)} ,
\end{equation}
where the second EOB potential $B(u;\nu) = g_{rr}^{\rm eff}$ is related to $A$ and the potential $\bar D (u;\nu)$ mentioned above via
$$
A \, B \, \bar D \equiv 1 \, .
$$

Along circular orbits $p_r^2$ vanishes, as does $Q(u,j,p_r;\nu)$ when one is working within the {\it standard} formulation of the EOB Hamiltonian (see below for more details). Circular orbits are obtained by extremizing the simple effective potential $A(u;\nu) (1+j^2 \, u^2)$ with respect to $u$, at fixed $j$. This leads to the following relation (valid along circular orbits) between $j^2$ and $u$:
\begin{equation}
\label{F19}
j^2 (u) = - \frac{A' (u)}{(u^2 \, A(u))'} = - \frac{A'(u)}{2u \, \widetilde A(u)},
\end{equation}
where a prime denotes differentiation with respect to $u$, and we have introduced the shorthand notation
\begin{equation}
\label{F20}
\widetilde A (u) \equiv A(u) + \frac{1}{2} \, u \, A'(u) \, .
\end{equation}
[For notational brevity we hereafter ignore the $\nu$-dependence of $A(u,\nu)$.]
Note that Eq.~(\ref{F19}) yields the simple result
\begin{equation}
\label{F21}
1+j^2 (u) \, u^2 = \frac{A(u)}{\widetilde A (u)} \, .
\end{equation}
Inserting the latter result into the effective Hamiltonian, Eq.~(\ref{Heff}), yields  yet another simple result:
\begin{equation}
\label{F22}
\widehat H_{\rm eff} (u) = \frac{A(u)}{\sqrt{\widetilde A (u)}} \, .
\end{equation}
At this stage Eqs.~(\ref{F19}) and (\ref{F22}) give the {\it exact} functional relations between $u$ and the energy and angular momentum of circular orbits. Note that the specific binding energy $e = (H - M)/\mu$ reads, in terms of the above notation,
\begin{equation}
\label{F22bis}
e \equiv \frac{H-M}{\mu} = \frac{1}{\nu} \, (h-1) = \frac{1}{\nu} \left( \sqrt{1+2\nu (\widehat H_{\rm eff} - 1)} - 1 \right) \, .
\end{equation}

In order to obtain the corresponding exact functional relation between the frequency parameter $x$ and $e$ and $j$ we need to relate $x$ to $u$. The latter link simply follows from the general Hamiltonian equation $\Omega = \partial H / \partial J$, which explicitly reads (along circular orbits)
\begin{equation}
\label{F23}
M \Omega (u) = \frac{j(u) \, u^2 \, A(u)}{h(u) \, \widehat H_{\rm eff} (u)}.
\end{equation}
Squaring this result, and substituting from Eqs.\ (\ref{F19}) and (\ref{F22}), yields the simpler looking relation
\begin{equation}
\label{F24}
M^2\Omega^2 (u) = \left( -\frac{1}{2} \, \frac{A' (u)}{h^2 (u)} \right) u^3.
\end{equation}
Recalling the definition of $x$ in Eq.~(\ref{defx}), we simply have $(M\Omega)^2 \equiv x^3$ so that (\ref{F24}) yields the following {\it exact} link between $u$ and $x$:
\begin{equation}
\label{F25}
x(u) = u \left( \frac{-\frac{1}{2} \, A'(u)}{h^2(u)} \right)^{1/3} \, .
\end{equation}

Note that, up to this stage, we have made no approximation. In other words, given an explicit expression for the main EOB potential $A(u;\nu)$, one can, by using EOB theory, write the {\it exact} functions $e(u)$, $j(u)$ and $x(u)$ along the sequence of circular orbits. Note also that we are considering here the full sequence of stable or unstable circular orbits. (See \cite{Damour:2009sm}, and below,  for the exact condition defining the ISCO separating stable orbits from unstable orbits.)

As, in this work, we are interested in GSF expansions at order $O(\nu) = O(q)$, let us now expand the above results in powers of $\nu$. Using Eq.~(\ref{defa}), it is trivial to obtain the functions $e(u)$ and $j(u)$ to order $O(\nu)$ in terms of the function $a(u)$ [see, e.g., Eq.~(4.19) of \cite{Damour:2009sm} for the expansion of $j(u)$]. However, an extra complication comes from the need to also expand the function $x(u)$ to order $O(\nu)$. This result was first obtained in Ref.\ \cite{Damour:2009sm}, Eq.~(4.21). Here, we are mainly interested in the $O(\nu)$ inverse relation $u(x)$, which reads [Eq.~(4.22) in \cite{Damour:2009sm}]
\begin{equation}
\label{F26}
u(x) = x \left[ 1+\frac{1}{6} \, \nu \, a'(x) + \frac{2}{3} \, \nu \left( \frac{1-2x}{\sqrt{1-3x}} - 1 \right) + O(\nu^2) \right] \, .
\end{equation}

Finally, inserting (\ref{F26}) into the $O(\nu)$ expansions of $e (u)$ and $j(u)$ leads to the following $O(\nu)$ expansions of the {\it composed functions} $e (u(x))$ and $j(u(x))$:
\begin{eqnarray}
\label{F27}
e(u(x)) &=& e_0 (x) + \nu \left[ \alpha_E (x) \, a'(x) + \beta_E (x) \, a(x) + \gamma_E (x) \right] \\ \nonumber
&\equiv & e_0 (x) + \nu e_{SF} (x),
\end{eqnarray}
\begin{eqnarray}
\label{F28}
j(u(x)) &=&  j_0(x) + \nu \left[ \alpha_j (x) \, a'(x) + \beta_j (x) \, a(x) + \gamma_j (x) \right]  \\ \nonumber
&\equiv &  j_0 (x) + \nu j_{SF} (x),
\end{eqnarray}
where 
\begin{equation}
\label{F270}
e_0 (x) \equiv \frac{1-2x}{\sqrt{1-3x}} - 1 ,
\end{equation}
\begin{equation}
\label{F280}
j_0(x) \equiv \frac{1}{\sqrt{x(1-3x)}},
\end{equation}
and where the coefficients $\alpha_E$, $\beta_E$, $\gamma_E$, $\alpha_j$ , $\beta_j$ and $\gamma_j$ are given by
\begin{equation}
\label{F29}
\alpha_E (x) = -\frac{1}{3} \, \frac{x}{\sqrt{1-3x}},
\end{equation}
\begin{equation}
\label{F30}
\beta_E (x) = \frac{1}{2} \, \frac{1-4x}{(1-3x)^{3/2}},
\end{equation}
\begin{equation}
\label{F31}
\gamma_E (x) = -  e_0 (x) \times \left[ \frac12  e_0 (x) + \frac{x}{3} \, \frac{1-6x}{(1-3x)^{3/2}} \right] ,
\end{equation}
\begin{equation}
\label{F32}
\alpha_j (x) = x^{-3/2} \, \alpha_E (x) = -\frac{1}{3} \, \frac{1}{\sqrt{x(1-3x)}},
\end{equation}
\begin{equation}
\beta_j (x) = -\frac12 \frac1{x^{1/2} (1-3x)^{3/2}},
\end{equation}
\begin{equation}
\gamma_j (x) = - \frac13  \frac{1-6x}{x^{1/2} (1-3x)^{3/2}} \times  e_0 (x).
\end{equation}
 Note in passing that the relation $d e [u(x)] = x^{3/2} d j[u(x)]$ holds exactly (and, in particular, at each order
 in $\nu$), and implies many relations between the various coefficient functions  $\alpha_E$, $\beta_E$, $\gamma_E$, $\alpha_j$ , $\beta_j$, $\gamma_j$. The simplest  of these relations is  the link $\alpha_E(x) = x^{3/2} \alpha_j(x)$ indicated above.

\subsection{Global fits for the binding energy and angular momentum}

Equations (\ref{F27}) and (\ref{F28}) link the functions $e_{SF} (x)$ and $j_{SF} (x)$ to the function $a(x)$ and its derivative $a'(x)$. Our global analytic model (\ref{fit}) hence translates to global analytic models for $e_{SF} (x)$ and $j_{SF} (x)$. Since the resulting analytic expressions are quite cumbersome we will not present them here but instead content ourselves with a plot of the results. 

First, however, it is useful to consider the asymptotic behavior of $e_{SF} (x)$ and $j_{SF} (x)$ at the two ends of the domain $0<x<\frac{1}{3}$. In the weak-field regime $x\ll 1$, our models, of course, reproduce the known PN behavior:
\begin{equation}
e_{SF} (x\ll 1) = \frac{1}{24}\, x^2 +O(x^3), 
\end{equation}
\begin{equation}
j_{SF} (x\ll 1) = \frac{1}{6}\, x^{1/2} +O(x^{3/2}). 
\end{equation}
Near the LR, we use Eq.\ (\ref{aLR}) in conjunction with (\ref{F27}) and (\ref{F28}) to obtain
\begin{equation}
e_{SF} (x\to \frac{1}{3}) \sim \left(\frac{1}{27}-\frac{1}{12}\zeta\right)z^{-2} \approx -\frac{5}{108}\, z^{-2},
\end{equation}
\begin{equation}
j_{SF} (x\to \frac{1}{3}) \sim \left(\frac{1}{3\sqrt{3}}-\frac{\sqrt{3}}{4}\zeta\right) z^{-2} \approx -\frac{5}{12\sqrt{3}}\, z^{-2},
\end{equation}
where we also used $\zeta\approx 1$. Hence, both $e_{SF}$ and $j_{SF}$ are expected to diverge {\it quadratically} in $1/z=(1-3x)^{-1}$ at the LR. 

To plot our analytic models for the binding energy and angular momentum it is convenient to introduce the rescaled quantities 
\begin{equation}\label{tildeESF}
\hat e_{SF}\equiv e_{SF}\times z^2\times (x^2/24)^{-1},
\end{equation}
\begin{equation}\label{tildejSF}
\hat j_{SF}\equiv j_{SF}\times z^2\times (\sqrt{x}/6)^{-1},
\end{equation}
which attain the regular values $\hat e_{SF}(0)=\hat j_{SF}(0)=1$ as well as $\hat e_{SF}(1/3)\approx -10$ and $\hat j_{SF}(1/3)\approx -\frac{5}{2}$. The functions $\hat e_{SF}(x)$ and $\hat j_{SF}(x)$ are plotted in Fig.\ \ref{fig:Ej}. Inspecting the plot, we note that both GSF corrections $\hat e_{SF}(x)$ and $\hat j_{SF}(x)$ turn from positive in the weak field to negative in the strong field. The transition occurs at $x\approx 0.0247$ ($r\approx 40.49M$) for $e_{SF}$ and at $x\approx 0.0435$ ($r\approx 22.99M$) for $j_{SF}$.

\begin{figure}[Htb]
\includegraphics[width=10cm]{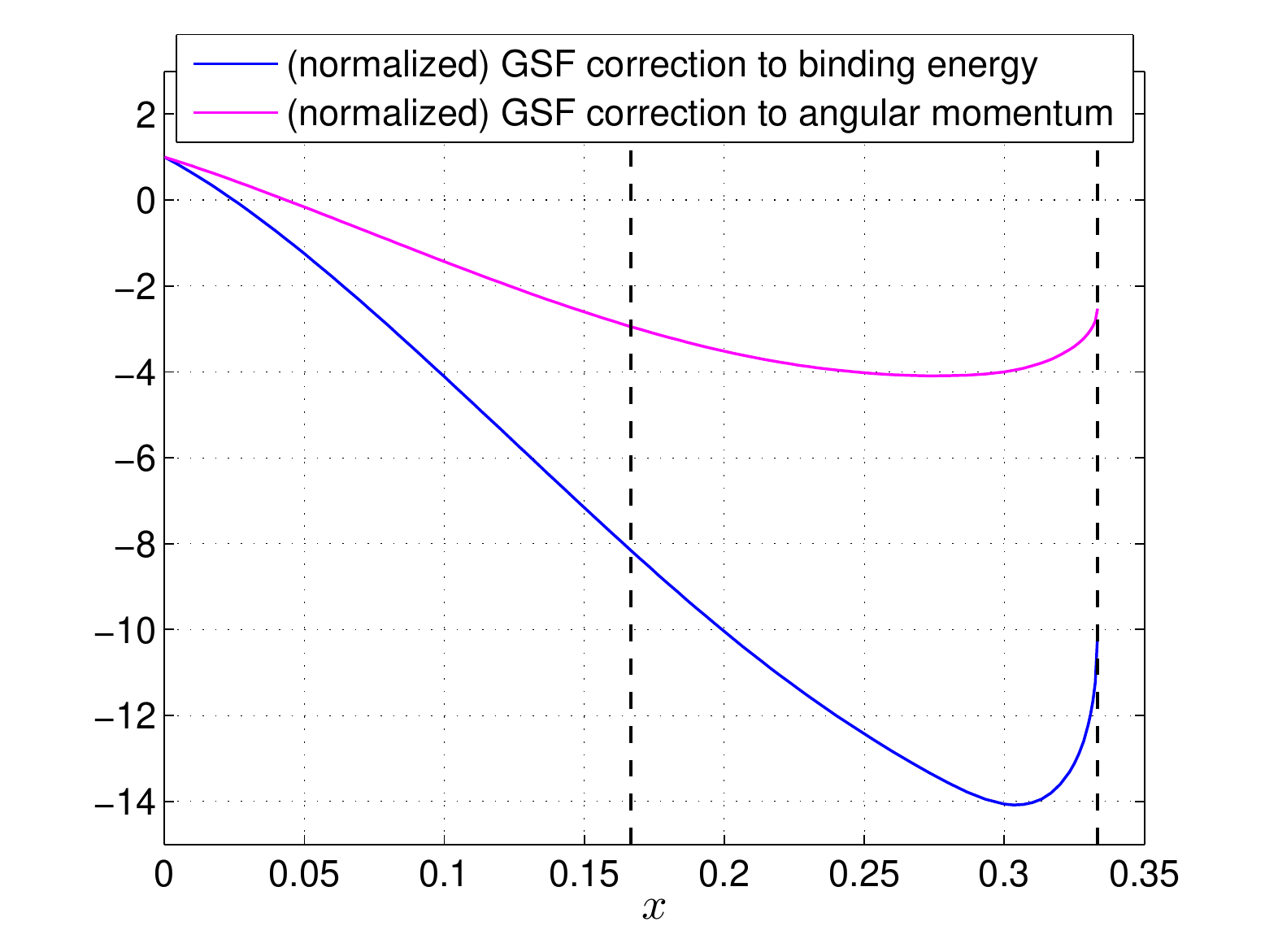}
\caption{The GSF corrections to the binding energy $e(u(x))$ and angular momentum $j(u(x))$. We show here the rescaled functions $\hat e_{SF}(x)$ (lower curve) and $\hat j_{SF}(x)$ (upper curve) defined in Eqs.\ (\ref{tildeESF}) and (\ref{tildejSF}). The curves shown represent the analytic functions obtained by inserting  our global analytic fit for $a(x)$, Eq.\ (\ref{fit}), into Eqs.\ (\ref{F27}) and (\ref{F28}) .
}
\label{fig:Ej}
\end{figure}
\section{Accurate determination of the $O(\nu)$ correction to the ISCO frequency}\label{Sec:ISCO}

Le Tiec {\it et al.}~pointed out in Ref.\  \cite{Tiec:2011dp} that the link they had established between the functions $e(x)$ and  $z_{SF}(x)$ provides an efficient method for calculating the GSF-induced [$O(\nu)$] shift in the value of the ISCO frequency---an important strong-field benchmark. This shift was first calculated in Ref.\ \cite{ISCOLetter} (with a crucial gauge correction introduced later in  \cite{Damour:2009sm}) by analyzing small-eccentricity perturbations of circular orbits. Full details of this calculation (and a slightly more accurate result) were presented in \cite{Barack:2010tm}. Le Tiec {\it et al.}~suggested calculating the ISCO shift by minimizing the binding energy function $e(x)$. This seems potentially advantageous from the computational point of view, because the function $z_{SF}(x)$ [from which $e(x)$ is determined] is derived from GSF computations along the sequence of strictly {\it circular} orbits. These are substantially simpler and less demanding than GSF computations along eccentric orbits, even for the small eccentricities considered in \cite{ISCOLetter,Barack:2010tm}. [However, below we comment that Le Tiec {\it et al.}'s method is essentially equivalent (computationally) to the {\it second} method used in \cite{ISCOLetter,Barack:2010tm}, in which the GSF is computed along circular orbits with a certain fictitious source term containing derivatives of the particle's energy-momentum.] 

The calculation by Le Tiec {\it et al.}~in Ref.\  \cite{Tiec:2011dp}, based on the circular-orbit GSF data for $r\geq 5m_2$ available to them at the time, produced a value in full agreement with the results of \cite{ISCOLetter,Damour:2009sm,Barack:2010tm}, and with a slightly improved accuracy---see Table \ref{table:ISCOshift}. This agreement also lent support to the assertion made in Ref.\ \cite{Tiec:2011ab} that the link between $e(x)$ and $z_{SF}(x)$ is valid not only through 3PN order (as explicitly proven by them) but also in the strong-field regime. 

\begin{table}[htb] 
\begin{tabular}{|l|l|}
\hline\hline
Source & $C_{\Omega}$ \\
\hline\hline
Barack \& Sago \cite{ISCOLetter}; Damour \cite{Damour:2009sm}	& 1.2513(6) \\
Barack \& Sago \cite{Barack:2010tm}								& 1.2512(4) \\
Le Tiec, Barausse \& Buonanno \cite{Tiec:2011dp}				& 1.2510(2) \\
This work														& 1.25101546(5) \\
\hline\hline
\end{tabular}
\caption{Value of the parameter $C_\Omega$ describing the GSF correction to the ISCO frequency [see Eq.\ (\ref{COmega})]. Parenthetical figures show the uncertainty in the last displayed decimals. 
}
\label{table:ISCOshift}
\end{table}

Following Refs.\ \cite{Damour:2009sm,Tiec:2011dp}, we parametrize the $O(\nu)=O(q)$ correction to the ISCO frequency by the dimensionless parameter  $C_{\Omega}$, such that
\begin{equation}\label{COmega}
(m_1+m_2)\Omega_{\rm isco}=6^{-3/2}\left[1+C_{\Omega}\nu + O(\nu^2)\right],
\end{equation}
where $\Omega_{\rm isco}$ is the {\it physical frequency} of the ISCO, defined with respect to an ``asymptotically flat'' time $t$, and where $6^{-3/2}$ is the dimensionless frequency of the unperturbed (geodesic) ISCO around a Schwarzschild black hole. Reference \cite{Damour:2009sm}  obtained an analytical expression for  $C_{\Omega}$ in terms of the values of $a(x)$ and its first two derivatives at the ISCO:
\begin{equation}\label{Ca}
C_\Omega = \frac{3}{2} \, \mathsf a(1/6) +1 -\frac{2\sqrt{2}}{3} ,
\end{equation} 
where $\mathsf{a}(1/6)$ denotes the combination (see next section)
\begin{equation}\label{sfa1/6}
\mathsf{a}(1/6) = a(1/6) + \frac{1}{6} a'(1/6) + \frac{1}{18} a''(1/6) .
 \end{equation}
On the other hand, by considering the minimum of the function $e(x)$,  
Ref.\ \cite{Tiec:2011dp} obtained an analytical expression for $C_{\Omega}$ in terms
of the values of the first two derivatives of the redshift function  $z_{SF}(x)$ at the ISCO:
\begin{equation}\label{Cz}
C_\Omega = \frac{1}{2} + \frac{1}{4\sqrt{2}} \left[ \frac{1}{3} z''_{SF}(1/6) - z'_{SF}(1/6) \right] . \end{equation}
It is easily checked that the link   (\ref{F13}) (found in Ref.\ \cite{Barausse:2011dq}) between 
 $a(x)$ and $z_{SF}(x)$ transforms Eq.\ (\ref{Ca}) into  Eq.\ (\ref{Cz}).

Our goal here is to obtain a more accurate value for $C_{\Omega}$ based on our new, improved GSF data. Given Eqs.\ (\ref{Ca}) and  (\ref{sfa1/6}),  this task amounts to accurately determining $a(x)$ and its first and second derivatives [or, equivalently, the perturbation $h_{uu}^{R,L}(x)$ and its first and second derivatives] at $x=\frac{1}{6}$. We comment that in the method referred to as {\em second} in Refs.\ \cite{ISCOLetter,Barack:2010tm} (which is modelled upon the scalar-field analysis of Ref.\ \cite{DiazRivera:2004ik}) one essentially also requires the second derivative of the metric perturbation from circular orbits, although in this method one derivative is taken with respect to the {\it field} point (to construct the GSF) and only the second is taken with respect to the orbital radius (to consider the effect of a small-eccentricity variation). From a computational point of view, we hence expect both methods (Le Tiec {\it et al.}'s \cite{Tiec:2011dp} and the second method \cite{ISCOLetter,Barack:2010tm}) to perform at a roughly equivalent level. Our improved accuracy in $C_{\Omega}$ will come primarily from using much more accurate numerical data based on the efficient frequency-domain code of Ref.\ \cite{Akcay:2010dx}. 

As a first attempt, we may simply use our global analytic model(s) for $a(x)$ to read off estimated values for $a(1/6)$, $a'(1/6)$ and $a''(1/6)$, and hence for $\mathsf{a}(1/6)$ and $C_\Omega$. For example, our accurate 16-parameter model ($\#14$ in Table \ref{Table:fit model}) and simpler 8-parameter model (\ref{fitsimple}) give, respectively, 
\begin{eqnarray}\label{sfaglobal}
\mathsf{a}(1/6) &=& 0.7958829\ldots  \quad \text{(using model \#14)}, \nonumber \\
\mathsf{a}(1/6) &=& 0.7958860\ldots  \quad \text{(using model \#24)},
\end{eqnarray}
and
\begin{eqnarray}\label{Cglobal}
C_\Omega &=& 1.2510153\ldots  \quad \text{(using model \#14)}, \nonumber \\
C_\Omega &=& 1.2510199\ldots  \quad \text{(using model \#24)}.
\end{eqnarray}
Both values of  $C_\Omega$ are  consistent with the results of \cite{ISCOLetter,Barack:2010tm, Tiec:2011dp}. However, placing an error bar on our predictions for $\mathsf{a}(1/6)$ and $C_\Omega$ requires a more careful analysis. Also, it is reasonable to expect that more reliable values for $\mathsf{a}(1/6)$ and $C_\Omega$ could be extracted from local analysis of the data near $x=\frac{1}{6}$ rather than relying on global fits. We proceed by presenting such a local analysis. 
 
First, let us give some consideration to the question of the optimal functions for the local analysis near  $x=\frac{1}{6}$. Naively, one may expect either $a(x)$ or $\hat a_E(x)$ [or $a_E(x)$, or even $z_{SF}(x)$ itself] to be equally suitable for a local fit near the ISCO, because all these functions are perfectly regular there. However, one should recall that the rate of convergence of the Taylor expansion in  $x-\frac{1}{6}$ about the ISCO (which can be measured by its  radius of convergence), depends on the global smoothness of the function. For instance, while $a_E(x)$ and $\hat a_E(x)$ are both bounded and continuous (by construction) not only on the {\it closed} interval on $0\leq x \leq \frac{1}{3}$ but even in larger intervals, say  $0\leq x \leq \frac{1}{2}$, the function $a(x)$ itself [as well as $\hat a(x)= a(x)/(2 x^3)$] blows up at the LR like $(1-3x)^{-1/2}$. The function $z_{SF}(x)$ blows up even faster: like $(1-3x)^{-1}$. If we assume that the functions we are dealing with can be analytically continued in the complex $x$ plane, the radius of convergence of the Taylor expansion around some point $x_0$ can (generally)  be estimated to be equal to the distance separating $x_0$ from the nearest singularity, in the complex plane, of the considered function \footnote{We are aware that our reasonings are non rigorous, especially in view of the residual weak logarithmic singularity $\sim z \ln (z^2)$ that seems to affect $a_E(x)$.  However, we have confirmed our expectations  through numerical experiments.} .
This suggests that the Taylor expansion (around $x_0$) of all functions having a singularity at $x_{\rm sing}=\frac13$ [such as $a(x)$, $\hat a(x)$ or $z_{SF}(x)$] will have a radius of convergence equal to $| x_{\rm sing} - x_0| = |\frac13 - x_0|$.  For $x_0 = \frac16$, this yields a radius of convergence equal to $\frac16$, i.e a Taylor series around the ISCO converging roughly like 
$ \sum_n    (6 (x- \frac16))^n$.  By contrast, if we assume, for instance, that the nearest (complex) singularity of the functions $a_E(x)$ and $\hat a_E(x)$ is beyond $x_{\rm sing}=\frac12$, this suggests that the radius of convergence of the ISCO expansion of these functions will be {\it larger} than $ |\frac12 - \frac16|= \frac13$, corresponding to a series converging roughly like
$ \sum_n    (3 (x- \frac16))^n$  (i.e.\ much faster than for the LR-singular functions).
This suggests that the use of a LR-regular function, such as  $\hat a_E(x)$,  should allow for a more accurate determination of the local $a(x)$ derivatives than the direct use of a LR-singular one, such as $a(x)$ or   $z_{SF}(x)$.

We have checked this expectation by comparing the relative performances (with respect to local fits) of several functions related to $a(x)$. Namely, we considered local fits for the following functions: $ \{a_i(x); i=1, \ldots, 5 \}=\{ a, a^s, a_E, \hat{a}^s, \hat{a}_E \} $, where $ a^s(x) \equiv s(x) a(x) = \sqrt{1-3x}\: a(x) $ and $ \hat{a}^s(x) \equiv a^s(x)/(2x^3)$. Our analysis began by selecting a subset of $a_i(x)$ data (for each $i$) around $x=\frac{1}{6}$. We chose 10 data points on each side of $x=\frac{1}{6}$, which, together with the $x=\frac{1}{6}$ point itself, comprise a 21-point subset in the range $x \in [ 2/15, 0.2 ]$, corresponding to $r/m_2 \in [ 5.0, 7.5]$. We least-square fitted each of these $a_i(x)$ datasets to polynomials in $\tilde{x}\equiv \left(x-\frac{1}{6}\right)$ of degrees varying from 7 to 13, as in Eq.(\ref{chi2}). We judged the quality of each fit by looking at the value $\chi^2_{\rm min}$/DoF resulting from each fit. Let $Q_i(N)$ denote the value of $\chi^2_{\rm min}$/DoF corresponding to fitting $a_i(x)$ to a polynomial of degree $N$. In Table \ref{table_of_chi_squares}, we present the values of $Q_i(N)$ for $i=1, \ldots, 5 $ (corresponding to the $a_i$'s taken in the order of the set defined above) and $ N = 7, \ldots, 13 $.
\begin{table}[htb]  \label{table_of_chi_squares}
    \begin{tabular}{ l | l l l l l } \hline
$N$ &  $a_1=a$    &  $a_2=a^s$	&  $a_3=a_E$	&  $a_4=\hat{a}^s$	&  $a_5=\hat{a}_E$ \\
  \hline\hline
7 &	$1.165\times 10^4$	&	133.5	&	672.9	&	2.419	&	7.408	\\			
8 &			111.1	&	2.906	&	6.862	&	2.188	&	2.200	\\				
9	&		3.118	&	2.387	&	2.399	&	2.387	&	2.388	\\
10	&		2.674	&	2.626	&	2.629	&	2.623	&	2.624	\\	
11	&		2.831	&	2.833	&	2.832	&	2.847	&	2.843	\\	
12	&		3.185	&	3.186	&	3.186	&	3.151	&	3.160	\\
13		&	3.432	&	3.414	&	3.432	&	3.313	&	3.328	\\
    \hline\hline
     \end{tabular}  \caption{The values for the ``quality of the fit'' $Q_i(N)$ for the various $a_i(x)$ used. $N$ is the degree of the polynomials used in the fits. The values in columns 2 through 5 are the $\chi^2_{\rm min}$/DoF values for these polynomial fits of degree $N$. For example, $Q_3(7)$ is the $\chi^2_{\rm min}$/DoF value obtained from a $7^{\mathrm{th}}$-degree polynomial fit to $a_E(x)$ (with DoF$=21-8=13$).}  
\end{table}

The results in Table \ref{table_of_chi_squares} confirm that the use of LR-regular functions [i.e., $a_2(x)$ through $a_5(x)$] is beneficial in the sense that fewer parameters (i.e., lower values of $N$) are required to obtain a $\chi^2$/DoF of $O(1) $. Among the LR-regular functions, $a_4(x)=\hat{a}^s(x)$ stands out as being optimal in that it already reaches $\chi^2$/DoF $ = 2.419$ for $N=7$ while for this value of $N$ all the other $a_i(x)$ fare worse, and importantly much worse in the case of the unregularized original function $a(x)$, which has $\chi^2$/DoF $ > 10^4$ for $N=7$ (and needs at least $N = 9 $ to be considered a good fit). Note also that the second-best function is  $a_5(x)=\hat{a}_E(x)$. 

Our strategy, therefore, is to use the function $\hat{a}^s(x)$ for our local analysis at the ISCO. 
Based on the values presented in Table \ref{table_of_chi_squares} above, we use $a_4(x)=\hat{a}^s(x)$ with $N=8$ to compute $a(1/6)$, $a'(1/6)$, $a''(1/6)$, and $a'''(1/6)$ (the latter will be needed later) by analytic differentiation of the best-fit model multiplied by $2 x^3/\sqrt{1-3x}$ [to
translate back from $\hat{a}^s(x)$ values to $a(x)$ values]. This yields the following results:
\begin{eqnarray}
a(1/6) & = & 0.0260941094800(93), \qquad a'(1/6) =  0.6164354346(12) \nonumber \\
a''(1/6) & = & 12.00689379(28), \qquad a'''(1/6) = 204.788188(53). \label{aISCO} 	
\end{eqnarray}
Here the 2-digit error bars refer to the last 2 decimals of each quantity. These errors have been obtained from the covariance matrix of the polynomial regression. For instance, the above procedure gives for $a'(1/6)$ an estimate of the form $ c_0 \tilde{a}_0 + c_1 \tilde{a}_1$, where $\tilde{a}_0, \tilde{a}_1$ are the coefficients in the fitting polynomial $P_{\mathrm{fit}}(\tilde{x}) = \tilde{a}_0 + \tilde{a}_1 \tilde{x} + \ldots + \tilde{a}_N \tilde{x}^N $, and $ c_0, c_1 $ are coefficients obtained by using the chain rule in differentiating $ a(x) = 2 x^3 \hat{a}^s(x)/\sqrt{1-3x} $ at $x=1/6$. This yields a squared error on $a'(1/6)$ given by $ \sigma_{a'}^2 = c_0^2 \sigma_{00} + 2 c_0 c_1 \sigma_{01} + c_{1}^2 \sigma_{11} $, where $\sigma_{ij}$ are the elements of the covariance matrix coming out of the least-square fit. (Note that here, we are treating the data points as random Gaussian variables centered on the values listed in Tables \ref{table:data1} and \ref{table:data2}, with the variance given by the errors in the tables.)

Inserting the values of Eq.\ (\ref{aISCO}) into Eq.\ (\ref{sfa1/6}) 
we obtain
\begin{equation}\label{sfa1}
\mathsf a (1/6) =  0.795883004(15)  \quad \text{[from local fit for $\hat{a}^s(x)$]},
\end{equation}
corresponding to  [using Eq.\ (\ref{Ca})]
\begin{equation}\label{COmega1}
C_{\Omega}=1.251015464(23) \quad \text{[from local fit for $\hat{a}^s(x)$]}.
\end{equation}
Note that the error bar on $\mathsf a (1/6)$ (and therefore on $C_{\Omega}$) is dominated by the error on $a''(1/6)$.
As above, the errors in these quantities have taken into account the correlations described by the covariance matrix. Actually, we find that these correlations are rather mild, the largest of which equaling $\sigma_{02}/\left[\sigma_{00}\sigma_{22}\right]^{1/2}=-0.777$ when normalized. As a check, we repeated this analysis with our second best LR-regular fitting function, namely $\hat{a}_E(x)$ for $N=8$. The results agreed with the ones listed in Eq.\ (\ref{aISCO}) well within the error bars indicated there. For example, the central value for $a''(1/6)$ obtained from a local fit to $\hat{a}_E(x)$ is $12.00689372$. We also used the non-LR-regular function $a(x)$ to repeat the above local analysis using now $N=10$ (see Table \ref{table_of_chi_squares} for why). As expected, it led to larger errors, but the central values it gave agreed with the ones listed above within the (larger) error bars entailed by the use of $a(x)$. For example, the central value for $a''(1/6)$ obtained from a local fit to $a(x)$ is $12.00689374(54)$. 

In addition, we also used 7- and 9-point stencil (midpoint) methods, applied to the functions $a_i(x)$ defined above to extract the same derivatives independently. The results for $a(1/6)$ and its derivatives from these stencil methods also agreed with the above results within the error estimates corresponding to the stencil methods [which happen to be substantially larger, especially as the order of the derivative increases, than those given by the local fits to $\hat{a}^s(x)$ or $\hat{a}_E(x)$]. For example, the 7-point stencil method applied directly to the unrenormalized function $a(x)$ gives $a''(1/6)=12.006890(7)$, where the error was computed considering that the stencil method estimates the derivatives as a weighted sum of data points, each of which is, as before, treated as a Gaussian random variable with the variance equal to the error listed in Tables \ref{table:data1} and \ref{table:data2}.




As a partially independent check on the value of $C_\Omega$, we repeated the above analysis working with the variable $z_{SF}(x)$. We first constructed a dataset for $z_{SF}(x)$ using Eq.\ (\ref{z1F}) [with (\ref{F12})], and then combined a local-fit procedure (with $ N $ between 9 and 11) with 7- and 9-point stencils applied to $z_{SF}(x)$ to estimate the values $z'_{SF}(1/6)$ and $z''_{SF}(1/6)$. 
We obtained
\begin{equation}
z'(1/6)=3.10379963(1), \quad\quad
z''(1/6)=22.056551(6) ,
\end{equation}
where the errors were now estimated from the dispersion between the local fits and the stencil estimates.
Hence, using Eq.\ (\ref{Cz}) yields
\begin{equation}\label{COmega3}
C_{\Omega}=1.2510155(4) \quad \text{[from local fit and stencils for $z_{SF}(x)$]}.
\end{equation}

Summarizing, all of the above results are consistent with each other within their own errors. We {\it a priori} consider that it is likely that the most accurate results are those obtained from using the first method above i.e. local fits to the LR-regular $\hat{a}^s(x)$. Indeed, by looking at the difference between the data points and the fits for $\hat{a}^s(x)$, one sees that they fluctuate in sign in a quasi-random manner across the entire data set. This suggests that the local fit is an effective way of averaging out these fluctuations over the 21 data points around the ISCO. However, the minimum $\chi^2$/DoF for the local fit is somewhat above unity (namely $2.188$ for the best fit used above). [Similarly, for our preferred global fit the minimum $\chi^2$/DoF was $4.77$]. This hints that, as already mentioned, our error bars on the data points have been somewhat underestimated. To be on the conservative side, we simply suggest that all our errors bars be {\it uniformly doubled}. In particular, this means that we recommend using as our preferred final results for $\mathsf{a}(1/6)$ and $C_\Omega$ the following values:
 \begin{equation}\label{ISCOsfa}
\mathsf a (1/6) =  0.795883004(30)  ,
\end{equation}
\begin{equation}\label{ISCOshift}
C_{\Omega}=1.251015464(46).
\end{equation}
%
Our final result (\ref{ISCOshift}), which in rounded numbers reads $1.25101546(5)$, is fully consistent with the value currently available in the literature, {\it and it adds 4 significant digits to it}. See Table \ref{table:ISCOshift} for a comparison.
%
%
%
%
%
\section{On determining the EOB potential $\bar d(x)$}\label{Sec:dfit}

In this section we discuss the determination of the $O(\nu)$ piece $\bar d(u)$ of the second EOB potential $\bar D(u;\nu)$, defined through 
\begin{equation}
\bar D(u;\nu)=1+\nu \bar d(u)+ O(\nu^2).
\end{equation}
[Here, as usual, we use $\bar d(u)$ to denote a {\it functional} form.] We present numerical results for $\bar d(u)$ on $u\leq \frac{1}{6}$, i.e., outside the ISCO as well as on the ISCO itself, where a certain subtlety occurs. These results are obtained from a combination of numerical data and analytic fits. We then comment on the extension of the function $\bar d(u)$ beyond the ISCO. Our discussion extends upon and improves the similar discussion presented in Ref.\ \cite{Barausse:2011dq}.

Reference \cite{Damour:2009sm} obtained a relation involving $\bar d(u)$, the function $a(u)$ (and its first and second derivatives), and the function $\rho(u)$ describing the $O(\nu)$ precession effect in slightly eccentric orbits (at the circular-orbit limit). The function $\rho(u)$  is defined for stable circular orbits through
\begin{equation}
\left(\frac{\Omega_r}{\Omega}\right)^2=1-6x+\nu \rho(x) +O(\nu^2),
\end{equation}
where $\Omega_r$ is the $t$-frequency of radial oscillations about the circular motion, and $\Omega$ is the usual azimuthal $t$-frequency of the circular orbit. As  discussed in \cite{Damour:2009sm}, the definition of $\rho(x)$ can be extended to include {\it unstable} circular orbits (i.e., to the entire regime $0<x<\frac{1}{3}$ where timelike circular orbits exist) by replacing the squared radial frequency $\Omega_r^2$ with (minus) the appropriate squared Lyapunov exponent associated with the growth rate of perturbations of the unstable orbit.
The said relation between the functions $\bar d(u)$, $a(u)$ and $\rho(u)$ is given by \cite{Damour:2009sm}
\begin{equation}\label{rhoda}
\rho(u)= 4u\left(1-\frac{1-2u}{\sqrt{1-3u}}\right) +{\mathsf a}(u)+(1-6u)\bar d(u) ,
\end{equation}
where we have introduced [consistent with Eq.\  (\ref{sfa1/6}) above]  the shorthand notation 
\begin{equation}\label{acomb}
{\mathsf a}(u)\equiv a(u)+u a'(u)+\frac{1}{2} u(1-2u)a''(u).
\end{equation}
This relation is valid over the entire range where timelike circular orbits exist, i.e.\ for $0<u\leq \frac{1}{3}$. 

The function $\rho(u)$ was computed numerically in Ref.\ \cite{Barack:2011ed} for $u<\frac{1}{6}$, and an analytic fit for it over the corresponding domain was obtained in Refs.\ \cite{Barack:2010ny,LeTiec:2011bk}.
This, in conjunction with an analytic fit for $a(u)$, allows one to obtain the function $\bar d(u)$ on $0<u< \frac{1}{6}$ via Eq.\ (\ref{rhoda}).  Reference \cite{Barausse:2011dq} proposed computing $\bar d(u)$ simply through  solving Eq.\ (\ref{rhoda}) with respect to  $\bar d(u)$:
\begin{equation}\label{rhosing}
\bar d(u)=\frac{1}{1-6u}\left[ \rho(u)-{\mathsf a}(u)+4u\left(\frac{1-2u}{\sqrt{1-3u}}-1\right)\right].
\end{equation}
Note, however, that this expression is formally singular at the ISCO, where $1-6u=0$. This singularity should in principle be removable  (as also discussed in \cite{Barausse:2011dq}), but the presence of the divergent factor $(1-6u)^{-1}$ makes it difficult to evaluate $\bar d(u)$ numerically in the immediate neighbourhood of the ISCO. Thus, the expression (\ref{rhosing}), as it stands, is in practice ill-suited for describing the behavior of $\bar d(u)$ across the ISCO (where this function is expected to be perfectly smooth). 
 
To overcome this difficulty we can use an ISCO-local analysis, as we have done in the preceding section.
An expression for $\bar d(1/6)$ can be obtained simply by  evaluating the $u$-derivative of  Eq.\ (\ref{rhoda}) at $u=\frac{1}{6}$ [or, equivalently, by using de l'H$\hat{\rm o}$pital's rule in Eq.\ (\ref{rhosing})], assuming $\bar d(u)$ is smooth across the ISCO. One finds
\begin{equation}\label{dISCO}
\bar d(1/6)=-\frac{1}{6}\left[\rho'(1/6)-{\mathsf a}'(1/6)+
\frac{8\sqrt{2}}{3}-4
\right],
\end{equation}
which was first derived in Ref.\ \cite{Damour:2009sm}. Note that ${\mathsf a}'(1/6)$ involves all first {\it three} derivatives of $a(u)$ at $u=\frac{1}{6}$:
\begin{equation} 
\label{sfa'}
{\mathsf a}'(1/6)=2a'(1/6)+\frac{1}{3}a''(1/6)+\frac{1}{18} a'''(1/6).
\end{equation}
Hence, calculating $\bar d(1/6)$ requires knowledge of $a'(1/6)$, $a''(1/6)$, $a'''(1/6)$
and $\rho'(1/6)$. Highly accurate values for the former three quantities were given in Eq.\ (\ref{aISCO}) above. 
These values give ${\mathsf a}'(1/6)=16.612290(3)$. 

To obtain the value $\rho'(1/6)$ we fit to a sample of $\rho(x)$ data points just outside the ISCO, as was done in Ref.\ \cite{Barack:2010ny} but using a denser sample of data for better accuracy. 
We obtain
\begin{equation}
\rho'(1/6)=12.70(5).
\end{equation}
This is consistent with the estimated value of 12.66 quoted in Ref.\ \cite{Barack:2010ny}. [We prefer here to compute $\rho'(1/6)$ based on an ISCO-local analysis, rather than extract this value from any of the {\it global} fit models of Refs.\ \cite{Barack:2010ny,LeTiec:2011bk}, since the local analysis is likely to produce a more accurate result.]
With these values, Eq.\ (\ref{dISCO}) gives
\begin{equation}\label{dISCOval}
\bar d(1/6)=0.690(8).
\end{equation}
The uncertainty in this result is entirely dominated by the numerical error in $\rho'(1/6)$; unfortunately, the latter value comes from eccentric-orbit GSF calculations, which are of a relatively limited accuracy. 

Table \ref{table:d} lists all $\rho(x)$ data available to date, combining results from Refs.\ \cite{Barack:2010ny} and \cite{LeTiec:2011bk}. For each data point we display the corresponding value of $\bar d(x)$, computed via Eq.\ (\ref{rhosing}) with our accurate 16-parameter model for $a(x)$ ($\#14$ in Table \ref{Table:fit model}). At the ISCO itself we quote the value obtained above. Figure \ref{fig:d} shows a plot of the $\bar d(x)$ data. Note that the error bars on $\bar d(x)$ are predominantly due to the numerical error in the $\rho$ data, which is larger than the model error in our $a(x)$ fit [itself based on very high accuracy $a(x)$ data]. For that reason, it is ``safe'' to use our analytic fit formula for $a(x)$ rather than the actual $a(x)$ data, as we have opted for here.

\begin{table}[htb] 
    \begin{tabular}{ llll }
    \hline
$r/m_2$ 			&\ \ \  $x$   		&\ \ \ 	$\rho$	& \ \ \   $\bar d$  \\
  \hline\hline
80 					& 0.0125			& 0.0024117(9)  & 0.0010406(9) 	\\
57.142857			& 0.0175$\ldots$	& 0.0048913(6)  & 0.0021230(6) 	\\
50					& 0.02				& 0.006494(2)   & 0.002829(2) 	\\
44.4444				& 0.02250$\ldots$	& 0.008351(2)   & 0.003652(2)\\
40					& 0.025				& 0.010470(1)   & 0.004598(1)\\
36.363636			& 0.0275$\ldots$	& 0.0128610(8)  & 0.005674(1)\\
34.2857				& 0.0291$\ldots$	& 0.0146099(8)  & 0.0064658(9)\\
30					& 0.0333$\ldots$	& 0.0195438(4)  & 0.0087235(5)\\
25					& 0.04				& 0.0291863(3)  & 0.0132258(4)\\
20					& 0.05				& 0.0479916(5)  & 0.0223171(7)\\
19					& 0.0526$\ldots$	& 0.053862(3)  & 0.025233(5)\\
18					& 0.0555$\ldots$	& 0.060857(2)  & 0.028753(3)\\
17					& 0.0588$\ldots$	& 0.069279(4)  & 0.033055(6)\\
16					& 0.0625			& 0.079537(2)  & 0.038386(2)\\
15					& 0.0666$\ldots$	& 0.092199(3)  & 0.045101(6)\\
14					& 0.0714$\ldots$	& 0.108061(2)  & 0.053718(4)\\
13.5				& 0.0740$\ldots$	& 0.117534(3)  & 0.058966(6)\\
13.25				& 0.0754$\ldots$	& 0.122734(2)  & 0.061880(4)\\
13					& 0.0769$\ldots$	& 0.128280(3)  & 0.065016(5)\\
12.75				& 0.0784$\ldots$	& 0.134204(3)  & 0.068396(6)\\
12.5				& 0.08				& 0.140536(5)  & 0.072041(9)\\
12.25				& 0.0816$\ldots$	& 0.147316(5)  & 0.075979(9)\\
    \hline\hline
    \end{tabular}
\quad\quad\quad\quad\quad
    \begin{tabular}{ llll }
    \hline
$r/m_2$  			& \ \ \ $x$  			&\ \ \  $\rho$	& \ \ \   $\bar d$  \\
  \hline\hline
12					& 0.0833$\ldots$	& 0.154578(3)  & 0.080229(5)\\
11.75				& 0.0851$\ldots$	& 0.162386(2)  & 0.084856(5)\\
11.5				& 0.0869$\ldots$	& 0.170784(4)  & 0.089887(8)\\
11.25				& 0.0888$\ldots$	& 0.179837(5)  & 0.09538(1)\\
11					& 0.0909$\ldots$	& 0.189605(3)  & 0.101369(7)\\
10.75				& 0.0930$\ldots$	& 0.200170(3)  & 0.107933(6)\\
10.5				& 0.0952$\ldots$	& 0.211643(5)  & 0.11520(1)\\
10.25				& 0.0975$\ldots$	& 0.224075(4)  & 0.12315(1)\\
10					& 0.1				& 0.237610(4)  & 0.131940(9)\\
9.75				& 0.102$\ldots$		& 0.252391(5)  & 0.14174(1)\\
9.5					& 0.105$\ldots$		& 0.268565(5)  & 0.15267(1)\\
9.25				& 0.108$\ldots$		& 0.286271(5)  & 0.16482(1)\\
9					& 0.111$\ldots$		& 0.305750(5)  & 0.17849(2)\\
8.75				& 0.114$\ldots$		& 0.327230(6)  & 0.19390(2)\\
8.5					& 0.117$\ldots$		& 0.351000(6)  & 0.21141(2)\\
8					& 0.125				& 0.406767(6)  & 0.25423(3)\\
7.5					& 0.133$\ldots$		& 0.47651(1)   & 0.31129(5)\\
7.4					& 0.135$\ldots$		& 0.492527(7)  & 0.32490(4)\\
7					& 0.142$\ldots$		& 0.56528(1)   & 0.38986(8)\\
6.8					& 0.147$\ldots$		& 0.607693(9)  & 0.42991(7)\\
6.5					& 0.153$\ldots$		& 0.68059(1)   & 0.5024(2)\\
6					& 0.166$\ldots$		& 0.8340103(2)   & 0.690(8)\\
\hline\hline
\end{tabular}
\caption{
Available numerical data for $\rho(x)$ (collected from \cite{Barack:2011ed,Barack:2010ny,LeTiec:2011bk}), and quasi-numerical values for the $O(\nu)$ EOB function $\bar d(x)$, obtained from Eq.\ (\ref{rhosing}) based on the $\rho$ data in conjunction with our accurate 16-parameter model for $a(x)$ ($\#14$ in Table \ref{Table:fit model}). The value of $\rho$ at the ISCO was determined from the simple relation $\rho(1/6)=\frac{2}{3}C_{\Omega}$, using the accurate value we have obtained for $C_\Omega$ in Eq.\ (\ref{ISCOshift}) above. The value of $\bar d$ at the ISCO is quoted from Eq.\ (\ref{dISCOval}).
Note that the $r$ values in this table are {\it exact}, while the $x$ values are computed as the inverse of these exact values. }
\label{table:d}
\end{table}

\begin{figure}[Htb]
\includegraphics[width=10cm]{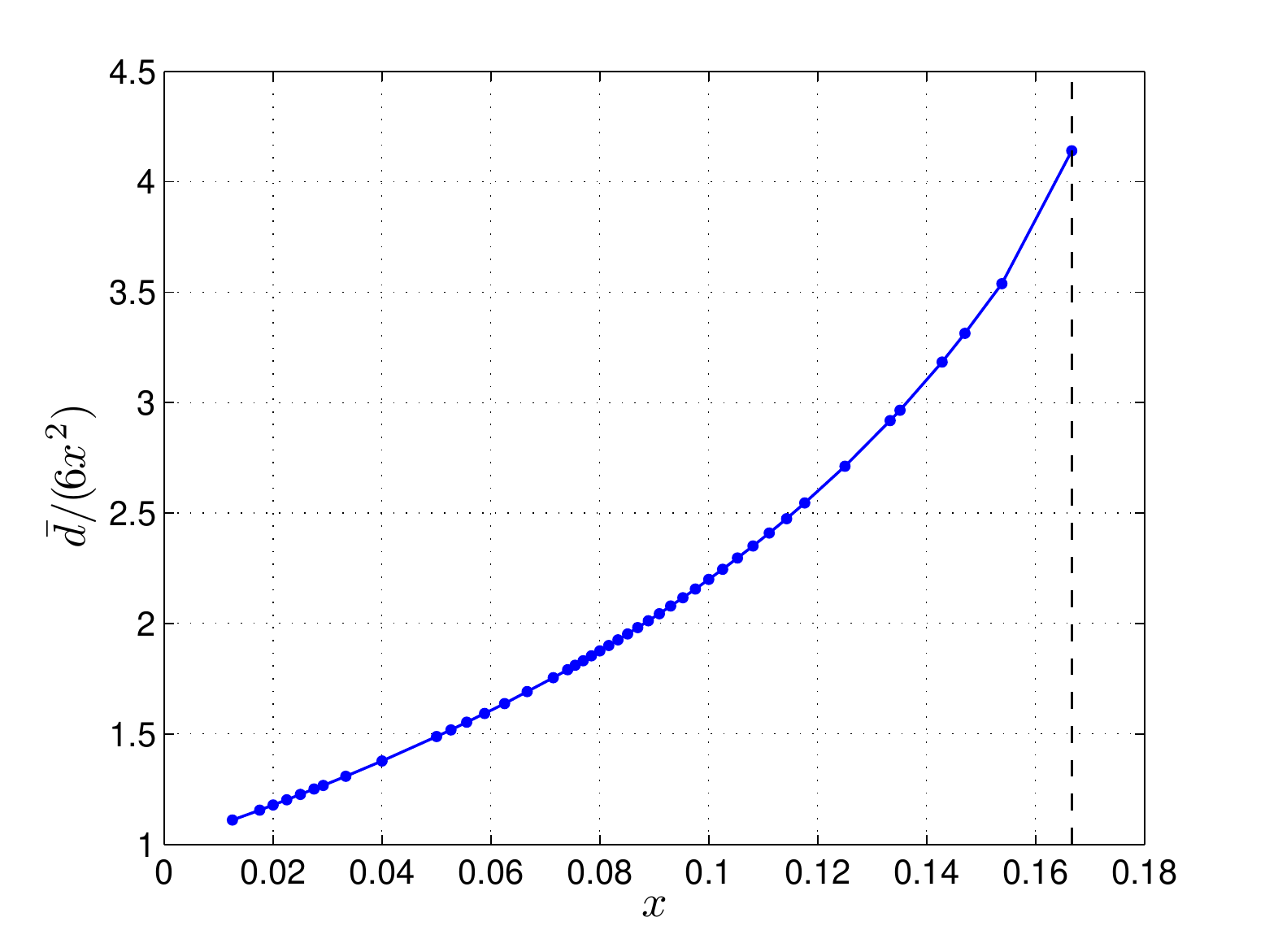}
\caption{The $O(\nu)$ EOB function $\bar d(x)$. We plot here the quasi-numerical data from Table \ref{table:d}, normalized by the leading-order PN term $6x^2$ for convenience. 
}
\label{fig:d}
\end{figure}

The data in Table \ref{table:d} can be used as a basis for an analytic fit model for $\bar d(u)$ over $0<u\leq \frac{1}{6}$, e.g., using the methods of Ref.\cite{Barack:2010ny}. We leave this to future work; we expect that the new, frequency-domain GSF method of Refs.\ \cite{Akcay:2010dx,Warburton:2011fk} could soon provide much more accurate data for $\rho(u)$, that will enable a more reliable and accurate fitting. It is of  importance to extend the computation of $\rho(u)$ {\it beyond} the ISCO, in order to facilitate the computation of $\bar d(u>1/6)$. This should be possible in principle based on the existing GSF computational framework, although the details are yet to be worked out and implemented. 

Such developments would allow one, in particular, to study the behavior of $\bar d(u)$ at the LR. It is interesting to speculate about this behavior based on the form of Eq.\ (\ref{rhosing}) and what we already know about the LR-behavior of $a(u)$. From Eqs.\ (\ref{aLR}) and (\ref{acomb}) we have ${\mathsf a}(u) \sim \frac{3\zeta}{32}(1-3u)^{-5/2}$ as $u\to \frac{1}{3}$ (with $\zeta\approx 1$). Hence, if the divergence of $\rho(u)$ at the LR is weaker than  $\propto(1-3u)^{-5/2}$, we must have that $\bar d(u)$  diverges as  $\bar d(u\to 1/3)\sim \frac{3\zeta}{32} (1-3u)^{-5/2}$.  We will indicate below an argument suggesting that  $\bar d(u)$ indeed has a strong divergence $\propto  (1-3u)^{-5/2}$.

\section{Light-ring behavior and implications for EOB theory}
\label{Sec:photosphere}


A striking result of our sub-ISCO GSF computation is the finding that, as one approaches the LR, i.e. as $x \to \frac13$ or  $u \to \frac13$, $h^{R,L}_{uu}(x)$ blows up proportionally to $(1-3x)^{-3/2}$, while, correspondingly, $a(u)$ has an {\it  inverse square-root}
 singularity:  $a(u)\approx 0.25 (1-3u)^{-1/2}$. This (apparent) singular behavior {\it a priori} raises an issue concerning the domain of validity  of the EOB formalism, or  that of the GSF expansion.  In previous EOB work, it was always tacitly assumed that all the functions parametrizing the EOB Hamiltonian, i.e. $A(u;\nu)$, $\bar D(u;\nu)$ and  $Q(u,p_{\varphi},p_r;\nu)$, were smooth functions of  $u= M/r_{\rm EOB}$,  from the weak-field value $u=0$ up to, at least,
  the Schwarzschild horizon value $u=\frac12$.  A smooth behavior of the EOB radial potentials
  across $u=\frac13$ seems {\it a priori} necessary for allowing the EOB formalism to describe, for instance,
 the head-on (or near head-on) coalescence of a (large mass ratio) binary  down to
 $r_{\rm EOB} \approx 2 M$. 
 Is the observed singular behavior $a(u)\approx 0.25 (1-3u)^{-1/2}$ 
 a signal that something pathological happens in the EOB formalism around the radius $r_{\rm EOB} \approx 3 M$, or is it an artefact of  formally trusting the first [$O(\nu^1)$] term in the GSF expansion beyond its physical domain of validity?
 Let us start discussing this issue by  considering  the physical origin of the LR singularities
 in  $h^{R, L}_{uu}(x)$ and $a(u)$.

 \subsection{Physical origin of the light-ring divergences  of   $h^{R, L}_{uu}(x)$ and $a(u)$}
 
We first explain the singular behavior $h^{R,L}_{uu}(x) \propto  (1-3x)^{-3/2}$  in terms of  heuristic technical considerations. [The weaker singular behaviour  $a(u)\propto (1-3u)^{-1/2}$ then follows from the structure of  Eq.\ (\ref{F18}).]  Note that we are interested in the {\it regularized} field $h^{R,L}_{uu}(x^{\lambda})$, which is obtained from the full, retarded perturbation $h^{L}_{uu}(x^{\lambda})$ by subtracting the Detweiler--Whiting S-field $h^{S, L}_{uu}(x^{\lambda})$ (and evaluating the result on the particle's circular orbit). We will consider in turn the LR behavior of the full and singular fields, and argue that the singular behavior of $h^{R,L}_{uu}(x)$ is inherited from the {\it full} field (while the S-field remains bounded at the LR). 

Consider first the full Lorenz-gauge metric perturbation $h^{L}_{\mu \nu}(x^{\lambda})$, which is sourced by the stress-energy tensor of particle 1, i.e. $T^{\mu \nu}(x^{\lambda}) = m_1  (-g)^{-1/2}  \int d\tau \, u^{\mu}  u^{\nu} \delta(x^{\lambda} - y^{\lambda}(\tau))=   (-g)^{-1/2} \int  m_1 u^{\mu}  dy^{\nu} \delta(x^{\lambda} - y^{\lambda}(\tau))$, where $\tau$, $y^{\lambda}(\tau)$ and $u^{\lambda}\equiv dy^{\lambda}/d\tau$ are, respectively, the particle's proper time, trajectory and four-velocity. As the particle approaches the LR (along a sequence of circular orbits), the (non-vanishing) components of its
  4-velocity  $u^{\mu}$  (in any frame at rest with respect to the background  Schwarzschild frame of $m_2$) tend towards infinity proportionally to  $u^0= dt/d\tau=(1-3x)^{-1/2}$.
 Therefore,  the (non-vanishing) components of   $T^{\mu \nu}$ too will tend to infinity proportionally to
 $(1-3x)^{-1/2}$. In other words, we can write  $T^{\mu \nu}= (1-3x)^{-1/2}  {\hat T}^{\mu \nu}$ 
 where all the components of the ``renormalized'' stress-tensor ${\hat T}^{\mu \nu}$ stay bounded as  particle 1 tends to the LR. Correspondingly, we can write $h^{L}_{\mu \nu}=(1-3x)^{-1/2}{\hat h}_{\mu \nu}$, where the ``renormalized'' metric perturbation ${\hat h}_{\mu \nu}$ is sourced by ${\hat T}^{\mu \nu}$, so that ${\hat h}_{\mu \nu}$ can be written as the convolution of a suitable tensorial Green function with  ${\hat T}_{\mu \nu}$. The latter
 convolution might introduce an additional, milder singular behavior  in the LR limit \footnote{This milder singular behavior might be connected with the sub-leading $ z \log z$ terms we have discussed above.}, but it is unlikely to alter the leading-order power-law blow-up $\propto (1-3x)^{-1/2}$.  Then, the value of the redshift-related scalar $h^{L}_{uu}$, which contains 
 two extra factors  $  (1-3x)^{-1/2}$ coming from the two contractions with the four-velocity (we assume here the ``constant'' off-worldline extension of the four-velocity discussed in Sec.\ \ref{Subsec:modesum}), is expected to blow up near the LR proportionally to  $(1-3x)^{-3/2}$. 
 
Let us next consider the S-field $h^{S, L}_{uu}(x^{\lambda})$. Near the particle, the trace-reversed counterpart of this field, $\bar{h}^{S,L}_{\mu \nu}(x^{\lambda}) \equiv h^{S,L}_{\mu \nu}(x^{\lambda}) - \frac12 h^{S,L}(x^{\lambda})g_{\mu \nu}(x^{\lambda})$, has the leading-order form $\bar{h}^{S,L}_{\mu \nu}(x^{\lambda}) \approx 4 \,  m_1 u_{\mu}  u_{\nu}/\epsilon$, where $\epsilon$ is the invariant orthogonal geodesic distance between $x^{\lambda}$ and the worldline. When contracting this local expression with $u^{\mu} u^{\nu}$,  the two factors of $u^{\mu}$ disappear (by virtue of $u^{\mu}  u_{\mu} =-1$) and we are left with $\bar{h}^{S,L}_{\mu \nu} u^{\mu} u^{\nu} \approx 4 \, m_1/\epsilon$. Now, $\epsilon$ scales proportionally to $u^{\mu}$ near the LR, so that $\bar h^{S, L}_{uu}$, and hence also $h^{S, L}_{uu}$, finally scales proportionally to the inverse of $u^{\mu}$ near the LR, i.e., $h^{S, L}_{uu}\propto (1-3x)^{+1/2}$. Note that the above ``cancellation'' of factors $u^{\mu}$ (due to $u^{\mu}  u_{\mu} =-1$) does not occur in the case of the full field $h^{L}_{uu}(x^{\lambda})$, which is obtained by a global integral that generally does not yield a result proportional to $u^{\mu} u^{\nu}$.
   
In summary, we find that the  LR behavior of the regularized difference field   $h^{R, L}_{uu}(x^{\lambda})=  h^{L}_{uu}(x^{\lambda}) - h^{S, L}_{uu}(x^{\lambda})$ is  dominated by the LR behavior of $h^{L}_{uu}(x^{\lambda})$, i.e., it is naturally expected to blow up near the LR proportionally to $(1-3x)^{-3/2}$. We can say that the blow ups  $h^{R, L}_{uu} \propto  (1-3x)^{-3/2}$, and correlatively [according to Eq.\ (\ref{F18})]  $a(u)  \propto (1-3u)^{-1/2}$, as particle 1 tends to the LR
 are simply rooted in the corresponding power-law blow-up of the components of the
 4-velocity near the LR: $u^{\mu} \propto (1-3x)^{-1/2}$.  
 
 Having understood this simple technical  origin of the LR behavior, we can reformulate it in a physically more transparent way. Instead of  parametrizing 
(as is usually done)  the strength of  first-order GSF  effects by means of the  rest-mass $m_1$,
one can say that the source of the perturbation is better measured by  the conserved
 energy of the small mass, say ${\cal E}_1$.  Indeed, we recall that $ {\cal E}_1$ is  given by a hypersurface integral of the contraction of the stress-energy tensor with the time-translation Killing vector $k^{\mu} \partial/\partial x^{\mu}= \partial/\partial x^0$
 i.e. $ {\cal E}_1 = - \int k^{\mu} T_{\mu}^{\nu}  dS_{\nu}$. We note also that
 $ {\cal E}_1 = - m_1 g_{\mu \nu} k^{\mu} u^{\nu}$,  clearly exhibiting the fact that $ {\cal E}_1 $
 measures the eventual growth of the components of $u^{\mu} $.
 In the case of circular orbits, it
 takes the simple form  $ {\cal E}_1= m_1 E_{\rm circ}(u_0)$
 where $u_0 = m_2/r_0$ is the background gravitational potential at the considered orbital
radius $r_0$, and where we used the notation
 \begin{equation} \label{defEcirc}
E_{\rm circ}(u) \equiv  \frac{1-2u}{\sqrt{1-3u}}.
\end{equation}
For later conceptual clarity we have here added a subscript ``circ'' to the function $E(u)$,
already defined in Eq.\ (\ref{E}).

We conclude that a physically transparent way of interpreting the blow-up of $a(u)$ near
the LR is to say that  the first-order GSF correction $a(u) \propto m_1/m_2$ actually
grows proportionally to the ratio of the conserved energy $ {\cal E}_1= m_1 E_{\rm circ}(u)$ of the small mass
to the large mass $m_2$ : i.e. $a(u) \sim q \, E_{\rm circ}(u)$.
We have already used this reformulation above to factor out the singular function $E_{\rm circ}(u)$
from the GSF-data-derived function $a(u)$. As for   $h^{R, L}_{uu}$,  its stronger blow-up
near the LR  is equivalent to saying that  $h^{R, L}_{uu}(x) \sim z^{-1} a(x) \sim q \, z^{-1}  E_{\rm circ}(x)$,
with $z=1-3x$ as usual.

  \subsection{Physical domain of validity of GSF results.}
  
 Before discussing the impact on the EOB formalism of the LR-divergent  behaviors
 of   $h^{R,L}_{uu}(x)$ and $a(u)$,  let us address the issue of the physical domain of validity
 of the GSF approximation.  We already mentioned (in Sec.\ \ref{Subsec:modesumLR}) the evident
 condition that a first-order [$O(\nu^1)$] GSF calculation makes sense only if the conserved
 energy of the small mass,  $ {\cal E}_1$, is parametrically smaller than that of the large mass,
 $ {\cal E}_2 \approx m_2$. In the context of circular orbits,  this leads to the  necessary condition  
\begin{equation} \label{condition1}
 {\cal E}_1= m_1 \, E_{\rm circ}\left( \frac{m_2}{r_0} \right) \ll m_2 \, ,
\end{equation}
where $E_{\rm circ}(u)$ is the function defined in Eq.\ (\ref{defEcirc}). 

Actually, this necessary condition is not sufficient for the consistency of the GSF expansion.
 Indeed, though we do not yet have a  {\it second-order} GSF calculation of  $h^{R,L}_{uu}$
and $a(u)$, one can physically estimate that second-order GSF effects will (at least approximately) modify the zeroth-order (geodesic) expression $ {\cal E}_1= m_1 \, E_{\rm circ}( \frac{m_2}{r_0} ) $ used in the condition  (\ref{condition1})  above for the energy of $m_1$ by including the
back reaction of $m_1$ on the background metric.  Therefore,  one expects that a 
more accurate version of the above necessary condition will roughly read
\begin{equation} \label{condition2}
m_1\,  E_{\rm circ}\left( \frac{m_2 + c \, {\cal E}_1}{r_0}  \right) \ll m_2,
\end{equation}
where we have modified the zeroth-order gravitational potential $u_0 = m_2/r_0$ by replacing
$m_2$ by $m_2 + c  \, {\cal E}_1$, where $c$ is some constant of order unity. However, if we now
look at the crucial square-root contained in the singular denominator of condition (\ref{condition2}), it reads
\begin{equation} 
\sqrt{1 - 3  \frac{m_2}{r_0} - 3 \, c\,  \frac{ \, {\cal E}_1}{r_0} }= \sqrt{1 - 3  \frac{m_2}{r_0} -  \, c\, \frac{3 m_2}{r_0}  \frac{ \, {\cal E}_1}{m_2} }.
\end{equation}
Near the LR, the hopefully more accurate condition  (\ref{condition2}) (which approximately takes
into account second-order GSF effects) is  consistent
with the first-order GSF condition  (\ref{condition1}) only if we have
\begin{equation}  \label{condition2bis}
 \frac{  {\cal E}_1}{m_2} \ll z_0 \equiv 1 - 3  \frac{m_2}{r_0} ,
\end{equation}
i.e.
\begin{equation}  \label{condition2ter}
\nu \ll  (z_0)^{3/2}.
\end{equation} 
This is much stronger (near the LR, i.e. when $z_0 \to 0$) 
than the condition  (\ref{condition1}) which corresponded to 
  $ q\, E_{\rm circ}(u_0)  \ll 1$, or $ q \approx \nu \ll \sqrt{z_0}$. We can also note
that the approximate inclusion (following the pattern used above) of third-order, and higher-order,
GSF effects do not {\it a priori} seem to require stronger constraints on $\nu$.  Indeed, our treatment
above essentially consisted of considering a first-order fractional 
modification of the ``effective background mass''  (say near the LR,
where $m_2/r_0 \sim 1$ does not introduce an independent small parameter) of the type
$m_2  \to m_2 [1 + c_1 \nu E_{\rm circ}(u_0)]$. We can generalize this treatment by considering higher-order
GSF contributions (proportional to higher powers of  $ {\cal E}_1/{m_2}$)
leading to a replacement of the type 
$m_2  \to m_2 [1 + c_1 \nu E_{\rm circ}(u_0)  +  c_2  (\nu E_{\rm circ}(u_0))^2 + \cdots]$. However, the condition
 (\ref{condition2bis})  is such that all the higher-order terms $ c_n (\nu E_{\rm circ}(u_0))^n, \ (n \geq 2) $
 are consistently smaller than the first-order one $ c_1 \nu E_{\rm circ}(u_0)$.
 
 Let us also remark in passing that another way to understand the necessity of the stronger
 consistency condition  (\ref{condition2bis}) is to notice that it is tantamount to requiring
 
 \begin{equation}  \label{condition2quat}
 h^{R, L}_{uu} \sim z_0^{-1}   \frac{  {\cal E}_1}{m_2}  \ll 1 .
\end{equation} 
 This makes sense because $ h^{R, L}_{uu}$ yields the first-order GSF perturbation of the
 proper time of particle 1.  More precisely,  the regularized proper time of particle 1 reads
 $$ d\tau_R = \sqrt{- (g^{(0)}_{\mu \nu} + h^{R,L}_{\mu \nu}) dx^{\mu} dx^{\nu}} = d\tau^{(0)} \sqrt{1- h^{R, L}_{uu}}  .$$  Clearly, it makes sense to expand the squareroot in powers of $q$ only if
 the actual magnitude of  $ h^{R, L}_{uu}$ is small compared to 1.

Summarizing so far: first-order GSF results near the LR are {\it a priori} physically meaningful only in
a limit where the ratio $\nu / (z_0)^{3/2}$ tends to zero.

 \subsection{Singular light-ring behavior as a coordinate singularity in the EOB phase space}
 
  
The findings of the previous subsection imply that the physical implications of the 
mathematically divergent LR behavior  
$a(u) \sim (1-3u)^{-1/2}$  of the $O(\nu)$  piece of the
EOB radial potential  $A(u,\nu)= 1-2u + \nu \, a(u) +O(\nu^2)$
are less dramatic than they seem to be at first sight.  Indeed,  in the domain where we
can trust the derivation of this result, i.e. under the condition  (\ref{condition2ter}),  the
first-order GSF contribution to $A(u,\nu)$ remains small, namely $\nu a(u) \sim \nu/ \sqrt{z} \ll z \ll 1$. In addition,   the first-order GSF contribution to the $u$-derivative of  $A(u,\nu)$ also remains,
under the consistency condition  (\ref{condition2ter}),
much smaller than unity, namely $\nu a'(u) \sim \nu/ z^{3/2}  \ll 1$. On the other hand, if we
consider the {\it second  $u$-derivative} of $A(u,\nu)$, it will be of the form 
$A''(u,\nu) = \nu a''(u) + O(\nu^2) \sim \nu/z^{5/2} $, which increases so much near the LR
that the condition (\ref{condition2ter}) is compatible with arbitrarily large values of  $A''(u,\nu) $
as one approaches the LR.  

Therefore, it is {\it a priori} possible that the near-LR behavior  
$a(u) \sim (1-3u)^{-1/2}$ only
corresponds to some mild type of physical singularity at the LR, and that higher-order
effects in $\nu$ will smooth out the shape of the $A$ potential into some sort
of ``boundary layer'' near the LR. However, the appearance
of a formal {\it square-root} singularity makes it unclear how the function $a(u)$ can be extended
{\it across} the LR to radial arguments $u > 1/3$.  Indeed, the formal analytic continuation
of  $(1-3u)^{-1/2}$ leads to {\it imaginary} values of $a(u)$ when $a >1/3$. 

However, independently of the possible role of higher-order effects, we are faced with the mathematical fact that the $O(\nu)$ piece in $A(u, \nu)$, i.e. the value of the $\nu$-derivative $\partial A(u, \nu)/ \partial \nu$ at $\nu=0$ has a singularity  $a(u) \sim (1-3u)^{-1/2}$. Is this inescapable mathematical singularity signalling the presence of some real  singularity in the EOB formalism?
We think instead that it does not correspond to any physical singularity of
the EOB dynamics, but is simply a {\it coordinate singularity in phase-space},
which can be avoided by a suitable (symplectic) phase-space transformation. 
Indeed,  a somewhat similar, formally singular LR behavior of  the EOB $A(u)$ 
potential has already appeared in another  EOB work, in a problem where one can see 
how this apparent singularity can be avoided by a suitable phase-space transformation.
We are alluding here to a recent analysis by Bini, Damour and Faye  \cite{Bini:2012gu}  of tidal effects in comparable-mass binary systems, based on an effective action approach, completed by an EOB reformulation. Reference   \cite{Bini:2012gu} found that the  perturbative description of these effects leads, within the standard EOB description of circular orbits, to a radial potential of the form 
 \begin{equation}  \label{aT}
A(u, \nu, \mu_T) = A_{2pp}(u,\nu) + \mu_T \, a_T(u, \nu) + O(\mu_T^2), 
 \end{equation}  
 where the term $ \mu_T \, a_T(u,\nu)$ denotes the additional contribution,  coming from
tidal interactions, to the ``two-point-particle'' EOB radial potential $A_{2pp}(u,\nu) $ . Here   $ \mu_T $ symbolically denotes a generic tidal parameter (actually there is a sum
over a set of tidal parameters:  $ \mu_T  = \mu_2^A, \mu_2^B, \mu_3^A, \cdots$).
It was proven in Sec.\ VI of Ref.\ \cite{Bini:2012gu}  that, in the extreme-mass-ratio limit $\nu \ll 1$, the
(various) tidal contributions $a_T(u, \nu) $ are {\it singular at the LR}. More precisely, they formally
blow-up as $a_T(u, \nu=0) \sim (1-3u)^{-1}$ when $u \to \frac13$.
 Let us sketch how this singularity in $a_T(u)$ can be avoided by a suitable phase-space transformation that replaces it with an alternative regular description.

First, it should be noted that when comparing Eq.\ (\ref{aT}) and  the GSF-expanded result $A(u,\nu)= 1-2u + \nu \, a(u) +O(\nu^2)$ the analogy is between a perturbation expansion in powers of
$ \mu_T$ (``tidal expansion") and a perturbation expansion in powers of $\nu$ (GSF expansion).
To make the argument more crisp, let us actually set $\nu$ to zero in Eq.\ (\ref{aT}), i.e. consider tidal effects on a body of mass $m_1 \ll m_2$. [It is in this limit that the results of Ref.\ \cite{Bini:2012gu}  which we shall use below   could be rigorously proven.]  Let us now recall how Ref.\ \cite{Bini:2012gu}  derived the presence of a LR singularity in $a_T(u)$. This was done in essentially two steps: (i) computation of the additional effective action due to tidal effects, and of its Hamiltonian formulation; (ii) reformulation of this original Hamiltonian perturbation  as a contribution to the {\it standard} EOB Hamiltonian in the limit of circular motions. 
The result of the first  step is that tidal effects add to the squared effective Hamiltonian $(\widehat H_{\rm eff} (u,p_{\varphi},p_r))^2$ a new contribution which, for general orbits, is {\it quartic} in the (effective) momenta $p_{\mu}$, say
\begin{equation} \label{deltaTH2original}
\delta_T  \widehat H^2_{\rm eff} (u,p_{\varphi},p_r) = \mu_T \, \bar q(u,p_{\mu}) +  O(\mu_T^2),
\end{equation}
with
\begin{equation} \label{qT}
 \bar q(u,p_{\mu})  = C^{\mu \nu \kappa \lambda}(u) \,p_{\mu} \,p_{ \nu}\, p_{ \kappa}\,p_{ \lambda},
 \end{equation}
 where the tensor $C$ is a smooth function of $u$ (in particular it is regular at $u=\frac13$). 
 [In the above expressions the time component $p_0$ is meant to be replaced by (minus) the unperturbed effective Hamiltonian.]
 For instance, for the leading-order quadrupolar tidal effects the tensor $C_{\mu \nu \kappa \lambda}$ is proportional to a symmetrized version of $(1-2u) \,  R^{\alpha}_{\ \mu \beta \nu} R^{\beta}_{\ \kappa \alpha \lambda}$, where $R^{\alpha}_{\ \mu \beta \nu}$ is the background curvature tensor.
 
 It is the second step taken in Ref.\ \cite{Bini:2012gu} (namely the reformulation into the standard EOB Hamiltonian form) which actually introduced the singular behavior at   $u=\frac13$.  To explain this point, let us first recall that the {\it standard form} of the EOB Hamiltonian,   introduced in Ref.\ \cite{Damour:2000we}, consists of imposing some specific {\it restrictions} on the momentum dependence of the squared effective EOB Hamitonian. Namely, it should have the form
 \begin{equation}
\label{Heff2}
\widehat H_{\rm eff}^2 (u,p_{\varphi},p_r) = 
A(u;\nu) \left( 1 + p_{\varphi}^2 \, u^2 + \frac{p_r^2}{B(u;\nu)} + \widehat Q_{\rm restr}(u, p_{\varphi},p_r; \nu) \right) \, ,
\end{equation}
with the specific form $p_{\varphi}^2 \, u^2$ of the term quadratic in $p_{\varphi}$, and with the restriction that the mass-shell deformation term  $\widehat Q_{\rm restr}$  vanish quartically in the limit of small radial momentum:
 \begin{equation}\label{Qrestr}
\widehat Q_{\rm restr}(u,p_{\varphi},p_r \to 0) = O(p_r^4) \, .
 \end{equation}
Reference \cite{Damour:2000we} showed, at the 3PN accuracy, how such a restricted form can be reached by applying a suitable symplectic  transformation of EOB phase-space variables $(q^i, p_i) \to ({q'}^i, {p'}_i) $.  At the 3PN accuracy, it was found that $\widehat Q_{\rm restr}(u,p_{\varphi},p_r )$ did not depend upon $p_{\varphi}$, and was given by  $\widehat Q_{3PN}(u,p_r)= 2 (4-3\nu) \nu u^2 p_r^4$. We, however, expect that $\widehat Q_{\rm restr}$  will involve a dependence on $p_{\varphi}$ at higher PN orders (see the discussions in  \cite{Damour:2009sm} and the appendix of  \cite{Barausse:2011dq}).  Let us also recall that this standard EOB form  is well tuned to the description of near-circular orbits.
 For instance, as discussed in \cite{Damour:2009sm}, it  allows one to describe the dynamics of small-eccentricity orbits only in terms of the radial potentials $A(u)$ and $B(u)$.

Reference \cite{Bini:2012gu}  used the fact that, {\it for the special case of circular motions},  the value of the original first-order Hamiltonian perturbation  (\ref{deltaTH2original})   must coincide with the corresponding value in the transformed phase-space coordinates,  $(q',p')$.  In accord with the results we shall derive below in a GSF context, this led to the following link between the original, LR-regular
momentum-dependent perturbation (\ref{qT}) and the momentum-independent  tidal perturbation of the standard radial $A$ potential, entering (\ref{aT}):
\begin{equation} \label{aTqT}
\left[  a_T(u') \right]_{u'=u} = \left[ \frac{   \bar q(u,p_{\mu})   }{  1 + {p}_{\varphi}^2 \, u^2  } \right]_{\rm circ} .
\end{equation}
Here, we have added a prime on the radial variable appearing on the left-hand side as a reminder that the function $a_T(u)$ belongs to the transformed phase-space coordinates $(q',p')$. However, at the first order in $\mu_T$ we are considering, $u'$ can (and actually should) be identified with the
dummy variable $u$ used on the right-hand side (RHS). Of crucial importance in the result (\ref{aTqT}) is the fact that all the quantities on the RHS should be evaluated along the one-parameter sequence of {\it circular motions}.  This means that the momentum components $p_{\mu}$ on the RHS are to be replaced as follows: $- p_0$ is replaced by $E_{\rm circ}(u)$, Eq.\ (\ref{defEcirc}), while the tangential component of the spatial momentum, say $p_{\shortparallel} \equiv {p}_{\varphi} \, u$ is replaced by
\begin{equation} \label{jcirc}
p_{\shortparallel}^{\rm circ}(u) \equiv u \, p_{\varphi}^{\rm circ}(u) \equiv  \sqrt{\frac{u}{1-3u}}.
 \end{equation} 
Here, and in the following, we shall often use, without explicating the change of notation, {\it scaled} EOB variables, such as $p_{i}=p^{\rm phys}_{i}/\mu$ or $r=r^{\rm phys}/M$. We also remove the label EOB from the radial coordinate $r$ for notational simplicity.

Finally, as the (transformed) inverse radial variable $u$ tends to $\frac13$, we see that the link  (\ref{aTqT}) implies that the potential $a_T(u)$ blows up proportionally to
\begin{equation} \label{}
 a_T(u \to \frac13)  = \left[ \frac{ O( p_{\shortparallel}^4)  }{  1 + {p}_{\shortparallel}^2 } \right]_{\rm circ}  \sim (p^{\rm circ}_{\shortparallel}(u))^2 \sim (1-3u)^{-1}.
\end{equation}
The main point we wanted to emphasize  by explicating the appearance of the pole singularity in
 $a_T(u)$ (as $u \to \frac13$) found in Ref.\ \cite{Bini:2012gu}, was that its origin was the growth, for large values of the tangential momenta $p_{\shortparallel}$,  of the original, momentum-dependent Hamiltonian perturbation (\ref{deltaTH2original}) which was, to start with, perfectly regular near (and across) $u=\frac13$. As we shall see in more technical detail below (in the GSF case), it is actually the change of phase-space variables $(q,p) \to (q',p')$, needed to go to the  {\it restricted, standard} EOB Hamiltonian form, Eq.\  (\ref{Heff2}),  that is responsible for introducing a singularity in $a_T(u)$.

 Let us now apply the same reasoning to our GSF context, i.e. using $\nu$ as a perturbation parameter, instead of $\mu_T$ in the tidal case above. To clarify this application (as well as what was at work in the tidal example above) we shall also
 show how to explicitly construct the phase-space transformation, say  ${\cal T} :  (q^i, p_i) \to ({q'}^i_{\rm EOB}, {p'}_i^{\rm EOB}) $, needed to go to the  {\it restricted, standard} EOB Hamiltonian form   (\ref{Heff2}), parametrized by  two radial potentials $A(u)$ and $B(u)$, and  an EOB mass-shell deformation function  $\widehat Q_{\rm restr}(u,p_{\varphi},p_r)$ constrained to be  $O(p_r^4)$ when $p_r \to 0$. 
  In the discussion of Ref.\ \cite{Bini:2012gu} recalled above, the transformation  ${\cal T} $  was implicitly used, but its explicit form did not enter. 
 
Let us start from  the {\it unperturbed} (squared effective) EOB Hamiltonian 
\begin{equation}
\label{Heff20}
\widehat H^2_{\rm eff \, 0} (u,p_{\varphi},p_r) = 
A_0(u) \left( 1 + p_{\varphi}^2 \, u^2 + \frac{p_r^2}{B_0(u)}  \right) \, ,
\end{equation}
with $A_0(u) = 1- 2 u$,   $B_0(u) = (1- 2 u)^{-1}$, and consider that, in some original phase-space coordinates $(q^i,p_i)$ (say, coordinates directly related to a Lorenz-gauge calculation) GSF effects modify it by adding a new contribution of the general form
\begin{equation}
\label{deltanuHeff2}
\delta_{\nu}  \widehat H^2_{\rm eff} (u,p_{\varphi},p_r) = \nu \, \bar q(u,p_{\varphi},p_r) +  O(\nu^2).
\end{equation}
We assume that, in the original phase-space coordinates  $(q^i, p_i)=(r,\varphi, p_r, p_{\varphi})$ (for planar motions), the perturbed Hamiltonian in
Eq.\  (\ref{deltanuHeff2}) is a smooth function of   $(q^i, p_i)$, and, in particular, that no divergence occurs when $u$ crosses the value $\frac13$.  On the other hand, we allow for a general (unrestricted) dependence of $\bar q(u,p_{\varphi},p_r)$ on the momenta. In other words, the sum
$ \widehat H^2_{\rm eff \, 0} (u,p_{\varphi},p_r)   +  \delta_{\nu}  \widehat H^2_{\rm eff} (u,p_{\varphi},p_r) $ is not assumed to be of the standard EOB form, Eq.\  (\ref{Heff2}).

Let us now look for a symplectic phase-space transformation, say
$ {\cal T}: (q^i, p_i) \to ({q'}^i_{\rm EOB}, {p'}_i^{\rm EOB}) $, which simplifies the form of the original squared effective Hamiltonian, $\widehat H^2_{\rm eff \, 0} (u,p_{\varphi},p_r) +  \delta_{\nu}  \widehat H^2_{\rm eff} (u,p_{\varphi},p_r) $, by putting it in the {\it standard EOB form} of the type displayed in  Eq.\  (\ref{Heff2}), with the restriction (\ref{Qrestr}) . 
We shall  sketch here how one can formally construct  the needed symplectic transformation $\cal T$ within our GSF-perturbative context.  (In previous work, $\cal T$ was constructed within a PN-perturbative context.)  $\cal T$  can be taken to arise from a (perturbed) {\it generating function} of the form $g(r,p_{\varphi}, p_r)= \nu\, p_r \epsilon^r(r,p_{\varphi}^2,p_r^2) + O(\nu^2)$. [In preparation for some technical developments below, we use the space-time symmetries of the two-body dynamics to infer that the time-reversal-invariant, scalar quantity $\epsilon^r$ can be written as a function of $p_{\varphi}^2= ( \bold q \times \bold p)^2$ and $ p_r^2= ( \bold q \cdot \bold p)^2/r^2$.] The $O(\nu)$ phase-space transformation generated by $g(r,p_{\varphi}, p_r)$ is effected via a Poisson bracket, namely $\delta_g X = \{ X, g\}$, on any phase-space function $X(q,p)$. From a technical point of view, let us formally consider that all the dynamical functions of interest are expanded in powers of $p_r$, while keeping the dependence on $r$ and $p_{\varphi}$ exact. (This differs from the usual PN-perturbative construction of  $\cal T$  which essentially uses a multiple expansion in powers of $p_r, p_{\shortparallel}\equiv u p_{\varphi}$ and $u$.)  A simple calculation (using $\{ r, p_r\}=1$,   $ \{p_r,p_{\varphi} \}=0$, etc.) then shows that $\delta_g p_r=\{ p_r, g\}=O(p_r)$. In other words, the transformation $\cal T$ respects each order in the expansion in powers of $p_r$, namely $\delta_g p_r^n= O(p_r^n)$.
 For simplicity, we shall focus here on the terms of zeroth-order in $p_r$. These terms are already quite nontrivial. It can be verified that the reasoning indicated below can be straightforwardly
 extended to higher powers of $p_r$.
 
At zeroth-order in $p_r$, the original Hamiltonian perturbation  $\nu \, \bar q(u,p_{\varphi},p_r =0) +  O(\nu^2)$ depends on {\it two} independent phase-space variables, namely $u$ and $p_{\varphi}$. Our aim here is to show how a suitable generating function $g$ can reduce the general dependence of   $ \bar q (u,p_{\varphi},p_r =0)$  on $u$ and $p_{\varphi}$   to the special one entering  Eq.\  (\ref{Heff2}). 
 At zeroth order in $p_r$, the only relevant changes in phase-space variables are that in $r$ and $p_{\varphi}$. [The change in $\varphi$ is irrelevant as the relevant dynamical observables do not explicitly depend on $\varphi$.] A simplifying feature is that $\delta_g p_{\varphi}= \{  p_{\varphi}, g\}=0$ (because $\partial g/\partial \varphi=0$). Let us then consider the change in $r$: $\delta_g r=\{ r, g\}= \partial g/\partial p_r$. This is easily found to be
$\delta_g r= \nu \,\epsilon^r(r,p_{\varphi}^2,p_r^2) + O(p_r^2)+O(\nu^2)$.  In other words, to zeroth order in $p_r$, and to first-order in $\nu$ [i.e., modulo corrections of $O(p_r^2)+O(\nu^2)$] we have 
$ \delta_g r= \nu \,\epsilon^r(r,p_{\varphi}^2,0)$. Note that the change in radial coordinate is more general than a simple (configuration-space) coordinate transformation $\delta r = \xi^r (r)$ in that $\delta_g r$ depends on {\it both} $r$ and $p_{\varphi}$.  This {\it phase-space} dependence of $\delta_g r$ is crucial for allowing the transformation $\cal T$ to reduce the original contribution Eq.\  (\ref{deltanuHeff2}) to the standard canonical EOB form. For convenience, we shall work in the following with the corresponding change in $u=1/r$, i.e. $\delta_g u=-\delta_g r/r^2$, and denote it as $ \delta_g u= \nu \,\epsilon^u(r,p_{\varphi}^2)$ [modulo corrections of $O(p_r^2)+O(\nu^2)$].

The condition on $g$ is that it transforms the sum of Eq.\  (\ref{Heff20}) and Eq.\  (\ref{deltanuHeff2}) into the standard form Eq.\  (\ref{Heff2}), with some modified potentials $A(u)=A_0(u) + \nu a(u) +O(\nu^2)$,  $B(u)=B_0(u) + \nu b(u) +O(\nu^2)$, and some {\it restricted} $O(\nu)$ mass-shell term $\widehat Q_{\rm restr}$ satisfying Eq.\  (\ref{Qrestr}). Written explicitly, this condition means that
$\delta_g  \widehat H^2_{\rm eff} (u,p_{\varphi},p_r)$ must be equal to the difference between
(\ref{deltanuHeff2}) and a GSF perturbation of the standard EOB Hamiltonian (\ref{Heff2}). The latter GSF perturbation has the structure
\begin{equation}
\delta_{\nu} \widehat H_{\rm eff \, standard}^2 (u,p_{\varphi},p_r) = \nu a(u)  \left( 1 + p_{\varphi}^2 \, u^2 + \frac{p_r^2}{B_0(u)}  \right) - \nu b(u)  \frac{A_0(u) p_r^2}{B_0^2(u)} + O(p_r^4),
\end{equation}
where the contribution $ O(p_r^4)$ comes from the  {\it restricted} $O(\nu)$ mass-shell term $\widehat Q_{\rm restr}$.
 At lowest order in $p_r$, and after division by $\nu$, the condition  on $g$ reads (with $j\equiv  p_{\varphi}$ to ease the notation)
 \begin{equation}
 \frac1{\nu} \delta_g \left[ (1-2u) (1+ j^2 u^2) \right]=  \bar q(u,j,p_r=0)- a(u) \, (1+j^2u^2) .
 \end{equation}
 If we introduce the short-hand notations 
 $$\bar \epsilon(u,j^2) \equiv \frac{2 u (1-3u)}{1+j^2u^2} \epsilon^u(r,j^2)$$
  and 
  \begin{equation}
 \alpha(u,j^2)\equiv  \frac{\bar q(u,j,p_r=0)}{1+j^2u^2} ,
 \end{equation} 
  the latter condition explicitly reads (after dividing both sides  by $1+ j^2 u^2$)
 \begin{equation} \label{gcondition}
 \bar \epsilon(u,j^2) (j^2-j^2_{\rm circ}(u))= \alpha(u,j^2)- a(u),
 \end{equation} 
 where 
 \begin{equation} \label{j2circ}
 j^2_{\rm circ}(u) \equiv \frac1{u(1-3u)}
 \end{equation} 
denotes the function of $u$ which describes the value of  $j\equiv  p_{\varphi}$ along the sequence of circular orbits. [One must carefully distinguish the general, independent phase-space variable $j$ from the specific function $j_{\rm circ}(u)$.]

From the condition  (\ref{gcondition})  on $g$, one first deduces [by taking the limit $j^2\to  j^2_{\rm circ}(u)$ on both sides], that the ({\it a priori} unkown) value of the canonical perturbed EOB potential $a(u)$ corresponding to the original non-canonical perturbation in Eq.\  (\ref{deltanuHeff2}) is given by
 \begin{equation}\label{afromq}
 a(u)= \alpha(u,j^2_{\rm circ}(u)) = \frac{\bar q(u,j_{\rm circ}(u),0)}{1+j^2_{\rm circ}(u)u^2} .
  \end{equation} 

Then, we derive the value of the function  $\epsilon^r(r,p_{\varphi}^2,p_r^2=0)$ determining the
transformation $g$ (at lowest order in $p_r$) from
\begin{equation}\label{gfromq}
\frac{2  \epsilon^u(r,j^2)}{j^2_{\rm circ}(u) (1+j^2u^2)}\equiv \bar \epsilon(u,j^2) = \frac{\alpha(u,j^2)-  \alpha(u,j^2_{\rm circ}(u))} {j^2-j^2_{\rm circ}(u)} .
  \end{equation} 
Note that the last expression on the RHS defines, despite the appearance of a denominator that vanishes along the phase-space curve $j^2-j^2_{\rm circ}(u)$, a smooth  function of $u$ and $j^2$ along this seemingly singular curve. This follows from the fact that $\alpha$ is a smooth function of its second argument. [Here, we are using the fact that, if $f(x)$ is a smooth function of $x$, $g(x,y)\equiv (f(x)-f(y))/(x-y)$ is a smooth function of the two variables $x$ and $y$, even in the vicinity of the diagonal $x=y$.]

The result in Eq.\  (\ref{afromq}) is the GSF-perturbation analog of the (tidal-perturbation) result  of  Ref.\ \cite{Bini:2012gu}  cited in  Eq.\  (\ref{aTqT}) above. It is this result which explains the appearance of LR singularities in the ``standard'' EOB potential $a(u)$ when starting from a LR-regular original non-standard perturbed EOB Hamiltonian, Eq.\  (\ref{deltanuHeff2}). Indeed, if  the original $O(\nu)$ GSF perturbation in Eq.\  (\ref{deltanuHeff2}) is regular in phase-space (including near $u=\frac13$), but grows as $p_i^n $ when the components $p_i$ of the momenta get large, we see from Eq.\ (\ref{afromq}) that such a growth at large momenta in the original phase space will lead, after the transformation $\cal T$, to a corresponding growth of the purely radial function $a(u)$ as $u\to \frac13^-$ on its (transformed) $u$ axis of the type
\begin{equation}
a(u) \propto \frac{j^n_{\rm circ}(u)}{1 + u^2 j^2_{\rm circ}(u)} \sim ( j_{\rm circ}(u))^{n-2} \sim \frac1{(1-3u)^{(n-2)/2}} .
\end{equation}
Reciprocally, if we reason backwards, our construction above of  the $O(\nu)$  generating function $g$ can be used (as we shall explicitly discuss in the next subsection), when starting from the singular standard $O(\nu)$ EOB potential  $a(u) \sim (1-3u)^{-1/2}$, to transform it away, and to replace it by  a regular, (unrestricted) momentum-dependent  $O(\nu)$ contribution to the EOB $Q$ function.

We therefore conclude that our finding above of a LR singularity in the perturbed standard  $O(\nu)$ EOB potential $a(u)$ probably \footnote{At the level $p_r^0$ in an expansion in powers of $p_r$, our explicit procedure for constructing the generating function $g$ gives a {\it proof} of this conclusion. However, one would need to have control of the full $O(\nu)$ Hamiltonian describing the full, non-circular dynamics in order to prove that all its  (apparently) singular behaviour related
to the LR can be described by a regular  $O(\nu)$ contribution $ \bar q(u,p_{\varphi},p_r) $.}
 originates from an everywhere-regular unrestricted perturbed $O(\nu)$ effective Hamiltonian  $ \nu \, \bar q(u,p_{\varphi},p_r) $,  Eq.\  (\ref{deltanuHeff2}), which grows {\it cubically}  (i.e. $ \propto p_{\varphi}^3$) as  $p_{\varphi} \to \infty$. Note that
the new, transformed $u$ axis corresponds to the original phase-space variable $u'=u+ \nu \epsilon^u(r,p_{\varphi}^2) +O(\nu^2)$. Note also that the result  Eq.\  (\ref{gfromq}) formally determining the transformation $\cal T$ to the canonical EOB form becomes {\it ill-defined} as $u$ tends to $\frac13^-$. Worse, if we approximate [for large $j^2$ and large $j^2_{\rm circ}(u)$] $ \bar q(u,j^2)$ as $\sim j^3$, so that  $\alpha(u,j^2)\equiv \bar q(u,j,0)/(1+j^2u^2) \sim j$, we see from Eq.\  (\ref{gfromq}) that 
\begin{equation}
 \epsilon^u \sim j^2 j^2_{\rm circ}  \frac{j-j_{\rm circ}}{j^2- j^2_{\rm circ}} \sim \frac{ j^2 j^2_{\rm circ}}{j+j_{\rm circ}} .
\end{equation} 
Not only is this result blowing up as either $j_{\rm circ}$ or $j$ gets large, it is actually only well-defined {\it above the LR}, i.e. when $r>3$ or $u<\frac13$, because it contains the {\it square root} $j_{\rm circ}(u)= (u(1-3u))^{-1/2}$. Therefore, the transformation $\cal T$, and the corresponding  standard EOB potential $a(u)$, are (probably)  {\it only defined when $u<\frac13$} .

One can also check that the reasoning above can be extended to higher orders in $p_r$. In particular, at order $O(p_r^2)$ it determines the value of  the second ``standard'' EOB  potential $ \delta_{\nu} B(u) \equiv B(u;\nu) -B_0(u) = \nu b(u) + O(\nu^2)$. A preliminary study of the $O(\nu)$
contribution  to the standard $B$ potential indicates that it blows up like  $b(u) \propto (1-3u)^{-5/2}$ as $u\to \frac13$. As a consequence, $\bar d (u)$ will have the same type of divergence near the LR.

We can summarize our conclusion by an analogy.  For many years, researchers in general relativity have been mystified by what they called the ``Schwarzschild singularity'', namely the fact that the standard Schwarzschild metric is singular at $r=2M$, notably because $g_{rr} = (1-2M/r)^{-1}$ blows up there and then changes sign. It was only in the 1960s, notably through the work of Kruskal, that it became clearly understood that this ``$r=2M$ Schwarzschild singularity'' is a  singularity of the standard Schwarzschild coordinates, which can be gauged away by a suitable transformation of the spacetime coordinates, including a necessary mixing of space and time coordinates. Our conclusion is that the singularity $a(u) \sim (1-3u)^{-1/2}$ we found is, somewhat similarly, only  due to the a singularity of the  {\it standard phase-space coordinates} used in the EOB formalism.
This ``phase-space-coordinate singularity'' can be gauged-away by a suitable symplectic transformation, necessarily mixing coordinates and momenta.

\subsection{Impact of our findings on the EOB formalism}

What conclusions should we draw from our GSF sub-ISCO results for the EOB formalism?
One possible reaction would be to modify the standard EOB strategy that concentrates all the
near-circular dynamical information into the effective metric [parametrized by the two radial potentials $A(u)=A_0(u)+\nu a(u) + O(\nu^2)$ and  $B(u)=B_0(u)+\nu b(u) + O(\nu^2)$], and to
allow the third EOB function, $\widehat Q(u,p_{r}, p_{\varphi}; \nu)$, to participate in the description of near-circular orbits, by relaxing the constraint that $\widehat Q(u,p_{r}, p_{\varphi}; \nu)$ vanishes when $p_r \to 0$. (In the analogy with the ``$r=2M$ Schwarzschild singularity'' case, this reaction is the analogue of modifying the standard Schwarzschild coordinates by allowing the use of a more general spacetime gauge fixing.) Let us sketch how this could be done. For this purpose, it is convenient to introduce a special notation for the piece of the squared effective Hamiltonian contributed
by the $Q$ function. Let us denote
\begin{equation}
\bar Q(u, p_{\varphi},p_r; \nu)  \equiv A(u;\nu)  \widehat Q(u, p_{\varphi},p_r; \nu) 
\end{equation} 
so that we have a simple linear decomposition of the squared effective Hamiltonian:
\begin{equation}\label{Heff2bis}
\widehat H^2_{\rm eff} (u,p_{\varphi},p_r) = 
A(u;\nu) \left( 1 + p_{\varphi}^2 \, u^2 \right) + \frac{A(u;\nu)}{B(u;\nu)}\ p_r^2 + \bar Q(u, p_{\varphi},p_r; \nu) .
\end{equation}  
Note that we are no longer adding a subscript ``restr'' to $\widehat Q$ or $\bar Q$. Indeed, we are no longer imposing the constraint (\ref{Qrestr}), but we allow a general momentum dependence in  $\widehat Q$ and $\bar Q$.

Let us now consider the GSF expansion of the squared effective Hamiltonian in Eq.\  (\ref{Heff2bis}), corresponding to the GSF expansions of  $A$, $B$ and $\bar Q$, namely $A(u)=A_0(u)+\nu a(u) + O(\nu^2)$ and  $B(u)=B_0(u)+\nu b(u) + O(\nu^2)$ and, consistently with  Eq.\  (\ref{deltanuHeff2}),  $\bar Q(u, p_{\varphi},p_r; \nu)=\nu \, \bar q(u,p_{\varphi},p_r) +  O(\nu^2)$. It yields 
$\widehat H^2_{\rm eff} (u,p_{\varphi},p_r)= H^2_{\rm eff \, 0} (u,p_{\varphi},p_r) + \delta_{\nu} \widehat H^2_{\rm eff} (u,p_{\varphi},p_r)$, where the zeroth-order term is given in Eq.\  (\ref{Heff20}) above, and where the $O(\nu)$ perturbation is given by 
\begin{equation}
\label{deltanuHeff2bis}
\delta_{\nu}  \widehat H^2_{\rm eff} (u,p_{\varphi},p_r) = \nu \, a(u)  \left( 1 + p_{\varphi}^2 \, u^2 \right)+ \nu \,  \left(     \frac{a(u)}{B_0(u)}   -  b(u)  \frac{A_0(u) }{B_0^2(u)} \right) p_r^2 + \nu \, \bar q(u,p_{\varphi},p_r) +  O(\nu^2).
\end{equation}
As above, let us focus on the terms in the Hamiltonian which survive in the limit where $p_r \to 0$:
\begin{equation}
\label{deltanuHeff2biscirc}
\delta_{\nu}  \widehat H^2_{\rm eff} (u,p_{\varphi},p_r=0) = \nu \, a(u)  \left( 1 + p_{\varphi}^2 \, u^2 \right)+ \nu \, \bar q(u,p_{\varphi},p_r=0) +  O(\nu^2).
\end{equation}
This contrasts with the corresponding $p_r \to 0$ limit of the perturbation of the {\it standard}, restricted EOB Hamiltonian which would only contain the first contribution, linked to the perturbation of the standard $A$ potential. 

In previous sections, we (following, in particular,  Ref.\ \cite{Barausse:2011dq}) have interpreted the numerical GSF data by assuming that we were working within the context of a {\it standard} EOB Hamiltonian.  It was within this context that we found a standard $a(u)$ potential of the form $a_E(u)  E_{\rm circ}(u) $. In other words, we interpreted the GSF data in terms of the following  perturbation of the standard-gauge EOB Hamiltonian
\begin{equation}
\label{deltanu0}
\delta_{\nu}  \widehat H^2_{\rm eff \, standard} (u,p_{\varphi},p_r=0) = \nu \, a_E(u)  E_{\rm circ}(u)  \left( 1 + p_{\varphi}^2 \, u^2 \right)+  O(\nu^2).
\end{equation}
where $E_{\rm circ}(u)$ is the function of $u$ alone defined in Eq.\ (\ref{defEcirc}). 

Let us now discuss the many ways in which the latter,  LR-singular standard EOB Hamiltonian (\ref{deltanu0})  can be traded off for an everywhere-regular Hamiltonian of the general, non-standard form (\ref{deltanuHeff2biscirc}).  From  Eq.\ (\ref{gcondition}),  the criterion for two Hamiltonians to be equivalent (at zeroth order in $p_r$) modulo a symplectic transformation is simply that their numerical values agree along the sequence of circular motions, i.e. when $p_{\varphi} = p_{\varphi}^{\rm circ}(u)$. This criterion leaves many possibilities for transforming   (\ref{deltanu0})  into an equivalent, but LR-regular Hamiltonian.

The simplest way of doing so is to replace  the problematic factor $E_{\rm circ}(u)= H_{\rm{eff} \, 0}(u,p_{\varphi}^{\rm circ}(u),p_r=0)$ in (\ref{deltanu0})  by  the $\nu \to 0$ limit of the full (non circularly reduced) effective EOB Hamiltonian, i.e. the square-root of Eq.\ (\ref{Heff20}), namely
\begin{equation}
\label{Heff0}
\widehat H_{\rm eff \, 0} (u,p_{\varphi},p_r) \equiv 
\sqrt{A_0(u) \left( 1 + p_{\varphi}^2 \, u^2 + \frac{p_r^2}{B_0(u)}  \right) } \, .
\end{equation}
This leads, when considering for simplicity the $p_r=0$ hypersurface in phase space \footnote{Here we are setting $p_r$ to zero only to avoid discussing the fate of the $B$-type contribution. This is a much weaker constraint (in phase space) than to restrict oneself to the one-parameter sequence of circular motions. },  to the following first non-standard possibility for  replacing (\ref{deltanu0}):
\begin{equation}
\label{deltanu1}
\delta'_{\nu}  \widehat H^2_{\rm eff} (u,p_{\varphi},p_r=0) = \nu \, a_E(u)  H_{\rm{eff} \, 0}(u,p_{\varphi},p_r=0) \left( 1 + p_{\varphi}^2 \, u^2 \right)+  O(\nu^2).
\end{equation}
Considered as functions over phase space, the two perturbed Hamiltonians in
Eqs.\ (\ref{deltanu0}) and (\ref{deltanu1}) are very different functions. In particular, the new Hamiltonian Eq.\ (\ref{deltanu1})  is regular across $u=\frac13$ \footnote{Here we are neglecting the relatively minor lack of regularity of $a_E(u)$ due to the presence of a term $\sim (1-3u) \, \ln | 1-3u|$. Alternatively, one could replace the argument of the offending logarithm by $\sim 1/H_{eff \, 0}^2(u,p_{\varphi})$, so as to have a completely smooth  functional dependence on $u$.}, 
because it   is constructed from the ``regularized'' function $a_E(u)$, extrapolated beyond $u=\frac13$, as discussed in  Sec.\ III. In addition,  $\delta'_{\nu}  \widehat H^2_{\rm eff} (u,p_{\varphi},p_r=0)$
 vanishes at $u=\frac12$ at least as fast as $\sqrt{1-2u}$. [Actually, as we saw above that the regularized function $a_E(u)$
is likely to change sign near $u=\frac12$, the new Hamiltonian Eq.\ (\ref{deltanu1}), being proportional to $a_E(u)  \sqrt{1-2u}$, nearly vanishes as  $(1-2u)^{3/2}$ near $u=\frac12$.]

The $p_{\varphi}$ dependence of   Eq.\ (\ref{deltanu1})  is $\propto ( 1 + p_{\varphi}^2 \, u^2)^{3/2}$. This is consistent with our general conclusion above that {\it any} LR-regular form of the
$O(\nu)$ perturbed Hamiltonian must grow as  $p_{\varphi}^3$ for large momenta. However, it was argued in Ref.\ \cite{Damour:2000we}  that it is most natural for the $Q$ contribution that it be $O(p^4)$ for {\it small} momenta. It is easy to accommodate such a requirement by considering new reformulations of the naive possibility  Eq.\ (\ref{deltanu1}) within the ``equivalence class'' of perturbed Hamiltonians that numerically agree along the sequence of circular motions [i.e. when $p_r=0$, and $p_{\varphi}^2 = j_{\rm circ}^2(u)$].

Let us  consider the following phase-space function
\begin{equation}
k(u,p_{\varphi}) \equiv (1-2u) \frac{p_{\varphi}^2 u}{ 1 + p_{\varphi}^2 \, u^2}\, .
\end{equation}
It is easily seen that, along circular orbits [when  $p_{\varphi}^2 = j_{\rm circ}^2(u)$] the phase-space function $k(u,p_{\varphi})$ is numerically equal to 1. On the other hand, as a function of momenta it is $O(p^2)$ for {\it small} momenta. We can then define a reformulation of Eq.\ (\ref{deltanu0})  which is both regular at $u=\frac13$ and  $O(p^4)$ for {\it small} momenta, by multiplying Eq.\ (\ref{deltanu1})  by the square of $k(u,p_{\varphi}) $, say
\begin{equation}
\label{deltanu2}
\delta''_{\nu}  \widehat H^2_{\rm eff} (u,p_{\varphi},p_r=0) = \nu \,  a_E(u) \, \left(k(u,p_{\varphi}) \right)^2 \,  H_{\rm{eff} \, 0}(u,p_{\varphi},p_r=0) \left( 1 + p_{\varphi}^2 \, u^2 \right)+  O(\nu^2) ,
\end{equation}
giving, explicitly,
\begin{equation}
\label{deltanu2bis}
\delta''_{\nu}  \widehat H^2_{\rm eff} (u,p_{\varphi},p_r=0) = \nu \, a_E(u) \, u^2 \, (1-2u)^{5/2} \frac{p_{\varphi}^4}{\sqrt{ 1 + p_{\varphi}^2 \, u^2} }  +  O(\nu^2).
\end{equation}
Note that the global $p$ dependence of this new Hamiltonian is quite different from 
the previous one [namely $p^4/\sqrt{1+p^2}$ instead of $(1+p^2)^{3/2}$, where $p$ stands for $p_{\shortparallel}= u \, p_{\varphi}$], though (by consistency)
they both grow like $p^3$ for large momenta. Note also that this new Hamiltonian has a faster vanishing near $u=\frac12$, namely $\propto a_E(u) (1-2u)^{5/2}$.

More generally, one could use a phase-space transformation which trades off the GSF result  Eq.\ (\ref{deltanu0}) for a perturbed Hamiltonian such that
\begin{equation}
\label{deltanugen}
\delta^{\rm gen}_{\nu}  \widehat H^2_{\rm eff} (u,p_{\varphi},p_r=0) = \nu \,  f(u,k(u,p_{\varphi}) ) \,  H_{\rm{eff} \, 0}(u,p_{\varphi},p_r=0) \left( 1 + p_{\varphi}^2 \, u^2 \right)+  O(\nu^2) ,
\end{equation}
where $f(u,k)$ is any function such that $f(u,1)=a_E(u)$. This clearly leaves a lot of freedom in the definition of such a LR-regular version of the standard result  (\ref{deltanu0}). Note that the phase-space function $k(u,p_{\varphi})$ tends to the finite limit $(1-2u)/u$ as $p_{\varphi} \to \infty$, independently of being restricted to the sequence of circular orbits, so that the whole class of regular  perturbed Hamiltonians  Eq.\ (\ref{deltanugen}) grows as $p^3$ for large momenta.

One should not be surprised by the existence of such a large freedom in the formulation of  a regular, non-standard Hamiltonian. Indeed, given any specific, LR-regular Hamiltonian, its transform by an arbitrary regular symplectic transformation ${\cal T}_{\rm reg}$ will generate a new, regular Hamiltonian. The fact that, at zeroth order in $p_r$, such a regular ${\cal T}_{\rm reg}$ can introduce an arbitrary function of two variables   $f(u,k)$ [constrained only along a certain curve in the $(u,k)$ plane] is linked to the presence of the arbitrary function $ \epsilon^r(r,p_{\varphi}^2,p_r^2=0)$ in the $p_r\to0$ limit of a generating function $g(r,p_{\varphi}, p_r)= \nu\, p_r \epsilon^r(r,p_{\varphi}^2,p_r^2) + O(\nu^2)$. We leave to future work the detailed generalization of our considerations to higher orders in powers of $p_r$, and simply note that, anyway, such a generalization  involves [at $O(\nu)$]  an arbitrary function of {\it three} arguments [corresponding to $\epsilon^r(r,p_{\varphi}^2,p_r^2)$]. We just recall here that part of the success of the EOB formalism consists of finding good ways of trimming down this large gauge freedom to parametrize the dynamics in terms of the minimum number of relevant functions.

We have just explained how to make a full use of our sub-ISCO results, without being restricted by  their singular behavior at the LR, by relaxing the  ``standard gauge fixing'' of the EOB formalism. However, it should
be noted that such a modification of the current EOB formalism is really only needed if one wishes to
describe the dynamics of {\it ultra-relativistic} quasi-circular  orbits ($p \to \infty$) near $u=\frac13$. By contrast,  the original motivation for, and main use of, the current EOB formalism is to describe the dynamics of {\it mildly-relativistic} radiation-reaction-driven quasi-circular  orbits. Such orbits
stay close to the sequence of (stable) circular orbits down to the ISCO (i.e. for $0<u \lesssim \frac16$), and then strongly deviate from the sequence of {\it unstable} orbits that formally continue to exist when $\frac16 \lesssim u \lesssim \frac13$. Indeed, though 
the ``plunging motion'' that follows the radiation-reaction-driven quasi-circular inspiral remains approximately circular  (i.e. with $p_r^2 \ll p_{\shortparallel}^2$, see Fig.~1 in Ref.\ \cite{Buonanno:2000ef}), its path in phase-space $(q,p)$ {\it drastically} deviates from the phase-space location
of unstable circular orbits. In particular, the angular momentum $p_{\varphi}$ of a plunging orbit
stays approximately equal to its value $j_{\rm circ}^{\rm ISCO}= \sqrt{12} + O(\nu)$ when it crossed the ISCO, while the formal adiabatic sequence of  $p_{\varphi}$ values along the unstable circular orbits is given by the function $j_{\rm circ}(u)$ defined as the square root of the RHS of  Eq.\ (\ref{j2circ}).  In particular, as one gets near $u=\frac13$ the two phase-space points
$(q_{\rm plunge},p_{\rm plunge})$ and  $(q_{\rm circ},p_{\rm circ})$ become infinitely far apart. This infinite phase-space separation (in the $p_{\varphi}$ direction) is both  the cause (as we have seen above) of the divergence of the standard $a(u)$ as $u\to\frac13$,  and an indication that the latter divergence is not of direct physical relevance for describing (as the EOB formalims aims to do) the dynamics of  plunging orbits. Actually, a proof of  the capability of the standard  
3PN-accurate EOB formalism (as defined in \cite{Damour:2000we}) to accurately describe the dynamics of (comparable-mass) {\it coalescing} black hole binaries {\it down to the light-ring}  has been recently given in Ref.\ \cite{Damour:2011fu}.  Figure 1 of the latter reference  shows in particular that the (uncalibrated) standard 3PN-EOB prediction for the $ {\cal E}(j)$ curve agrees remarkably well with the NR one {\it down to the radial location of the LR} (indicated as the leftmost vertical line in the figure). By contrast, the  $ {\cal E}(j)$ curve corresponding to the formal adiabatic sequence of circular orbits starts to exhibit a strong, and increasing
deviation from $ {\cal E}^{NR}(j)$ after the crossing of the ISCO (see the dash-dotted line in the latter figure). 

In view of this effectiveness of the standardly gauge-fixed EOB formalism
for the description of the dynamics of mildly-relativistic  binary systems, it might be useful to set up
a {\it minimal way} of modifying the EOB formalism so as to incorporate our GSF sub-ISCO results.
[In our above analogy, this is like continuing to use  standard (or nearly standard) Schwarzschild coordinates when describing a system for which the coordinate singularity at $r=2M$ is not interfering with the physics one is interested in.]
Above we have discussed ways of {\it entirely trading off} the $O(\nu)$ piece of the radial $A$ potential for an equivalent momentum-dependent $Q$-type contribution. Actually, the latter
momentum-dependent $Q$-type contribution (growing $\propto p_{\varphi}^3$ for large momenta)
is only needed for describing ultra-relativistic motions, while the standard $A$-type contribution
is a simple and effective way of describing mildly-relativistic motions above the ISCO. 
One can then conceive of a mixed scheme, where the dynamics is described partly by a certain radial $ \nu \, a_0(u)$ potential, and partly by
a $\nu  \, \bar q$ contribution, with the $ \nu \, a_0(u)$ potential playing the leading role during the
inspiral, and the $\nu  \, \bar q$ contribution taking over only during the (late) plunge.
For instance, considering as above the terms remaining when $p_r \to 0$, if we constrain the $\nu \, \bar q$ contribution to have the same $p$-dependence as in (\ref{deltanu2bis}) , namely  $p^4/\sqrt{1+p^2}$, we can use
\begin{eqnarray}
\label{deltanu3}
\delta'''_{\nu}  \widehat H^2_{\rm eff} (u,p_{\varphi},p_r=0) &=& \nu \, a_0(u)  \left( 1 + p_{\varphi}^2 \, u^2 \right)+
\nonumber\\
&&
  \nu \,  \frac{p_{\varphi}^4}{\sqrt{ 1 + p_{\varphi}^2 \, u^2} }  \left( a_E(u) \, u^2 \, (1-2u)^{5/2} - a_0(u) \, u^2 \, (1-2u)^{3/2} (1-3u)^{1/2} \right) +  O(\nu^2).
\end{eqnarray}
Here one can choose $a_0(u)$,  which corresponds to a $\nu$-deformed radial potential $A(u;\nu) = A_0(u) + \nu \, a_0(u)$, at will. For instance, one could choose $a_0(u) = 2 \, u^3$, so that the
$p^4/\sqrt{1+p^2}$-type contribution in Eq.\ (\ref{deltanu3}) starts, when $u$ is small, proportionally to $u^6$, i.e. at the 3PN level [by contrast to Eq.\ (\ref{deltanu2bis}) which starts like $u^5 p_{\varphi}^4/\sqrt{1+u^2 p_{\varphi}^2}$, which corresponds to the 2PN level]. Alternatively, one could choose an $a_0(u)$ which stays very close to the ``exact'' standard one ($ a_E(u)  E_{\rm circ}(u)$) up to some value $u=u_0$, and then deviates from it when $u>u_0$, and stays regular across the LR.  Such a choice would ensure that the $p^4/\sqrt{1+p^2}$-type contribution in Eq.\ (\ref{deltanu3}) has a negligible effect when $u\leq u_0$, and starts modifying the dynamics only when $u>u_0$. For instance, one could choose a value of $u_0$ between the ISCO and the LR. This would allow one to make full use of our new strong-field results on $a(u)$ up to $u=u_0 < \frac13$, essentially without modifying the EOB formalism up to $u=u_0 $. [It is with this program in mind---of defining some simple, accurate $a_0(u)$ approximation to $a^{\rm standard}(u)$---that we have given (above) accurate estimates for the first three derivatives of  $a^{\rm standard}(u)$  at the ISCO.]
The LR-regularized $p^4/\sqrt{1+p^2}$-type contribution in Eq.\ (\ref{deltanu3})  would then only affect the end of the plunge which follows the crossing of the ISCO.

We leave to future work a study of the performances and relative merits of the various possible completions of the EOB formalism discussed above, as well as a discussion of the needed extra
terms of order $O(p_r^2)$ ($B$-type contributions) and  $O(p_r^4)$  (old, standard $Q_{\rm restr}$-type contributions). In this respect, we note that it would be very valuable to be able to use GSF data
on {\it plunging motions} to directly extract EOB-useful information about the plunge dynamics taking place
after the crossing of the ISCO. Alas, the current state of development of GSF theory (namely the lack of explicit $O(\nu^2)$ results) does not allow
one to extract gauge-invariant information from the calculation of the gauge-variant self-force
along a plunging orbit. In a related vein, we note that, even if we go back to the case of exactly circular orbits,
our current GSF calculation of the {\it first-order only}, $O(\nu)$, contribution to the (standard)
radial $A$ potential is quite insufficient for allowing one to construct an estimate of
the function $A(u;\nu)$ able to accurately describe the dynamics of comparable-mass binary systems. Our current best-bet knowledge of the function $A(u;\nu)$ for comparable-mass
(non-spinning) systems (i.e. $\nu \sim \frac14$) has been obtained by: (i) introducing \cite{Damour:2009kr}
a two-parameter family
of putative $A$-functions incorporating  current analytical (and GSF) knowledge, and then (ii) best-fitting,
for each available value of $\nu$, the corresponding EOB-predicted waveform to NR waveforms
\cite{Damour:2009kr,Buonanno:2009qa}. The results of these EOB/NR fits indicate that the
function $A(u;\nu)$ {\it cannot} be accurately described with only a {\it linear}  dependence on $\nu$, i.e. a function of the form $A(u;\nu)= 1-2u + \nu \, a(u)$. The latter fact was explicitly discussed in the Conclusions of Ref.\ \cite{Damour:2009sm}, especially around Eq.\ (8.2) there. Even if one disregards the indications from such EOB/NR fits about the need for $O(\nu^2)$ terms in $A(u,\nu)$,
our GSF results on the behavior of $a(u)$ near $u=\frac13$ give another hint about the need for and importance of such terms. Let us just quote two illustrative examples which involve mildly strong-field effects \footnote{We note that the results of Refs. \cite{LeTiec:2011bk} and \cite{Tiec:2011dp}, which find good NR/AR agreements when using only corrections 
linear in $\nu$, concern certain quantities (considered in restricted domains) for which  the EOB predictions can also
be approximately described by a $\nu$-linear approximation to $A$. Such particular cases tell us nothing about the behavior of other physical quantities, and/or more general (and notably, stronger-field) domains, where the non-linearities in $\nu$ become quite important. We recall in this respect that a remarkable fact of the (canonical) 3PN-accurate $A(u,\nu)$ potential is that, thanks to special cancellations among terms of order $\nu^2$ and $\nu^3$, it happens to be {\it exactly linear in $\nu$}. This means that the $O(\nu^2)$ contribution to $A(u,\nu)$ must be at least as small as $O(u^5)$ in a weak-field expansion}  in the dynamics of a small mass around a large mass.

First, there is the issue of the existence of an ISCO. We know that it exists in the geodesic limit $\nu \to 0$, and is located at $r^{\rm phys}=6 M$, i.e. $u=\frac16$.  We should {\it a priori} expect that such a 
mildly strong-field phenomenon continues to exist (when neglecting radiation-reaction effects) as the symmetric mass ratio increases to values of order $\frac14$. 
 In the (standardly gauge-fixed) EOB formalism, the condition for the existence of an ISCO (defined as the condition for the existence of an inflection point in the effective potential describing the radial motion) is  \cite{Damour:2009sm}
\begin{equation}
\label{lsocondition}
\Delta(u_{\rm ISCO},\nu) = 0 ,
\end{equation}
where $u_{\rm ISCO}$ is the looked-for inverse radius, and where the function $\Delta(u,\nu)$ is defined as
\begin{equation}
\label{}
\Delta(u,\nu) \equiv 2  A(u,\nu) A'(u,\nu)+ 4\,u\, (A'(u,\nu))^2- 2\,u\, A(u,\nu) \, A''(u,\nu) ,
\end{equation}
with the prime denoting $d/du$.  By mathematical continuity with the solution $u_{\rm ISCO}=\frac16$ which exists when $\nu=0$, the condition (\ref{lsocondition}) will certainly admit a solution of the form
$u_{\rm ISCO}=\frac16+O(\nu)$ in a neighborhood of $\nu=0$. We can now explore the physical need for terms of order $O(\nu^2)$ in the function $A(u;\nu)$ by studying the range of values of $\nu$ where a solution of  (\ref{lsocondition})  continues to exist under the assumption that the $A$ potential is assumed to be exactly {\it linear} in $\nu$, i.e. given by the formula  $A^{\rm lin}(u;\nu)= 1-2u + \nu \, a_{\rm GSF}(u)$ with our above-determined sub-ISCO $a$ function, $a_{GSF}(u)=a_E(u) \, E_{\rm circ}(u)$. 
We find that such a solution exists only for a rather small neighborhood $(0,\nu_{\rm max})$ of $\nu=0$ with $\nu_{\rm max} \approx 0.108$. 
Note  in passing that a radial $A$ potential of the type  $A^{\rm lin}(u;\nu)= 1-2u + \nu \, a_{\rm GSF}(u)$ also has the unpleasant physical consequence of predicting the existence of  some {\it stable} rest position, at some radius $r$, with fixed values of $\theta$ and $\varphi$. Indeed, the  effective radial potential $  \widehat H^2_{\rm eff} (u, p_{\varphi}; \nu)$, considered as a function of the radial coordinate $u=1/r$, for any fixed angular momentum $p_{\varphi}$, reduces in the case where  $p_{\varphi}=0$ to 
simply $A^{\rm lin}(u;\nu)= 1-2u + \nu \, a_{\rm GSF}(u)$, which exhibits, for any non-zero value of $\nu$,  a {\it minimum} at some value of $u < \frac13$ (because $a_{\rm GSF}(u) \to + \infty$ as $u \to \frac13^-$). 

Let us give a second illustrative example for the undesired physical consequences of keeping only the term  linear in $\nu$ in the $A$ potential. It concerns another mildly strong-field effect.
In the geodesic limit, one of the unstable circular orbits plays a somewhat preferred role. It is the one located at $u=\frac14$
(i.e. $r^{\rm phys}=4M$). It has a (specific) angular momentum $p_{\varphi}=4$, and a {\it vanishing} binding energy, i.e. $E=1$. This marginally bound orbit is the end point of the special zero-binding zoom-whirl orbit which starts, in the infinite past, with zero kinetic energy at infinity (but with the non-zero angular momentum $p_{\varphi}=4$) and ends up, in the infinite future, ``whirling indefinitely'' around the large mass.  As in the case of the ISCO, one would {\it a priori} expect that such a 
mildly strong-field phenomenon will continue to exist (when neglecting radiation-reaction effects) as the symmetric mass ratio increases to values of order $\frac14$.  The condition for such a zero-binding circular orbit to exist has been written down (in the EOB formalism) in Ref.\ \cite{Damour:2009sm}. It reads 
\begin{equation}
\label{zcondition}
Z(u_*,\nu) = 0 ,
\end{equation}
where $u_*$ is the looked-for radius, and where the function $Z(u,\nu)$ is defined as
\begin{equation}
\label{}
Z(u,\nu) \equiv A(u,\nu) + \frac12 u \, A'(u,\nu) - A^2(u,\nu) \, .
\end{equation}
We know that the condition (\ref{zcondition}) admits the solution $u_*=\frac14$ when $\nu=0$. By mathematical continuity, it will certainly admit a solution of the form
$u_*=\frac14+O(\nu)$ in a neighborhood of $\nu=0$. Like in the case of the ISCO, we can now explore  the range of values of $\nu$ where a solution of  (\ref{zcondition})  continues to exist under the assumption that the $A$ potential is assumed to be exactly {\it linear} in $\nu$. We find that this leads to a much stronger constraint on the magnitude of $\nu$ than in the case of the ISCO. Namely, we find that a solution of  (\ref{zcondition}) exists only for a very small neighborhood $(0,\nu'_{\rm max})$ of $\nu=0$ with $\nu'_{\rm max} \approx 0.035$.  For larger values of $\nu$ the  growth of $a_{\rm GSF}(u)$ in the interval  $\frac14 <u<\frac13$ prevents the continued existence of a zero-binding circular orbit (as well as of the corresponding zero-binding zoom-whirl orbit). 

These striking physical consequences  of  neglecting $O(\nu^2)$ terms in $A(u,\nu)$ suggest that the $O(\nu^2)$ contribution to $A(u,\nu)$ is also divergent near $u=\frac13$, but has a {\it negative} sign.   In addition, one expects it to diverge proportionally to $(1-3u)^{-2}$ so that  the {\it ratio} between  the $O(\nu^2)$ contribution and the  $O(\nu^1)$ one is of order  $\nu/(1-3u)^{3/2}$,
as suggested by the  previously derived  consistency condition (\ref{condition2ter}).
We leave to future work
a detailed study of the higher-order GSF contributions to  $A(u,\nu)$.

\section{Conclusions and outlook}\label{Sec:conclusions}

We have computed  the conservative piece of the gravitational self-force
(GSF) acting on a particle of mass $m_1$ as it moves along any (stable or unstable) circular geodesic orbit around a
Schwarzschild black hole of mass $m_2\gg m_1$. Our main results and conclusions
are as follows.

(1)  We numerically computed  the function $h^{R, L}_{u u}(x) \equiv h^{R,
L}_{\mu \nu} u^{\mu} u^{\nu}$, where $h^{R, L}_{\mu \nu}(\propto m_1)$ is
the regularized metric perturbation in the Lorenz gauge, $u^{\mu}$ is the
four-velocity of $m_1$ in the background Schwarzschild metric of $m_2$,
and $x\equiv [Gc^{-3}(m_1+m_2)\Omega]^{2/3}$ is a dimensionless measure of
 the orbital frequency $\Omega$. Our results are collected in Tables \ref{table:data1}
and \ref{table:data1} in Appendix \ref{AppA}. The fractional  accuracy of our numerical results
ranges between $10^{-10}$ and  $10^{-8}$ for most data points, and  never
gets worse than   $10^{-5}$ (except at a single point, closest to the LR).
 Our results improve on previous calculations both in accuracy (except for
very small values of $x$, $ x< 1/200$), and
in range. In particular, our work is the first to explore the 
unstable orbits between $x=\frac15$ [slightly below the innermost stable
circular orbit (ISCO) located at $x=\frac16$], and the light ring (LR), located at
$x=\frac13$.

(2)  We particularly studied the behavior of $h^{R, L}_{u u}(x)$ just
outside the LR at $x=\frac13$ (i.e., $r= 3 G m_2/c^2$), where the circular
orbit becomes null.  We found that  $h^{R, L}_{u u}(x)$ blows up like
$h^{R, L}_{u u}(x \to \frac13)\sim -  \frac 12 \frac{m_1}{m_2} \zeta
(1-3x)^{-3/2}$, where $\zeta \approx 1$. We argued that the divergence of
$h^{R, L}_{u u}(x \to \frac13)$   can be understood from the divergent
behavior of (some of) the components of the four-velocity $u^{\mu}$  near
the LR.

(3) Using a recently discovered link \cite{Barausse:2011dq} between $h^{R,
L}_{u u}(x)$
and the piece $a(u)$, linear in the symmetric mass ratio $\nu \equiv m_1
m_2/(m_1+m_2)^2$, of the main radial potential  $A(u,\nu)= 1-2u + \nu \,
a(u) +O(\nu^2)$
of the effective one body (EOB) formalism, we computed from our GSF data
 the EOB function $a(u)$ over the entire
domain $0<u<\frac13$. Our results for the function $a(u)$  improve on
previous calculations both in accuracy, and in range. In particular, our
work is the first to explore the behavior of the EOB potential
$a(u)$ as $u \to \frac13$.
We found that $a(u)$ {\it diverges} like $a(u)\sim \frac14 \zeta 
(1-3u)^{-1/2}$ (where $\zeta \approx 1$) at the light-ring limit, $u \to
\left(\frac{1}{3}\right)^-$.

(4)  We then considered the energy-rescaled function $a_E(u) \equiv
a(u)/E(u)$, where $E(u)=(1-2u)/\sqrt{1-3u}$ is the (specific) relativistic
energy of $m_1$ in the background Schwarzschild black hole of mass $m_2\gg
m_1$.  This  energy-rescaled function has a finite limit as $u\to\frac13$,
but seems to have a weak singularity $\sim c_0 +  (1-3u) (c_1^{\rm log}
\ln |1-3u| + c_1)$ there. We  gave several high-accuracy global analytical
representations of $a_E(u)$ that incorporate all the presently known
post-Newtonian (PN)  analytical information about it, and essentially
reproduce all our numerical results within their numerical errors. We
think that our analytical models of $a_E(u) $ give  a reasonably accurate
representation of the behavior of  that function even {\it beyond} the
range ($0<u<\frac13$) where GSF data can compute it, say in the range
$\frac13 <u \lesssim \frac12$. See notably the curves for models 13, 14
and 19 in Fig.\ \ref{fig:global}, which show the Newton-rescaled function $\hat a_E(u)=
a_E(u)/ 2 u^3$. In other words, we think that our GSF calculations give
us, for the first time, valuable information about the truly strong-field
regime $u = GM/c^2 r \lesssim \frac12$.

(5)  Using our accurate analytical fits of the EOB potential $a(u)$, we
computed global analytical representations of the $O(\nu)$ pieces in the
functions giving the total energy and total angular momentum of a binary
system in terms of the frequency parameter $x$. We found that these  $O(\nu)$ functions
have rather strong (negative) divergences near the LR, namely $\sim - \tilde c
\nu (1-3x)^{-2}$ (with positive constants $\tilde c$).

(6) The GSF-induced, $O(\nu)$, shift in the value of the orbital frequency
of the innermost stable circular orbit (ISCO) has been a touchstone for
comparing various analytical descriptions of binary dynamics. Using a
multi-pronged analysis of our accurate new data, we have been able to
improve the computation of the  $O(\nu)$ ISCO shift   {\it by four orders
of magnitude}---see Table \ref{table:ISCOshift}.  We have also expressed our improved result in
terms of the combination $\mathsf a (1/6)$, Eq.\ (\ref{sfa1/6}), of
derivatives of the EOB potential, thereby providing a direct, accurate way
of calibrating the EOB formalism in the $\nu \to 0$ limit---see Eq.\ (\ref{ISCOsfa}). 
In addition, for further helping the construction of  $O(\nu)$-accurate
EOB Hamiltonians, we have given accurate
numerical estimates of the values of $a(u)$ and its first {\it three}
derivatives at the ISCO point $u=\frac16$: Eq.\ (\ref{aISCO}).

(7)  In previous work we used GSF data on slightly eccentric orbits to
compute a certain linear combination
of $a(u)$ and its first two derivatives, involving also the $O(\nu)$ piece
of a second EOB radial potential ${\bar D}(u)= 1 + \nu \, {\bar d} (u)
+O(\nu^2)$.  Combining these results with our new accurate  global
analytic representation of $a(u)$, we numerically
 computed $\bar d (u)$ on the interval $0<u\leq \frac{1}{6}$.

(8)  Our finding of an inverse-square-root singularity $a(u) \propto
(1-3u)^{-1/2}$ in the $O(\nu)$ EOB potential seems to put in question  the
domain of validity of the GSF expansion, and/or that of the EOB formalism.
 We addressed both issues in detail.  First, we argued that $O(\nu)$ GSF
results are physically reliable near the LR only in a limit where the
ratio $\nu/(1-3u)^{3/2}$ tends to zero. This limit allows for the
unboundedness, near $u=\frac13$, of second (and higher) $u$-derivatives of
 the EOB potential $A(u,\nu)$, and thereby signals the presence of some
type of singularity of the (standard)  EOB formalism at  $u=\frac13$ .
However, we argued that  the (mathematical) singularity  $a(u) \propto
(1-3u)^{-1/2}$ we found is only a {\it spurious singularity}, due to the
use, in the current, {\it standard} EOB formalism, of some specific way of
fixing the phase-space gauge freedom.  [The $a(u) \propto (1-3u)^{-1/2}$
singularity is a phase-space analog of, e.g., the $g_{r r} =
(1-2M/r)^{-1}$ ``Schwarzschild coordinate singularity'' at $r=2M$.] We
explicitly showed (at lowest order in $p_r$) how to ``gauge-away'' the
singularity $a(u) \propto (1-3u)^{-1/2}$ by relaxing the standard,
phase-space gauge-fixing conditions of the EOB Hamiltonian (namely by
allowing the third, $Q$ EOB potential to grow $\propto p_{\varphi}^3$ when
$p_r=0$ and $ p_{\varphi} \to \infty$). In addition, we exhibited {\it
minimal} ways of modifying the current EOB gauge-fixing, which are
appropriate when dealing with radiation-driven inspiralling and coalescing
binaries, rather than with highly unbound ultra-relativistic circular
orbits near the LR.  In order to globally construct these modifications of
the current EOB formalism, it is essential to make use of our finding
that, after factoring $E(u)$ out of $a(u)$,  one ends up with a function
that is continuous at $u=\frac13$, and can be naturally extended to {\it
larger} values of $u$.

Finally, let us make the following remarks about some of the future
research directions that suggest themselves to complete our results.

(a) First, it would be useful to improve the accuracy of the GSF data near the LR, in order to allow a better characterization of the behavior there. This would likely require a reformulation of the mode-sum scheme to achieve a more rapid convergence of the multipole mode-sum near the LR, where the standard high-$l$ behavior is no longer applicable. 

(b) It would be interesting to extend the GSF computation of the
precession of slightly eccentric orbits so as to extend the range of
determination of the second EOB potential, $\bar d (u)$, from the current
range $0<u<\frac16$ to the full range  $0<u<\frac13$ where it is, in
principle, computable.

(c) Several aspects of our work have emphasized the need for an
understanding of higher-order terms in the GSF expansion, notably terms
$O(\nu^2)$.  This provides a motivation for pushing more effort in this
direction.

(d)  Our work has also emphasized the importance of being able to extract
gauge-invariant dynamical information about plunging orbits.

(e) Finally, it would be interesting to study the performances and
relative merits of the various non-standard EOB schemes whose necessity is
suggested by our work.

\section*{Acknowledgements} 
We are grateful to Alessandro Nagar for valuable advice on model fitting using computer algebra.
SA and LB acknowledge support from STFC through grant numbers PP/E001025/1 and
ST/J00135X/1. LB also acknowledges support from the European Research Council under grant No.\ 304978. NS acknowledges support by the Grant-in-Aid for Scientific Research (No.\ 21244033).

\appendix

\section{Numerical data} \label{AppA}

We tabulate here the complete set of GSF numerical data used in our analysis. We show numerical values for the Lorenz-gauge quantities $h_{uu}^{R,L}(x)$ sampled at (generally) equal intervals in $x$ in the range $1/150\leq x\leq 1/3$, corresponding to $3m_2<r\leq 150m_2$. For practical reasons we split the data between Table \ref{table:data1} ($x<1/6$) and Table \ref{table:data2} ($x\geq 1/6$). The tables also show the corresponding values of the $O(\nu)$ EOB potential $a(x)$ derived from $h_{uu}^{R,L}(x)$ using Eq.\ (\ref{F18}). Our data for $x>1/5$ is new. A subset of the $x\leq 1/5$ data already appeared in the literature \cite{Detweiler:2008ft,SBD} but at significantly lower accuracy. Reference \cite {Blanchet:2010zd}, which was concerned primarily with the  PN domain, presented a sample of very high accuracy data for the weak-field range $1/500\leq x\leq 1/200$. As we are interested here in the {\it global} behavior (in particular in the strong-field domain), we do not include these high-accuracy large-$r$ points in our sample to avoid statistical bias in our $\chi^2$ analysis.  All of our data points are consistent with previously published results within the respective error bars (where quoted). [To see the agreement with  Refs.\ \cite{Detweiler:2008ft} or \cite{Blanchet:2010zd} one needs to use our Eq.\ (\ref{FtoL}) in order to convert between the Lorenz-gauge values given in our tables and the flat-gauge values given in those sources.]

To allow for a meaningful $\chi^2$ analysis in the present work, it was important for us to obtain a reliable estimate of the numerical error in the data points. Our methods for error estimation are described in detail in Refs.\ \cite{Barack:2010tm} and \cite{Akcay:2010dx}. For most data points the error is by far dominated by the uncertainty in the value of the analytically fitted large-$l$ tail contribution to the regularized metric. Essentially, we use some of the our large-$l$ numerical data points to fit a power-law model using several plausible models (varying over the number of power-law terms and the number of data points used for the fit), and use the variance of the results as a rough measure of the tail-fit error. See \cite{Barack:2010tm,Akcay:2010dx} for more details. We expect this procedure to give us the actual error to within a factor $\sim 2$ or so. (This is indeed confirmed by our $\chi^2$ analysis: we find that the value of $\chi^2$/DoF settles at around 3-4 and does not reduce any further upon adding model parameters.)

The parenthetical figures in Tables \ref{table:data1} and \ref{table:data2} correspond to the error estimates coming from the above procedure. For instance, $0.02693868484(1)$ stands for $\sim 0.02693868484\pm 10^{-11}$. In this example, $\pm 10^{-11}$ describes our ``best guess'' for the numerical error, although a more conservative approach (taking into account the uncertainty in the error itself) would perhaps set this at $\pm 2\times 10^{-11}$. Note that the tables also quote (in the fifth column) more ``precise'' values for the numerical errors, given to 3 places right of the decimal point. Strictly speaking, this extra information is, of course, meaningless, given the factor $\sim 2$ uncertainty in the errors; we present it here only for the purpose of allowing interested readers to fully reproduce our $\chi^2$ analysis.

\begin{table}[htb] 
    \begin{tabular}{ cccccc }
    \hline
300$x$ &  $z$ &    $r/m_2$ &        $q^{-1} h_{uu}^{R,L}(x)$   & Num.\ err. &  $ a(x) $  \\
  \hline\hline
2.0  &  0.98 & 150.000000 & $ -0.02693868484(1) $ 	& $1.000\times 10^{-11}$& $ 0.000000628989(5) $  \\  
3.0  &  0.97 & 100.000000 & $ -0.04061834870(1) $ 	& $1.000\times 10^{-11}$& $ 0.000002183516(5) $  \\  
4.0  &  0.96 & 75.000000  & $ -0.05444418654(1) $ 	& $1.000\times 10^{-11}$& $ 0.000005318948(5) $  \\  
5.0  &  0.95 & 60.000000  & $ -0.06842101467(3) $ 	& $3.367\times 10^{-11}$& $ 0.00001066749(2) $  \\  
6.0  &  0.94 & 50.000000  & $ -0.08255394376(4) $ 	& $4.464\times 10^{-11}$& $ 0.00001891471(2) $  \\  
7.0  &  0.93 & 42.857143  & $ -0.09684839467(3) $ 	& $3.365\times 10^{-11}$& $ 0.00003079997(2) $  \\  
8.0  &  0.92 & 37.500000  & $ -0.11131011555(2) $ 	& $2.282\times 10^{-11}$& $ 0.00004711690(1) $  \\  
9.0  &  0.91 & 33.333333  & $ -0.12594520104(2) $ 	& $2.497\times 10^{-11}$& $ 0.00006871439(1) $  \\  
10.0  &  0.90 & 30.000000 & $ -0.14076011159(2) $ 	& $2.255\times 10^{-11}$& $ 0.00009649701(1) $  \\  
11.0  &  0.89 & 27.272727 & $ -0.15576169693(2) $ 	& $2.492\times 10^{-11}$& $ 0.00013142683(1) $  \\  
12.0  &  0.88 & 25.000000 & $ -0.17095721902(5) $ 	& $5.184\times 10^{-11}$& $ 0.00017452421(2) $  \\  
13.0  &  0.87 & 23.076923 & $ -0.18635437743(3) $ 	& $2.506\times 10^{-11}$& $ 0.00022686906(1) $  \\  
14.0  &  0.86 & 21.428571 & $ -0.20196133845(3) $ 	& $2.648\times 10^{-11}$& $ 0.00028960291(1) $  \\  
15.0  &  0.85 & 20.000000 & $ -0.21778676312(3) $ 	& $2.688\times 10^{-11}$& $ 0.00036392975(1) $  \\  
16.0  &  0.84 & 18.750000 & $ -0.23383984256(3) $ 	& $2.569\times 10^{-11}$& $ 0.00045111905(1) $  \\  
17.0  &  0.83 & 17.647059 & $ -0.25013033092(3) $ 	& $2.619\times 10^{-11}$& $ 0.00055250677(1) $  \\  
18.0  &  0.82 & 16.666667 & $ -0.26666858464(3) $ 	& $3.039\times 10^{-11}$& $ 0.00066949804(1) $  \\  
19.0  &  0.81 & 15.789474 & $ -0.28346560363(3) $ 	& $2.900\times 10^{-11}$& $ 0.00080356947(1) $  \\  
20.0  &  0.80 & 15.000000 & $ -0.30053307633(4) $ 	& $4.293\times 10^{-11}$& $ 0.00095627173(2) $  \\  
21.0  &  0.79 & 14.285714 & $ -0.31788342798(3) $ 	& $2.637\times 10^{-11}$& $ 0.00112923221(1) $  \\  
22.0  &  0.78 & 13.636364 & $ -0.33552987395(4) $ 	& $4.052\times 10^{-11}$& $ 0.00132415814(2) $  \\  
23.0  &  0.77 & 13.043478 & $ -0.35348647703(4) $ 	& $3.569\times 10^{-11}$& $ 0.00154283972(1) $  \\  
24.0  &  0.76 & 12.500000 & $ -0.37176820994(3) $ 	& $3.197\times 10^{-11}$& $ 0.00178715358(1) $  \\  
25.0  &  0.75 & 12.000000 & $ -0.39039102352(4) $ 	& $3.700\times 10^{-11}$& $ 0.00205906652(1) $  \\  
26.0  &  0.74 & 11.538462 & $ -0.40937192102(3) $ 	& $3.419\times 10^{-11}$& $ 0.00236063948(1) $  \\  
27.0  &  0.73 & 11.111111 & $ -0.42872903910(3) $ 	& $3.237\times 10^{-11}$& $ 0.00269403185(1) $  \\  
28.0  &  0.72 & 10.714286 & $ -0.44848173631(3) $ 	& $3.194\times 10^{-11}$& $ 0.00306150609(1) $  \\  
29.0  &  0.71 & 10.344828 & $ -0.46865068987(4) $ 	& $3.505\times 10^{-11}$& $ 0.00346543262(1) $  \\  
30.0  &  0.70 & 10.000000 & $ -0.48925800172(4) $ 	& $3.820\times 10^{-11}$& $ 0.00390829530(1) $  \\  
31.0  &  0.69 & 9.677419 & $ -0.51032731469(4) $ 	& $3.919\times 10^{-11}$& $ 0.00439269707(1) $  \\  
32.0  &  0.68 & 9.375000 & $ -0.53188393996(4) $ 	& $3.507\times 10^{-11}$& $ 0.00492136623(1) $  \\  
33.0  &  0.67 & 9.090909 & $ -0.55395499707(4) $ 	& $4.085\times 10^{-11}$& $ 0.00549716304(1) $  \\  
34.0  &  0.66 & 8.823529 & $ -0.57656956839(4) $ 	& $3.895\times 10^{-11}$& $ 0.00612308706(1) $  \\  
35.0  &  0.65 & 8.571429 & $ -0.59975886869(3) $ 	& $3.494\times 10^{-11}$& $ 0.00680228486(1) $  \\  
36.0  &  0.64 & 8.333333 & $ -0.62355643293(4) $ 	& $3.601\times 10^{-11}$& $ 0.00753805854(1) $  \\  
37.0  &  0.63 & 8.108108 & $ -0.64799832304(4) $ 	& $3.628\times 10^{-11}$& $ 0.00833387474(1) $  \\  
38.0  &  0.62 & 7.894737 & $ -0.67312335749(4) $ 	& $3.980\times 10^{-11}$& $ 0.00919337468(1) $  \\  
39.0  &  0.61 & 7.692308 & $ -0.69897336535(5) $ 	& $4.992\times 10^{-11}$& $ 0.01012038486(2) $  \\  
40.0  &  0.60 & 7.500000 & $ -0.72559346831(4) $ 	& $4.231\times 10^{-11}$& $ 0.01111892870(1) $  \\  
41.0  &  0.59 & 7.317073 & $ -0.75303239481(5) $ 	& $4.703\times 10^{-11}$& $ 0.01219323936(1) $  \\  
42.0  &  0.58 & 7.142857 & $ -0.78134282910(5) $ 	& $4.585\times 10^{-11}$& $ 0.01334777348(1) $  \\  
43.0  &  0.57 & 6.976744 & $ -0.81058180205(5) $ 	& $4.826\times 10^{-11}$& $ 0.01458722644(1) $  \\  
44.0  &  0.56 & 6.818182 & $ -0.84081112697(6) $ 	& $6.060\times 10^{-11}$& $ 0.01591654886(2) $  \\  
45.0  &  0.55 & 6.666667 & $ -0.87209788852(6) $ 	& $5.738\times 10^{-11}$& $ 0.01734096473(2) $  \\ 
46.0  &  0.54 & 6.521739 & $ -0.90451499141(5) $ 	& $4.813\times 10^{-11}$& $ 0.01886599135(1) $  \\  
47.0  &  0.53 & 6.382979 & $ -0.93814177780(6) $ 	& $5.528\times 10^{-11}$& $ 0.02049746125(1) $  \\  
48.0  &  0.52 & 6.250000 & $ -0.97306472296(5) $ 	& $4.990\times 10^{-11}$& $ 0.02224154634(1) $  \\  
49.0  &  0.51 & 6.122449 & $ -1.00937822184(7) $ 	& $6.524\times 10^{-11}$& $ 0.02410478457(2) $  \\  
    \hline\hline
    \end{tabular}
    \vspace{-3mm}
\caption{
Numerical data (part I). Each row displays data for a circular geodesic with a particular Schwarzschild radius $r$, given in the third column. The first and second columns show the corresponding values of $x=m_2/r$ and $z=1-3m_2/r$. [The relation $x=m_2/r$ holds only at $O(\nu^0)$, but here we may ignore higher-order corrections because $x$ and $r$ are used as arguments (independent variables) for quantities which are already $O(\nu)$, namely $h_{uu}^{R,L}(x)$ and $\nu a(x)$.] The fourth and last columns give, respectively, our numerical results for the Lorenz-gauge quantity $h_{uu}^{R,L}$  [see Eq.\ (\ref{huu})] and for the $O(\nu)$ EOB potential $a(x)$ [see Eq.\ (\ref{F18})].  In the fifth column we give estimates of the absolute numerical errors in the $h_{uu}^{R,L}$ data. These error values are the ones used in our $\chi^2$ analysis, and we give them here in full (showing several insignificant digits) in order to allow readers to reproduce this analysis accurately.  Our actual error estimates for the individual data points (which we expect to be only accurate to within a factor $2$ or so) are expressed in the form of parenthetical figures in the fourth and last columns, showing our best estimate of the uncertainty in the last displayed decimals.}
\label{table:data1}
\end{table}

\begin{table}[htb] 
    \begin{tabular}{cccccc}
    \hline
$300x$ &  $z$ &    $r/m_2$ &        $h_{uu}^{R,L}(x) $   & Num.\ err.  &  $ a(x) $   \\
  \hline\hline
50.0  &  0.50 & 6.000000 & $ -1.0471854796(1) $ 	& $1.054\times 10^{-10}$& $ 0.02609410950(3) $  \\  
51.0  &  0.49 & 5.882353 & $ -1.08659952251(6) $ 	& $6.245\times 10^{-11}$& $ 0.02821688301(2) $  \\  
52.0  &  0.48 & 5.769231 & $ -1.12774434980(7) $ 	& $7.185\times 10^{-11}$& $ 0.03048093197(2) $  \\  
53.0  &  0.47 & 5.660377 & $ -1.17075624628(7) $ 	& $7.489\times 10^{-11}$& $ 0.03289458866(2) $  \\  
54.0  &  0.46 & 5.555556 & $ -1.21578528730(6) $ 	& $5.720\times 10^{-11}$& $ 0.03546673669(1) $  \\  
55.0  &  0.45 & 5.454545 & $ -1.26299706191(6) $ 	& $6.359\times 10^{-11}$& $ 0.03820686140(1) $  \\  
56.0  &  0.44 & 5.357143 & $ -1.31257466418(8) $ 	& $7.999\times 10^{-11}$& $ 0.04112510844(2) $  \\  
57.0  &  0.43 & 5.263158 & $ -1.36472098296(8) $ 	& $7.670\times 10^{-11}$& $ 0.04423234741(2) $  \\  
58.0  &  0.42 & 5.172414 & $ -1.4196613648(1) $ 	& $9.639\times 10^{-11}$& $ 0.04754024627(2) $  \\  
59.0  &  0.41 & 5.084746 & $ -1.47764670375(8) $ 	& $7.896\times 10^{-11}$& $ 0.05106135426(2) $  \\  
60.0  &  0.40 & 5.000000 & $ -1.5389570493(1) $ 	& $1.137\times 10^{-10}$& $ 0.05480919704(2) $  \\  
61.0  &  0.39 & 4.918033 & $ -1.60390583535(9) $ 	& $9.043\times 10^{-11}$& $ 0.05879838596(2) $  \\  
62.0  &  0.38 & 4.838710 & $ -1.6728448488(1) $ 	& $1.087\times 10^{-10}$& $ 0.06304474249(2) $  \\  
63.0  &  0.37 & 4.761905 & $ -1.7461701015(1) $ 	& $1.107\times 10^{-10}$& $ 0.06756544251(2) $  \\  
64.0  &  0.36 & 4.687500 & $ -1.8243287924(1) $ 	& $1.355\times 10^{-10}$& $ 0.07237918263(2) $  \\  
65.0  &  0.35 & 4.615385 & $ -1.9078276091(1) $ 	& $1.351\times 10^{-10}$& $ 0.07750637433(2) $  \\  
66.0  &  0.34 & 4.545455 & $ -1.9972426612(1) $ 	& $1.363\times 10^{-10}$& $ 0.08296936903(2) $  \\  
67.0  &  0.33 & 4.477612 & $ -2.0932314430(1) $ 	& $1.371\times 10^{-10}$& $ 0.08879272321(2) $  \\  
68.0  &  0.32 & 4.411765 & $ -2.1965473024(1) $ 	& $1.373\times 10^{-10}$& $ 0.09500350907(2) $  \\  
69.0  &  0.31 & 4.347826 & $ -2.3080570549(1) $ 	& $1.375\times 10^{-10}$& $ 0.10163168283(2) $  \\  
70.0  &  0.30 & 4.285714 & $ -2.4287625435(1) $ 	& $1.401\times 10^{-10}$& $ 0.10871052135(2) $  \\  
72.0  &  0.28 & 4.166667 & $ -2.7026089937(1) $ 	& $1.445\times 10^{-10}$& $ 0.12437313326(2) $  \\  
74.0  &  0.26 & 4.054054 & $ -3.0299861727(1) $ 	& $1.496\times 10^{-10}$& $ 0.14234657312(2) $  \\  
76.0  &  0.24 & 3.947368 & $ -3.4275063529(1) $ 	& $1.471\times 10^{-10}$& $ 0.16308580174(2) $  \\  
78.0  &  0.22 & 3.846154 & $ -3.9189791853(2) $ 	& $1.627\times 10^{-10}$& $ 0.18718609087(2) $  \\  
80.0  &  0.20 & 3.750000 & $ -4.5395895523(2) $ 	& $1.609\times 10^{-10}$& $ 0.21544503763(2) $  \\  
82.0  &  0.18 & 3.658537 & $ -5.3432356852(6) $ 	& $6.309\times 10^{-10}$& $ 0.24896018744(6) $  \\  
84.0  &  0.16 & 3.571429 & $ -6.416105450(1) $ 		& $1.404\times 10^{-9}$& $ 0.2892884360(1) $  \\  
86.0  &  0.14 & 3.488372 & $ -7.903439422(2) $ 		& $1.905\times 10^{-9}$& $ 0.3387190693(1) $  \\  
88.0  &  0.12 & 3.409091 & $ -10.066672801(2) $ 	& $1.559\times 10^{-9}$& $ 0.40077307330(9) $  \\  
90.0  &  0.10 & 3.333333 & $ -13.4187934749(9) $ 	& $9.284\times 10^{-10}$& $ 0.48120301414(5) $  \\  
91.0  &  0.09 & 3.296703 & $ -15.849341504(3) $ 	& $2.644\times 10^{-9}$& $ 0.5312203677(1) $  \\  
92.0  &  0.08 & 3.260870 & $ -19.093469318(8) $ 	& $8.388\times 10^{-9}$& $ 0.5902619091(3) $  \\  
93.0  &  0.07 & 3.225806 & $ -23.580398048(7) $ 	& $7.359\times 10^{-9}$& $ 0.6612773504(3) $  \\  
94.0  &  0.06 & 3.191489 & $ -30.07746738(2) $ 		& $2.200\times 10^{-8}$& $ 0.7488226643(7) $  \\  
95.0  &  0.05 & 3.157895 & $ -40.08190371(5) $ 		& $4.853\times 10^{-8}$& $ 0.860429954(1) $  \\  
96.0  &  0.04 & 3.125000 & $ -56.8876661(5) $ 		& $4.832\times 10^{-7}$& $ 1.00975332(1) $  \\  
97.0  &  0.03 & 3.092784 & $ -89.13862(2) $ 		& $1.684\times 10^{-5}$& $ 1.2250733(3) $  \\  
97.5  &  0.025 & 3.076923 & $ -118.32926(2) $ 		& $1.775\times 10^{-5}$& $ 1.3763418(2) $  \\  
98.0  &  0.02 & 3.061224 & $ -167.13828(2) $ 		& $2.138\times 10^{-5}$& $ 1.5789875(2) $  \\  
1575/16 & 1/64 & 3.047619 & $ -244.5136(1) $ 		& $1.271\times 10^{-4}$& $ 1.828231(1) $  \\  
98.5  &  0.015 & 3.045685 & $ -260.3517(2) $ 		& $1.914\times 10^{-4}$& $ 1.872213(1) $  \\  
99.0  &  0.01 & 3.030303 & $ -484.6(5) $ 			& $5.398\times 10^{-1}$& $ 2.357(3) $  \\  

    \hline\hline
    \end{tabular}
    
\caption{Numerical data (part II), covering the sub-ISCO range $1/6\leq x<1/3$. The table is structured in the same way as Table \ref{table:data1}.}
\label{table:data2}
\end{table}

\end{document}